\documentclass[sigconf,nonacm]{acmart} 
\usepackage{balance}
\usepackage{mathtools}
\usepackage{enumitem}
\usepackage{bigstrut}
\usepackage{arydshln}
\usepackage{subfig}
\usepackage{color}
\usepackage{graphicx}
\usepackage{tikz-cd}
\usepackage{algpseudocode}

\usetikzlibrary{automata, positioning,arrows,shapes,decorations.pathmorphing}

\definecolor{mycolor}{rgb}{0.122, 0.435, 0.698}
\makeatletter
\newcommand{\mybox}[1]{%
  \setbox0=\hbox{#1}%
  \setlength{\@tempdima}{\dimexpr\wd0+13pt}%
  \begin{tcolorbox}[colframe=mycolor,boxrule=0.5pt,arc=4pt,
      left=6pt,right=6pt,top=3pt,bottom=3pt,boxsep=0pt,width=\@tempdima]
    #1
  \end{tcolorbox}
}

\newcommand{\introparagraph}[1]{\vspace{0.7mm} \noindent \textbf{\em #1.}}

\newcommand{\rebut}[1]{{\color{black}{#1}}} 

\newcommand{\rebuttal}[1]{{\color{black}{#1}}} 

\newcommand*{\defeq}{\stackrel{\text{def}}{=}}

\newcommand{\mL}{\mathcal{L}}

\newcommand{\mH}{\mathcal{H}}    

\newcommand{\mV}{\mathcal{V}}
\newcommand{\mE}{\mathcal{E}}
\newcommand{\mM}{\mathcal{M}}

\newcommand{\bS}{\mathbf{S}}

\newcommand{\cqneq}{CQ^{\neg}}

\newtheorem{claim}{Claim}[section]

\usepackage[longend,ruled,vlined, linesnumbered]{algorithm2e}
\usepackage{relsize,amsmath} 
\usepackage{bbm}
\usepackage{tikz}
\usepackage{forest}
\usepackage{multirow}
\usepackage[export]{adjustbox}
\usepackage{colortbl}

\usetikzlibrary{calc,shapes.multipart,chains,arrows}

\newcommand{\dom}{\mathsf{Dom}}

\newcommand{\AST}{\mathsf{AST}}
\providecommand{\bx}[0]{\mathbf{x}}
\providecommand{\ba}[0]{\mathbf{a}}
\providecommand{\bc}[0]{\mathbf{c}}

\newcommand{\subw}{\mathsf{subw}}
\newcommand{\CFG}{\textsf{CFG}}
\newcommand{\Rewrite}{\mathsf{Refactor}}

\newcommand{\BuildOracle}{\mathsf{BuildOracle}}

\newcommand{\RangeSum}{\mathsf{RangeSum}}
\newcommand{\RangeSumOracle}{\mathsf{RangeSumOracle}}
\newcommand{\derive}{\stackrel{*}{\Rightarrow}}
\newcommand{\yield}{\Rightarrow}

\newcommand{\ackfunc}[1]{\alpha(14 |#1|, |#1|)}

\providecommand{\arraysize}[0]{w}

\newcommand{\sfa}{\mathsf{a}}
\newcommand{\sfb}{\mathsf{b}}
\newcommand{\sfc}{\mathsf{c}}
\newcommand{\sfd}{\mathsf{d}}

\newcommand{\formula}{\psi}
\newcommand{\nonterm}{\ell}
\newcommand{\zerobf}{\mathbf{0}}
\newcommand{\onebf}{\mathbf{1}}
\newcommand{\db}{\mathcal{D}}

\newcommand{\dbsize}{|\mathcal{D}|}
\newcommand{\interval}{\omega}
\newcommand{\open}{\omega^-}
\newcommand{\close}{\omega^+}
\newcommand{\Oracle}{\mathcal{T}}

\providecommand{\bt}[0]{\mathbf{t}}
\providecommand{\bu}[0]{\mathbf{u}}
\providecommand{\bv}[0]{\mathbf{v}}

\newcommand{\Enum}{\mathsf{Enum}}

\newcommand{\absorb}[1]{\widehat{#1}}

\newcommand{\oeps}{\textsf{OEPS}}

\newcommand{\cqneg}{\mathsf{CQ}^{\neg}}
\newcommand{\faqneg}{\mathsf{FAQ}^{\neg}}
\newcommand{\nestfaqneg}{\mathsf{NestFAQ}^{\neg}}

\newcommand{\indicator}[1]{\mathbbm{1}_{{#1}}}
\newcommand{\bmm}{\mathsf{BMM}}

\newcommand{\mT}{\mathcal{T}}

\newcommand{\rmvertex}[2]{{#1}{[\setminus{#2}]}}
\newcommand{\pn}{\text{signed}}
\newcommand{\removeleaf}[2]{\langle {#1}, {#2} \rangle}
\newcommand{\reduced}[1]{[{#1}] }

\newcommand{\myparagraph}[1]{\noindent \textbf{#1}}
\newcommand{\card}[1]{|{#1}|}

\newcommand{\node}[1]{\mathsf{node}_{#1}}

\newcommand{\prevptr}[0]{\mathsf{prev}}
\newcommand{\nextptr}[0]{\mathsf{next}}

\newcommand{\rowoftwo}[2]{{${#1}$} & {${#2}$} }
\newcommand{\rowofthree}[3]{{${#1}$} & {${#2}$} & {${#3}$} }
\newcommand{\rowoffour}[4]{{${#1}$} & {${#2}$} & {${#3}$} & {${#4}$}}

\AtBeginDocument{%
  }

\begin{document}

\title{Conjunctive Queries with Negation and Aggregation: A Linear Time Characterization}


\author{Hangdong Zhao}
\affiliation{%
    \institution{University of Wisconsin--Madison}
    \city{Madison}
    \country{USA}
}
\email{hangdong@cs.wisc.edu}

\author{Austen Z. Fan}
\affiliation{%
  \institution{University of Wisconsin--Madison}
  \city{Madison}
  \country{USA}
  }
\email{afan@cs.wisc.edu}

\author{Xiating Ouyang}
\affiliation{%
  \institution{University of Wisconsin--Madison}
  \city{Madison}
  \country{USA}
  }
\email{xouyang@cs.wisc.edu}

\author{Paraschos Koutris}
  \affiliation{%
    \institution{University of Wisconsin--Madison}
    \city{Madison}
    \country{USA}
    }
  \email{paris@cs.wisc.edu}



\begin{abstract}
In this paper, we study the complexity of evaluating Conjunctive Queries with negation ($\cqneg$). 
First, we present an algorithm with linear preprocessing time and constant delay enumeration for a class of CQs with negation
called free-connex signed-acyclic queries. We show that no other queries admit such an algorithm subject to 
lower bound conjectures.  Second, we extend our algorithm to Conjunctive Queries with negation and aggregation over a
general semiring, which we call Functional Aggregate Queries with negation ($\faqneg$). Such an algorithm achieves constant delay enumeration for the same class of queries,
but with a slightly increased preprocessing time which includes an inverse Ackermann function. We show that this surprising appearance
of the Ackermmann function is probably unavoidable for general semirings, but can be removed when the semiring has specific structure. 
Finally, we show an application of our results to computing the difference of CQs.
\end{abstract}



%


\maketitle

\section{Introduction}
\label{sec:intro}

This paper focuses on the query evaluation problem for Conjunctive Queries with negation ($\cqneg$), a fundamental class of relational queries. We will think of a $\cqneg$ as having the following form:
\begin{align} \label{eq:cqneg}
Q(\bx_F) \leftarrow \bigwedge_{K \in \mE^+} R_K(\bx_K) \wedge  \bigwedge_{K \in \mE^-} \neg R_K(\bx_K)
\end{align}
where the variables are of the form $x_i$, $i \in [n] = \{1, 2, \dots, n\}$, $\mE^+, \mE^-$ are two sets of hyperedges that are subsets of $[n]$, and
$F \subseteq [n]$ are the free variables. We will call the triple $\mH = ([n], \mE^+, \mE^-)$  the {\em signed hypergraph} of $Q$.
We will consider only {\em safe queries}, where $\bigcup_{K \in \mE^+} K = [n]$.
When $\mE^-  = \emptyset$, then $Q$ is a Conjunctive Query (CQ) with the associated hypergraph $([n], \mE^+)$.

The complexity of query evaluation for CQs is well-understood. Yannakakis~\cite{Yannakakis81} first showed that a Boolean CQ (i.e., $F = \emptyset$) can be evaluated
on a database $\db$ of size $|\db|$ in time $O(|Q| \cdot |\db|)$ if the hypergraph $([n], \mE^+)$ is $\alpha$-acyclic ($|Q|$ denotes the size of the query). Further work~\cite{BaganDG07,BerkholzGS20} generalized this result to show that if $Q$ is {\em free-connex $\alpha$-acyclic} then the tuples in $Q(\db)$ can be enumerated with constant delay $O(|Q|)$ after a linear-time preprocessing step. Free-connex $\alpha$-acyclicity means that both $([n], \mE^+)$ and $([n], \mE^+ \cup \{F\})$ are $\alpha$-acyclic hypergraphs. It was also shown~\cite{BaganDG07} that this tractability result is tight under widely believed lower-bound conjectures.

\subsection{CQs with Negation}

Our first goal in this paper is to generalize the above classic result to the case where $\mE^- \neq \emptyset$. Prior work has looked into this problem, but without achieving a complete answer.

To explain the current progress, let us first consider the case of a Boolean $\cqneg$ and attempt to solve the (seemingly harder) problem of counting the number of valuations that satisfy the body of $Q$ with input a database $\db$, which we will denote as $\# Q(\db)$. We should note here that $\# Q$ is solvable in $O(|\db|)$ time if $Q$ is an $\alpha$-acyclic CQ. Brault-Baron~\cite{BBThesis} had the insight that we can compute $\# Q$ using the inclusion-exclusion principle. Indeed, let $Q_S$ be the Boolean CQ with hypergraph $([n], \mE^+ \cup S)$ for any $S \subseteq \mE^-$. Then we can write:
\begin{align} \label{eq:inc-excl}
\# Q(\db) = \sum_{S \subseteq \mE^-} (-1)^{|S|} \# Q_S(\db) 
\end{align}
Hence, if every $Q_S$ is $\alpha$-acyclic, then $\# Q$ (and thus $Q$) can be computed with data complexity $O(|\db|)$. This naturally leads to the notion of {\em signed acyclicity}, introduced in~\cite{BBThesis}: a Boolean $\cqneq$ is signed-acyclic if the hypergraph $([n], \mE^+ \cup S)$ is $\alpha$-acyclic for every $S \subseteq \mE^-$. Thus, a Boolean $\cqneg$ can be computed in linear time (data complexity) if it is signed-acyclic. Interestingly, if $\mE^+$ consists only of singleton hyperedges, then signed acyclicity is equivalent to $\beta$-acyclicity of the hypergraph $([n], \mE^-)$. However, there are two issues with applying the inclusion-exclusion approach. First, it has an exponential dependency on $|Q|$ and thus does not give a polynomial-time algorithm in combined complexity. Second, it cannot be used to provide any delay guarantees for the enumeration problem in non-Boolean queries. 
  
The second issue was partially addressed by Brault-Baron~\cite{Baron12,BBThesis}, who proposed an enumeration algorithm for any free-connex signed-acyclic $\cqneg$. However, this algorithm either achieves constant delay with a $O(|\db| \log^{|Q|} |\db|)$ preprocessing time, or achieves logarithmic delay with linear preprocessing time. The logarithmic factor is a consequence of the technique used, which translates a database instance to an instance over the Boolean domain.

Our first main result  shows that the translation to the Boolean domain is not necessary and in fact we can achieve both constant delay and linear time preprocessing for  free-connex signed-acyclic queries. Moreover, our algorithm has only a polynomial dependence on the size of the query.



\begin{theorem}\label{thm:boolean}
Let $Q$ be a free-connex signed-acyclic $\cqneg$. Then there is an algorithm that can enumerate the results of $Q(\db)$ with $O(|Q|^3 + {|Q|} \cdot |\db|)$ preprocessing time and $O(|Q|)$ delay.
\end{theorem}

\subsection{FAQ with Negation}

Our second goal is to study the evaluation of $\cqneg$ in the presence of aggregation. We do this by studying a more general problem, that of computing a $\cqneg$ under a general semiring, following the approach of \textsf{FAQ}s~\cite{DBLP:conf/pods/KhamisNR16}. More precisely, given a commutative semiring $\bS = (\boldsymbol{D}, \oplus, \otimes, \mathbf{0}, \mathbf{1})$, we define an $\faqneg$ as an expression of the form:
\begin{align} \label{def:faqneg}
\varphi(\bx_F) = \bigoplus_{\bx_{[n] \setminus F}} \; \bigotimes_{K \in \mE^+}  R_K(\bx_K) \otimes \bigotimes_{N \in \mE^-}  \overline{R}_N(\bx_N), 
\end{align}
Here, a {\em positive  factor} $R_K$ can be viewed as a table of entries of the form $\langle \bx_K, R_K(\bx_K)\rangle$ (where the weight of tuple $\bx_K$ is the value $R_K(\bx_K) \in \boldsymbol{D}$), and for entries not in the table, the weight is implicitly $\zerobf$. On the other hand, a {\em negative factor} $\overline{R}_N$ is a table of entries of the form $\langle \bx_K, R_N(\bx_N)\rangle$, and for entries not in the table, the weight is a default constant value $\mathbf{c}_N \neq \zerobf$. To recover the $\cqneg$ setting, we choose the Boolean semiring, and encode the values of the negative factor such that it is $\zerobf$ if the tuple is in the table, otherwise $\onebf$. Our semiring formulation is though much more general: the only difference between a negative and a positive factor in our setting is whether the "default" value of a tuple not in the table is $\zerobf$. This observation offers a novel angle to the semantics of negation, and is critical to our algorithms.
For $\faqneg$, we can show the following.

\begin{theorem}
\label{conj:main-upper}
{Let $\varphi$ be a free-connex signed-acyclic $\faqneg$ query over a commutative semiring $\bS$. Then there is an algorithm that can enumerate $\varphi$ with $O(|\varphi|^3 + {|\varphi|} \cdot |\db| \cdot \alpha(14 \cdot |\db|, |\db|))$ preprocessing time and $O(|\varphi|)$ delay. }
\end{theorem}

Here, $|\varphi|$ denotes the size of the query $\varphi$ and $\alpha(m, n)$ denotes the inverse Ackermann function, which grows extremely slowly as a bi-variate function of $m, n$. The appearance of this function in the runtime expression is surprising in our opinion. It occurs because the aggregation problem reduces to the well-studied problem of computing interval sums over an arbitrary semiring, called $\mathsf{RangeSum}$~\cite{ChazelleR91,Yao82}. If the semiring structure allows for linear preprocessing and constant-time answering for its $\mathsf{RangeSum}$ problem, then we can drop the Ackermann factor. Examples of such semirings are the Boolean semiring $(\{\textsf{true}, \textsf{false}\}, \vee, \wedge, \textsf{false}, \textsf{true})$, tropical semiring $(\mathbb{R}, \min, +, +\infty, 0)$, and semirings with additive inverse {(e.g. the counting ring over integers, i.e. $(\mathbb{Z}, +, \times, 0,1)$ which we use to count solutions)}.

\begin{theorem}
 {Let $\varphi$ be a free-connex signed-acyclic $\faqneg$ query over a commutative semiring $\bS$ with  additive inverse. Then there is an algorithm that can enumerate $\varphi$ with $O(|\varphi|^3 + {|\varphi|} \cdot |\db|)$ preprocessing time and $O(|\varphi|)$ delay.}
\end{theorem}



\subsection{Lower Bounds}

Our third goal is to match our linear-time upper bounds with lower bounds. In this direction, we show that under believable conjectures, any $\cqneg$ that is not free-connex $\pn$-acyclic does not admit an algorithm that can emit the first result in linear time (hence matching the upper bound). Our conditional lower bounds are somewhat weaker than the ones obtained for CQs because the presence of negation means that we cannot use the sparse version of some problems (e.g., detecting a triangle, or Boolean matrix multiplication). For $\faqneg$, we show stronger conditional lower bounds over the tropical semiring and counting ring based on weaker lower bound conjectures. Finally, we provide some evidence that the inverse Ackermann factor in the runtime for general semirings is unavoidable. In particular, we show that the query
$\varphi(x) = \bigoplus_{y} A(x) \otimes B(y) \otimes \overline{R}(x,y)$
corresponds to a variant of the offline $\mathsf{RangeSum}$ problem over a general $\oplus$ operator. Using this observation, we can modify a construction of Chazelle~\cite{ChazelleR91} to show a superlinear lower bound on the number of $\oplus$ operations necessary to compute $\varphi$.

\subsection{Difference of CQs}

Finally, we show that our algorithm for $\cqneg$ can be applied to obtain optimal algorithms with linear-time preprocessing and constant delay for the problem of computing the difference between two CQs of the form $Q_1 - Q_2$, which was recently studied by Hu and Wang~\cite{CQDiff}. 

%


\section{Preliminaries}
\label{sec:prelim}

\myparagraph{Hypergraphs.} 
A hypergraph is a pair $\mH = ([n], \mE)$ where $[n] = \{1, \dots, n\}$ is the set of vertices of $\mH$ and $\mE$ is a multiset\footnote{A multiset is a collection of elements each of which can occur multiple times} of hyperedges where each $K \in \mE$ is a nonempty subset of $[n]$. 

A signed hypergraph is a tuple $\mH = ([n], \mE^+, \mE^-)$ where $[n]$ is the set of vertices of $\mH$, $\mE^+$ and $\mE^-$ are two multisets of hyperedges where each hyperedge $K \in \mE^+$ (resp. $N \in \mE^-$) is a subset of $[n]$. We consider only {\em safe} signed hypergraphs, where every vertex in $[n]$ occurs in some hyperedge $K \in \mE^+$.

\medskip
\myparagraph{Semigroups and semirings.}
A triple $\bS = (\boldsymbol{D}, \oplus)$ is a semigroup if $\oplus$ is an associative binary operator over $\boldsymbol{D}$.
A tuple $\bS = (\boldsymbol{D}, \oplus, \otimes, \mathbf{0}, \mathbf{1})$ is a (commutative) semiring 
if $\oplus$ and $\otimes$ are commutative binary operators over $\boldsymbol{D}$ satisfying the following conditions:
(1) $(\boldsymbol{D}, \oplus)$ is a commutative monoid\footnote{In the usual semiring definition, we do not need the multiplicative monoid to be commutative.}with an additive identity, denoted by $\zerobf$; (2) $(\boldsymbol{D}, \otimes)$ is a commutative monoid with a multiplicative identity, denoted by $\onebf$; (3) $\otimes$ distributes over $\oplus$; and (4) For any element $e \in \boldsymbol{D}$, we have $e \otimes \zerobf = \zerobf \otimes e = \zerobf$.


\medskip
\myparagraph{Enumeration and Complexity.} 
In this paper, we study the enumeration problem for $\faqneg$ queries, $\Enum(\varphi, \db)$, which takes as input a $\faqneg$ query $\varphi$ and a database instance $\db$ and outputs a sequence of answers such that every {\em answer} in $\varphi(\db)$ is printed precisely once. 
An enumeration algorithm for $\Enum(\varphi, \db)$ may consist of two phases: 
\begin{itemize}
    \item (preprocessing phase) it constructs efficient data structures from $\varphi$ and $\db$; and
    \item (enumeration phase) it may access the data structures built during preprocessing, and emit the answers of $\varphi(\db)$ one by one, without repetitions.
\end{itemize}

We say that an enumeration algorithm enumerates with delay $O(\tau)$ if the time between the emission of any two consecutive answers (and the time to emit the first answer, and the time from the last answer to the end) is bounded by $O(\tau)$. In particular, we say that an enumeration algorithm is {\em constant-delay} if it enumerates with delay independent of the input database size $|\db|$. 

\medskip
\myparagraph{Model of computation.}
We adopt the {\it random-access machine (RAM)} as our computation model with $O(\log n)$-bit
words, which is standard in fine-grained complexity. The machine has read-only input registers and it contains the database and the query, read-write work memory registers, and write-only output registers. It is assumed that each register can store any tuple, and each tuple is stored in one register. The machine can perform all ``standard''~\footnote{This includes all arithmetic (e.g. $+, -, \div, *$) and logical operations \rebuttal{(see~\cite{HartmanisS74, SimonS92, GrandjeanJ22}).}}

In this paper, we consider the \emph{combined complexity} of $\cqneg$ and $\faqneg$, i.e. the complexity is measured in the size of both the query and the database instance.
For a $\faqneg$ query $\varphi$, we define $|\varphi|$ to be the sum of the arity of all factors in $\varphi$. 
For a database $\db$, we define $|\db|$ as sum of the size of all relations in $\db$.

\medskip
\myparagraph{Ackermann function.} The Ackermann function $A(m, n)$ for integers $m ,n \geq 0$ is recursively defined as follows:
\begin{enumerate}
\item $A(0, n) = n +1$;
\item $A(m+1, 0) = A(m, 1)$; and
\item $A(m+1, n+1) = A(m, A(m+1, n))$.
\end{enumerate}

It is known that the Ackermann function grows faster than any primitive recursive function and therefore is not itself primitive recursive. 
The inverse Ackermann function $\alpha(m, n)$ is defined as 
$$\alpha(m ,n) = \min \{ i \geq 1 : A(i, \lfloor \frac{m}{n} \rfloor) \geq \log_2(n)\}$$
and $\alpha(m ,n)$ grows very slowly. For example, we have $\alpha(n, n) < 5$ for any practical integer $n$.

\section{Signed Acyclicity}
\label{sec:pn-acyclic}

Before we introduce the definition of signed acyclicity, we first go over the notions of $\alpha$- and $\beta$-acyclicity.

A hypergraph $\mH = ([n], \mE)$ is $\alpha$-acyclic if there is a tree $\mT = (V(\mT), E(\mT))$ and a bijective function $\chi : \mE \rightarrow V(\mT)$ such that for every vertex $v \in [n]$, the set of nodes $\{\chi(K) \mid v \in K, K \in \mE\}$ induces a connected component in $\mT$. 
A vertex $v \in [n]$ is an $\alpha$-leaf of $\mH$ if the multiset $\{K \in \mE \mid v \in K\}$ contains a maximal element $K_v$ with respect to $\subseteq$. It is known that every $\alpha$-acyclic hypergraph has an $\alpha$-leaf~\cite{10.1145/2983573}.

A hypergraph $\mH = ([n], \mE)$ is $\beta$-acyclic if for any subset $\mE' \subseteq \mE$, $\mH' = ([n], \mE')$ is $\alpha$-acyclic.
A vertex $v \in [n]$ is a $\beta$-leaf of $\mH$ if the multiset $\{K \in \mE \mid x \in K\}$ can be linearly ordered by $\subseteq$. It is known that every $\beta$-acyclic hypergraph has a $\beta$-leaf~\cite{10.1145/2983573,brouwer1980super}.

We can now introduce the notion of signed acyclicity, slightly modified from~\cite{BBThesis} to take multisets into account.

\begin{definition}[Signed Acyclicity]
\label{defn:pn-acyclic}
A signed hypergraph $\mH = ([n], \mE^+, \mE^-)$ is \emph{signed-acyclic} if $([n], \mE^+ \cup \mE')$ is $\alpha$-acyclic for every multiset $\mE' \subseteq \mE^-$. 
\end{definition}

Similar to $\alpha$ and $\beta$-leaves, we can define $\pn$-leaves.

\begin{definition}[$\pn$-leaf]
\label{defn:pn-leaf}
Let $\mH = ([n], \mE^+, \mE^-)$ be a signed hypergraph. We say that $x \in [n]$ is a \emph{$\pn$-leaf} if there exists a hyperedge $U \in \mE^+$, called the pivot of $x$, such that: 
\begin{enumerate}
\item ($\alpha$-property) $K \subseteq U$ for every $K \in \mE^+$ that contains $x$; and
\item ($\beta$-property) the multiset $\{N \in \mE^- \mid x \in N, N \not\subseteq U\} \cup \{U\}$ can be linearly ordered by $\subseteq$ with $U$ being the minimal element.
\end{enumerate}
\end{definition}

The notion of a $\pn$-leaf degenerates to an $\alpha$-leaf when $\mE^- = \emptyset$ (since the $\beta$-property holds trivially) and reduces to a $\beta$-leaf when $\mE^+$ contains only singleton hyperedges (since the $\alpha$-property holds trivially and the pivot is a singleton set).
Recall that every $\alpha$-acyclic hypergraph has an $\alpha$-leaf and every $\beta$-acyclic hypergraph has a $\beta$-leaf. 
The next proposition should not be surprising.

\begin{proposition}
\label{prop:pn-acyclic-leaf}
Every signed-acyclic signed hypergraph has a $\pn$-leaf.
\end{proposition}

Next, we generalize the notion of an elimination sequence~\cite{DBLP:conf/pods/KhamisNR16} to signed hypergraphs.
Given $\mH$ and a $\pn$-leaf $v$ with a pivot $U$, we define $\removeleaf{\mH}{v}$ to be the hypergraph that $(i)$ removes from $\mH$ any (positive or negative) hyperedge $V$ that contains $v$ such that $V \subseteq U$, and $(ii)$ removes $v$ from any hyperedge that contains it. 
 
\begin{definition}[$\pn$-elimination sequence]
Let $\mH = ([n], \mE^+, \mE^-)$  be a  signed hypergraph. A vertex ordering $\sigma=(v_1, v_2, \dots, v_n)$ of all vertices in $[n]$ is a $\pn$-elimination sequence of $\mH$ if $v_i$ is a $\pn$-leaf of $\mH_i$ for every $i \in [n]$, where $\mH_n = \mH$, and $\mH_{j} = \removeleaf{\mH_{j+1}}{v_{j+1}}$ for $j = n-1, n-2, \dots, 1$.
\end{definition}

For a vertex ordering $\sigma = (v_1, v_2, \dots, v_{n-1}, v_n)$, we often denote $\sigma = \sigma' \cdot v_n$ where $\sigma' = (v_1, v_2,\dots,v_{n-1})$.
By definition, if $\sigma = \sigma' \cdot v$ is a signed-elimination sequence of a signed hypergraph $\mH$, then $\sigma'$ is a signed-elimination sequence of $\removeleaf{\mH}{v}$.

\begin{proposition}
\label{prop:removal-sequence}
Let $\mH = ([n], \mE^+, \mE^-)$ be a signed hypergraph. If $\mH$ is signed-acyclic, then 
\begin{enumerate}
\item $\mH$ has a $\pn$-elimination sequence; and
\item for every hyperedge $ \{u_1, u_2, \dots, u_k\}\in \mE^-$, $\mH$ has a $\pn$-elimination sequence of the form $(u_1, u_2, \dots, u_k, v_{k+1}, \dots, v_{n}).$
\end{enumerate}
\end{proposition}

\begin{example} \label{ex:pn:acyclic}
Consider the following $\cqneg$, which will function as our running example.
\begin{align*}
Q(x_1,x_2,x_3,x_4) & \gets  A(x_1,x_2,x_3) \land U(x_3, x_4) \land \\
& \lnot V(x_4) \land \lnot R(x_2,x_3,x_4) \land \lnot S(x_1,x_2,x_3,x_4)
\end{align*}
The vertex 4 is a $\pn$-leaf for the hypergraph $\mH$ of $Q$ with pivot the hyperedge $\{3,4\}$ that corresponds to the atom $U$.
The query corresponding to the resulting hypergraph $\removeleaf{\mH}{4}$ is:
$$Q'(x_1,x_2,x_3) =  A(x_1,x_2,x_3) \land U(x_3) \land \lnot R(x_2,x_3) \land \lnot S(x_1,x_2,x_3)$$
In fact, $Q$ is signed-acyclic with the $\pn$-elimination sequence $\sigma = (1,2,3,4)$.
\end{example}

We should remark here that our notion of $\pn$-leaf is equivalent to the notion of a leaf for signed hypergraphs as defined in~\cite{BBThesis} (we show this in \autoref{appendix:pn-acyclic}). However, for our purposes we need to define a slightly different elimination sequence, since the hypergraph after the removal of a vertex $x$ is defined differently.







\section{Enumeration of full $\cqneg$}
\label{sec:enumeration-full}

In this section, we present an algorithm that can enumerate the answers of a signed-acyclic full $\cqneg$ (where $F = [n]$) with constant delay after linear preprocessing time.
The correctness and runtime analysis of the algorithm is included in Appendix~\ref{appendix:enumeration-full-proof}.

Let $Q$ be a full signed-acyclic $\cqneg$ with a signed-acyclic signed hypergraph $\mH = ([n], \mE^+, \mE^-)$ and $\db$ a database instance. 
Let $\sigma = \sigma' \cdot v$ be a $\pn$-elimination sequence of $\mH$. 

Our preprocessing phase recursively eliminates a variable $v$ in the $\pn$-elimination sequence.
When eliminating $v$, the key idea is to construct in linear time a database $\db'$, and reduce the problem of computing the answers of $Q(\db)$ to computing the answers of $Q'(\db')$, where the signed hypergraph of $Q'$ is $\mH' = \removeleaf{\mH}{v}$ and has a $\pn$-elimination sequence $\sigma'$.
The reduction needs to ensure that $\Pi_{[n] \setminus \{v\}} Q(\db) = Q'(\db')$.
\footnote{For a tuple $\ba_N$ and $X \subseteq N$, we denote $\Pi_{X} \ba_N$ as the projection of $\ba_N$ on $X$.}
To achieve constant delay enumeration, we construct (also in linear time) a data structure over the domain of the variable $x_v$, that can, given an answer to $Q'(\db')$, extend it to an answer to $Q(\db)$ with constant delay. 

\begin{example}
\label{ex:preprocessing-idea}
We will continue with the query $Q$ in Example~\ref{ex:pn:acyclic}, the database $\db$ in Figure~\ref{fig:preprocessing}(a) and a $\pn$-elimination sequence $\sigma = (1,2,3,4)$. The result $Q(\db)$ is depicted in Figure~\ref{fig:preprocessing}(d).
After eliminating $x_4$, we obtain $Q'$ and the instance $\db'$ in Figure~\ref{fig:preprocessing}(b).
We have that $Q'(\db') = \{(\sfa_1, \sfb_1, \sfc_1), ( \sfa_2, \sfb_2, \sfc_2), ( \sfa_3, \sfb_3, \sfc_3)\}$.
The preprocessing phase reduces computing $Q(\db)$ to $Q'(\db')$ and produces a data structure as shown in Figure~\ref{fig:preprocessing}(c), such that given any tuple $\ba' \in Q'(\db')$, we may use the data structure to enumerate all answers of the form $(\ba', a_4)$ in $Q(\db)$, $a_4$ being a value of $x_4$.
\end{example}

\usetikzlibrary{decorations.pathreplacing}
\usetikzlibrary{calc}
\usetikzlibrary{positioning}
\usetikzlibrary{arrows,shapes,decorations.pathmorphing}

\definecolor{Gray}{gray}{0.9}

\begin{figure*}
\scalebox{.84}{
\centering
\begin{minipage}{0.4\textwidth}
\subfloat[A database instance $\db$.]{
\begin{tabular}{c c | c}
\begin{tabular}{l l l}
\multicolumn{3}{c}{$A$} \\
\hline
$x_1$ & $x_2$ & $x_3$ \\
\hline
\rowofthree{\sfa_1}{\sfb_1}{\sfc_1} \\
\rowofthree{\sfa_2}{\sfb_2}{\sfc_2} \\
\rowofthree{\sfa_3}{\sfb_3}{\sfc_3} \\
\rowofthree{\sfa_4}{\sfb_3}{\sfc_3} \\
& & \\
& & \\
& & \\
& & \\
& & \\
& & \\
\end{tabular}
&
\begin{tabular}{l l}
\multicolumn{2}{c}{$U$} \\
\hline
$x_3$ & $x_4$ \\
\hline
\rowcolor{Gray}
\rowoftwo{\sfc_1}{\sfd_1} \\
\rowoftwo{\sfc_1}{\sfd_2} \\
\rowoftwo{\sfc_2}{\sfd_1} \\
\rowcolor{white}
\rowoftwo{\sfc_2}{\sfd_3} \\
\rowcolor{Gray}
\rowoftwo{\sfc_2}{\sfd_4} \\
\rowoftwo{\sfc_2}{\sfd_5} \\
\rowoftwo{\sfc_3}{\sfd_2} \\
\rowcolor{white}
\rowoftwo{\sfc_3}{\sfd_3} \\
\rowcolor{Gray}
\rowoftwo{\sfc_3}{\sfd_4} \\
\rowoftwo{\sfc_3}{\sfd_5} \\
\end{tabular}
&
\begin{tabular}{l l l l}
& & & $V$ \\
\cline{4-4}
& & & $x_4$ \\
\cline{4-4}
& & & $\sfd_3$ \\
& \multicolumn{3}{c}{$R$} \\
\cline{2-4}
& $x_2$ & $x_3$ & $x_4$ \\
\cline{2-4}
\rowoffour{}{\textcolor{blue}{\sfb_3}}{\textcolor{blue}{\sfc_3}}{\textcolor{blue}{\sfd_2}} \\
\rowoffour{}{\textcolor{red}{\sfb_3}}{\textcolor{red}{\sfc_3}}{\textcolor{red}{\sfd_5}} \\
\multicolumn{4}{c}{$S$} \\
\hline
$x_1$ & $x_2$ & $x_3$ & $x_4$ \\
\hline
\rowoffour{\textcolor{violet}{\sfa_2}}{\textcolor{violet}{\sfb_2}}{\textcolor{violet}{\sfc_2}}{\textcolor{violet}{\sfd_4}} \\
\rowoffour{\sfa_4}{\sfb_3}{\sfc_3}{\sfd_4}
\end{tabular}
\end{tabular}
}
\quad
\subfloat[A new database instance $\db'$.]{
\begin{tabular}{c c | c}
\begin{tabular}{l l l}
\multicolumn{3}{c}{$A$} \\
\hline
$x_1$ & $x_2$ & $x_3$ \\
\hline
\rowofthree{\sfa_1}{\sfb_1}{\sfc_1} \\
\rowofthree{\sfa_2}{\sfb_2}{\sfc_2} \\
\rowofthree{\sfa_3}{\sfb_3}{\sfc_3} \\
\rowofthree{\sfa_4}{\sfb_3}{\sfc_3} 
\end{tabular}
&
\begin{tabular}{l l l}
\multicolumn{1}{c}{$U$} & & \\
\cline{1-1}
$x_3$ & &\\
\cline{1-1}
$\sfc_1$ & &\\
$\sfc_2$ & &\\
$\sfc_3$ & &\\
 \\
\end{tabular}
&
\begin{tabular}{l l l}
& \multicolumn{2}{c}{$R$} \\
\cline{2-3}
& $x_2$ & $x_3$  \\
\cline{2-3}
\\
\multicolumn{3}{c}{$S$} \\
\hline
$x_1$ & $x_2$ & $x_3$ \\
\hline
\rowofthree{\sfa_4}{\sfb_3}{\sfc_3} \\
\end{tabular}
\end{tabular}	
}
\end{minipage}%
\begin{minipage}{0.6\textwidth}
\subfloat[The list data structure $\mL_{4}$.]{
\begin{tikzpicture}[list/.style={rectangle,draw}, >=stealth, start chain,startnode/.style={
        draw,minimum width=0.75cm,minimum height=1.4cm}, bot/.style={}]   

\node[startnode] (c1) {$\sfc_1$};

\node[bot,left= of c1] (c1nullleft) {$\bot$};
\node[list,right=of c1] (d11) {$\sfd_1$};
\node[list,right=of d11] (d12) {$\sfd_2$};
\node[bot,right=of d12] (c1nullright) {$\bot$};

\path[<-] (c1nullleft.east) edge (c1.west);
\path[<->] (c1.east) edge (d11.west);
\path[<->] (d11.east) edge (d12.west);
\path[->] (d12.east) edge (c1nullright);

\node[startnode,below=0pt of c1] (c2) {$\sfc_2$};

\node[bot,left= of c2] (c2nullleft) {$\bot$};
\node[list,right=of c2] (d21) {$\sfd_1$};
\node[list,right=of d21] (d24) {\textcolor{violet}{$\sfd_4$}};
\node[list,right=of d24] (d25) {$\sfd_5$};
\node[bot,right=of d25] (c2nullright) {$\bot$};

\path[<-] (c2nullleft.east) edge (c2.west);
\path[<->] (c2.east) edge (d21.west);
\path[<->] (d21.east) edge (d24.west);
\path[<->] (d24.east) edge (d25.west);
\path[->] (d25.east) edge (c2nullright);

\node[startnode,below=0pt of c2] (c3) {$\sfc_3$};

\node[bot,left=of c3] (c3nullleft) {$\bot$};
\node[list,right=of c3] (d32) {\textcolor{blue}{$\sfd_2$}};
\node[list,right=of d32] (d34) {$\sfd_4$};
\node[list,right=of d34] (d35) {\textcolor{red}{$\sfd_5$}};
\node[bot,right=of d35] (c3nullright) {$\bot$};

\path[<-] (c3nullleft.east) edge (c3.west);
\path[<->] (c3.east) edge (d32.west);
\path[<->] (d32.east) edge (d34.west);
\path[<->] (d34.east) edge (d35.west);
\path[->] (d35.east) edge (c3nullright);


\path[dotted,<->] (d21.north east) edge[bend left] node[above] {\textcolor{violet}{$(\sfa_2, \sfb_2, \sfc_2)$}} (d25.north west);

\path[dotted,->] (c3.east) +(0,-0.2) edge[bend right] node[above] {\textcolor{black}{$(\sfa_4, \sfb_3, \sfc_3)$}} (c3nullright);

\path[densely dotted,<->] (c3.east) +(0,+0.2) edge[bend left] node[above] {$\textcolor{blue}{(\sfb_3, \sfc_3)}$} (d34.north west);
\path[densely dotted,->] (d34.north east) edge[bend left] node[above] {$\textcolor{red}{(\sfb_3, \sfc_3)}$} (c3nullright);



\end{tikzpicture}
} 
\quad \centering
\subfloat[The query result.]{
\begin{tabular}{l l l l}
\multicolumn{4}{c}{$Q(\db)$} \\
\hline
$x_1$ & $x_2$ & $x_3$ & $x_4$ \\
\hline
\rowoffour{{\sfa_1}}{{\sfb_1}}{{\sfc_1}}{{\sfd_1}} \\
\rowoffour{\sfa_1}{\sfb_1}{\sfc_1}{\sfd_2} \\
\rowoffour{\sfa_2}{\sfb_2}{\sfc_2}{\sfd_1} \\
\rowoffour{\sfa_2}{\sfb_2}{\sfc_2}{\sfd_5} \\
\rowoffour{\sfa_3}{\sfb_3}{\sfc_3}{\sfd_4} 
\end{tabular}
}
\end{minipage}%
}
\caption{Given the query $Q$ in Example~\ref{ex:preprocessing-idea}, the database instance in (a) and the signed-elimination sequence $\sigma = (1,2,3,4)$ as inputs, Algorithm~\ref{alg:full-scq-preprocessing} first produces the data structure $\mL_4$ in (c), and then constructs a new query $Q'$ in Example~\ref{ex:preprocessing-idea}, a new database as in (b), and the sequence $\sigma' = (1,2,3)$ as inputs to the recursive call.}
\label{fig:preprocessing}
\end{figure*}
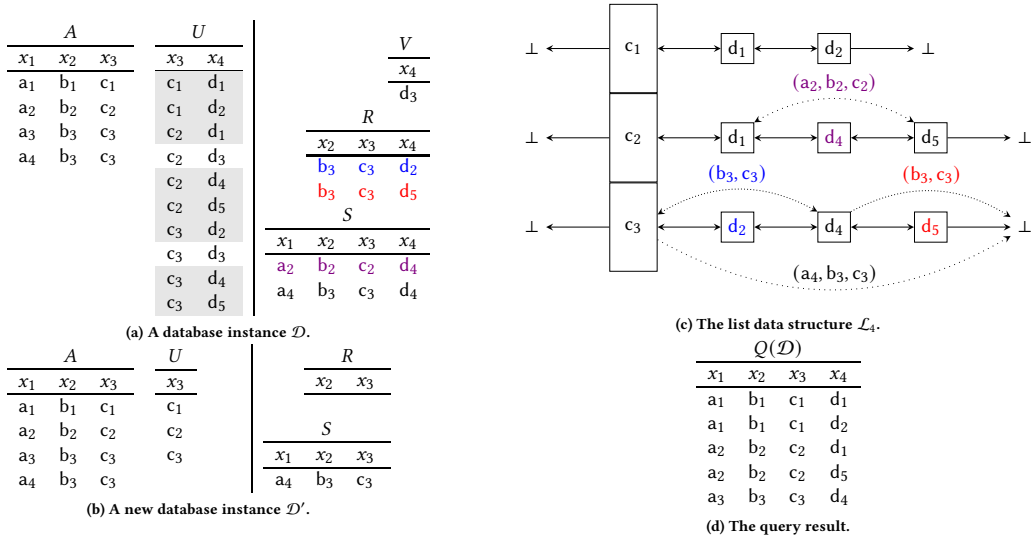

\subsection{Preprocessing phase}
\label{sec:preprocessing-cqneg}


Algorithm~\ref{alg:full-scq-preprocessing} describes the preprocessing phase that takes as input the signed hypergraph $\mH$, a database instance $\db$, and a signed elimination sequence $\sigma$ of $\mH$.
Let $U$ be a pivot hyperedge of the $\pn$-leaf $v$ in $\mH$ and $N_0 = \{v\} \subseteq U \subseteq N_1 \subseteq \dots \subseteq N_m$ the linear order of the negative hyperedges that contain $v$.
The preprocessing phase outputs a data structure $\mL_v$ that maps tuples of the form $\Pi_{U \setminus \{v\}} \ba_{U}$ to a doubly linked list of elements from the domain of $x_v$, with some extra tuple-labeled skipping links that we will introduce later.
We explain how the algorithm eliminates the variable $x_v$ in two steps: $\alpha$-step and $\beta$-step.

\begin{algorithm}[t]
\caption{$\mathsf{PreprocessFullCQ}(\mH, \db, \sigma)$}
\label{alg:full-scq-preprocessing}
\SetKwInOut{Output}{Construct}
\KwIn{signed hypergraph $\mH$, instance $\db$, $\pn$-elimination sequence $\sigma = \sigma' \cdot v$}
\Output{$\mL_v$ for every vertex $v$ in $\sigma$} 
\If{$\sigma$ is empty}{
	\textbf{terminate}
}
$U\leftarrow$ a pivot hyperedge for the $\pn$-leaf $v$ in $\mH$  \\
\ForEach(\Comment{$\alpha$-step}){$K \in \mE^+ \cup \mE^-$ such that $v \in K \subseteq U$}{
	\If{$K \in \mE^+$}{
		$R_U \leftarrow R_U \ltimes R_K$\\
	}
	\Else{
		$R_U \leftarrow R_U \setminus (R_U \ltimes R_K)$\\
	}
	\textbf{remove} relation $R_K$ from $\db$ \\
}
$\mL_v \leftarrow \mathsf{BuildList}(v, {U})$ \\
\textbf{replace} $R_U$ with $R_{U \setminus \{v\}} \gets \Pi_{U \setminus \{v\}} R_U$ \\
Let $N_1, \dots, N_m \in \mE^-$ s.t. $U\subseteq N_1 \subseteq \dots \subseteq N_m$ \Comment{$\beta$-step} \\
\ForEach{$i = 1,2,\dots,m$}{
$\mathsf{ExtendList}(\mL_v, N_i)$ \\
\textbf{replace} $R_{N_i}$ with $R_{N_i \setminus \{v\}}$ that contains all $\ba_{N_i \setminus \{v\} } \in \Pi_{N_i \setminus\{v\}} R_{N_i}$ such that $\Pi_{U \setminus \{v\}} \ba_{N_i \setminus \{v\} } \in \mL_v$ \\
$\text{ and } \mL_v[\Pi_{U \setminus \{v\}} \ba_{N_i \setminus \{v\} }].\mathsf{nextM}(\ba_{N_i \setminus \{v\} }) = \bot$ \\
}
$\mathsf{PreprocessFullCQ}(\removeleaf{\mH}{v}, \db, \sigma')$
\end{algorithm}

\begin{algorithm}[t]
\caption{$\mL_v.\mathsf{Iterate}(\ba_{W})$}
\label{alg:full-scq-traverse}
\KwIn{a tuple $\ba_{W}$ such that $(U \setminus \{v\}) \subseteq W$}
$\mathsf{currNode} \gets \mL_v[\Pi_{U \setminus \{v\}} \ba_W]$ \\
\While{$\mathsf{currNode}.\mathsf{nextM}(\Pi_{W\setminus \{v\}} \ba_W) \neq \bot$}{
	$\mathsf{currNode} \gets \mathsf{currNode}.\mathsf{nextM}(\Pi_{W\setminus \{v\}} \ba_W)$ \\
	\textbf{emit} $\mathsf{currNode}.\mathsf{val}$ 
}
\textbf{emit} $\bot$
\end{algorithm}

\subsubsection*{$\alpha$-step.} 
The algorithm first performs an $\alpha$-step (that mimics Yannakakis' algorithm) in which we simply remove tuples from $R_{U}$ that will not contribute to a query answer of $Q(\db)$ by filtering $R_U$ with any positive (or negative) atom $R_K$ with $v \in K$ and $K \subseteq U$.

Then, Algorithm~\ref{alg:buildlist} builds a hash table $\mL_v$ that maps each tuple $\ba_{U \setminus \{v\}} \in \Pi_{U \setminus \{v\}} R_{U}$ to a doubly linked list of the set $ \{a_{v} \mid (\ba_{U \setminus \{v\}}, a_{v}) \in R_{U}\}$, 
with a slight modification that every pointer is now parameterized/labeled with $\emptyset$ (see Algorithm~\ref{alg:buildlist} in the appendix). 
The extra label on the pointer allows us to add pointers with different labels in later steps.
If $Q$ contains no negative atoms, this data structure is sufficient to achieve constant delay enumeration by traversing the list from the head to the end.
Finally, $R_U$ is replaced with its projection $\Pi_{U \setminus \{v\}} R_{U}$.

\begin{example}
\label{ex:preprocessing-alpha}
For our running example $U(x_3, x_4)$ is chosen as the pivot, and the tuples highlighted in gray in Figure~\ref{fig:preprocessing}(a) denotes the tuples in $U$ that survive the semijoin step with the negative atom $\lnot V(x_4)$. The linked lists of the hash table correspond to the data structure in Figure~\ref{fig:preprocessing}(c) if we ignore the dotted edges.
\end{example}

\subsubsection*{$\beta$-step.} 
Assume $\ba \in Q^+(\db)$, where $Q^+(\bx_{[n]}) \gets \bigwedge_{K \in \mE^+} R_K(\bx_K)$.
When negative atoms are present, for $\ba$ to be an answer to $Q(\db)$, we also need that $\Pi_{N_i} \ba \notin R_{N_i}$ for every $i \in [m]$.
Therefore, we need to augment $\mL_v$  such that we can skip the enumeration of any value in the linked list that does not contribute to the answer. 
The linear order on the schemas of negative atoms allows us to construct this data structure in a dynamic fashion.
If only one negative atom $\lnot R_{N_1}$ is present (i.e., $m = 1$), for each tuple $\ba_{N_1}\in R_{N_1}$ we can add a {\em skipping link} labeled with $\Pi_{N_1 \setminus \{v\}} \ba_{N_1}$ that bypasses the node with value $\Pi_{\{v\}} \ba_{N_1}$ in the doubly linked list $\mL_v[\Pi_{U \setminus \{v\}} \ba_{N_1}]$.  
If $m = 2$, then there is another negative atom $\lnot R_{N_2}$ with $N_1 \subseteq N_2$.
For each tuple $\ba_{N_2} \in R_{N_2}$, we only need to add a skipping link labeled with $\Pi_{N_2 \setminus \{v\}} \ba_{N_2} $ that bypasses the node with value $\Pi_{\{v\}} \ba_{N_2}$ in the doubly linked list $\mL_v[\Pi_{U \setminus \{v\}} \ba_{N_2}]$ if that node has not been bypassed by a skipping link with label $\Pi_{N_1 \setminus\{v\}} \ba_{N_2}$ yet (i.e., $\Pi_{N_1} \ba_{N_2} \notin R_{N_1}$).
In general, for every tuple $\ba_{N_i} \in R_{N_i}$ such that $\Pi_{\{v\}} \ba_{N_i}$ has not been bypassed by any skipping link labeled with $\Pi_{N_j \setminus \{v\}} \ba_{N_i}$ (i.e., for every $1 \leq j < i \leq m$, $\ba_{N_j} \notin R_{N_j}$),
we add a skipping link labeled with $\Pi_{N_i \setminus \{v\}} \ba_{N_i}$ that bypasses $\Pi_{\{v\}} \ba_{N_i}$, as well as every skipping links labeled with $\Pi_{N_j \setminus \{v\}} \ba_{N_i}$ with $1 \leq j < i$ starting in $\Pi_{\{v\}}\ba_{N_i}$.

Algorithm~\ref{alg:full-scq-buildlist} implements this idea using the parameterized/labeled pointers to support the skipping links.
For any node in the doubly linked list, its next node given a tuple $\ba_W$ with $W \supseteq U \setminus \{v\}$ is fetched by following $\nextptr[\Pi_{N_i \setminus \{v\}} \ba_W]$ for the largest possible $i\in\{0,1,\dots,m\}$ in that node.
Recall that $N_0 = \{v\}$ and thus pointers $\nextptr[\emptyset]$ and $\prevptr[\emptyset]$ are always present for each node.
This operation is denoted as $\mathsf{nextM}(\ba_W)$, and we define $\mathsf{prevM}(\ba_W)$ similarly \rebuttal{(formally defined in Appendix~\ref{appendix:enumeration-full-proof})}.
To traverse $\mL_v$, we implement an iterator (Algorithm~\ref{alg:full-scq-traverse}) that takes a tuple $\ba_W$ with $W \supseteq U \setminus \{v\}$, and then traverses the linked list $\mL_v[\Pi_{U \setminus \{v\}} \ba_W]$ using $\mathsf{nextM}(\ba_W)$.

\begin{algorithm}[t]
\caption{$\mathsf{ExtendList}(\mL_v, {N_i})$}
\label{alg:full-scq-buildlist}
\KwIn{a data structure $\mL_v$, $N_i \in \mE^-$}
\SetKwInOut{Output}{Global variables}
\Output{$U \in \mE^+$, $N_1, N_2 \dots, N_i \in \mE^-$ s.t.~$v \in U$, $U \subseteq N_1 \subseteq N_2 \subseteq \dots \subseteq N_i$}
	\ForEach{$\ba_{N_i} \in R_{N_i}$}{
		\If{$\Pi_{U} \ba_{N_i} \notin R_U$ or $\exists j \in \{1,2,\dots,i-1\}$~such~that~$\Pi_{N_j} \ba_{N_i} \in R_{N_j}$}
		{
			\textbf{continue}
		}
		$\mathsf{currNode} \leftarrow   v\in \mL_{v}[\Pi_{U \setminus \{v\}} \ba_{N_i}]$ with $v.\mathsf{val} = \Pi_{\{v\}} \ba_{N_i}$ \\
		$\mathsf{prevNode} \leftarrow \mathsf{currNode}.\mathsf{prevM}(\ba_{N_{i} \setminus \{v\}})$ \\
		$\mathsf{nextNode} \leftarrow \mathsf{currNode}.\mathsf{nextM}(\ba_{N_{i} \setminus \{v\}})$ \\
		$\mathsf{prevNode}.\nextptr[\ba_{N_i \setminus \{v\}}] \leftarrow \mathsf{nextNode}$ \\
		\If{$\mathsf{nextNode} \neq \bot $}{
		$\mathsf{nextNode}.\prevptr[\ba_{N_i \setminus \{v\}}] \leftarrow \mathsf{prevNode}$ }
	}
\end{algorithm}

\begin{example}
Algorithm~\ref{alg:full-scq-buildlist} takes as input the basic linked list data structure and all tuples in the negative atoms $\lnot R(x_2,x_3,x_4)$ and $\lnot S(x_1,x_2,x_3,x_4)$ in Figure~\ref{fig:preprocessing}(a) to produce the data structure in Figure~\ref{fig:preprocessing}(c). Note that the skipping link bypassing $\sfd_4$ with label $(\sfa_4, \sfb_3, \sfc_4)$ also needs to bypass both skipping links labeled by $\textcolor{blue}{(\sfb_3, \sfc_3)}$ and $\textcolor{red}{(\sfb_3, \sfc_3)}$ that bypass $\textcolor{blue}{\sfd_2}$ and $\textcolor{red}{\sfd_5}$ respectively.
\end{example}

Finally, for each negative atom $\lnot R_{N_i}$, we need to keep precisely the subset of $\Pi_{N_i \setminus \{v\}} R_{N_i}$ so that we can \emph{avoid} emitting an answer to $Q'(\db')$ that cannot be extended to an answer to $Q(\db)$.
This is done by keeping the tuple $\ba_{N_i \setminus \{v\}} \in \Pi_{N_i \setminus \{v\}} R_{N_i}$ only if 
 $\Pi_{U \setminus \{v\}} \ba_{N_i \setminus \{v\}} \in \mL_v$, but the tuple cannot be extended to any answer to $Q(\db)$, i.e., the traversal on $\mL_v$ using $\ba_{N_i \setminus \{v\}}$ does not lead to any value of $x_v$.

\begin{example}
\label{ex:reduction}
After removing $x_4$ from relations $A(x_1,x_2,x_3, x_4)$ and $U(x_3,x_4)$, we end up with relations $A(x_1,x_2,x_3)$ and $U(x_3)$ in Figure~\ref{fig:preprocessing}(b).
The query containing only the positive atoms of $Q'$ would return $\{(\sfa_1, \sfb_1, \sfc_1), (\sfa_2, \sfb_2, \sfc_2), (\sfa_3, \sfb_3, \sfc_3), (\sfa_4, \sfb_3, \sfc_3)\}$,
but we wish to skip enumerating $(\sfa_4, \sfb_3, \sfc_3)$, since it cannot be extended to an answer to $Q(\db)$. Indeed, the traversal on the linked list $\mL_4[\sfc_3]$ will simply follow the black skipping link $(\sfa_4, \sfb_3,\sfc_3)$ to reach the end of the list. We achieve this by adding $(\sfa_4, \sfb_3, \sfc_3)$ to $S(x_1,x_2,x_3)$ so that this tuple is not returned in the recursive call. 
\end{example}


\subsection{Enumeration phase}
\label{sec:enumeration-cqneg}

The enumeration phase is shown in Algorithm~\ref{alg:full-scq-enumerate-iterative}.
It follows the reverse order of $\sigma = \sigma' \cdot v$ to first recursively enumerate a tuple $\ba'_{\sigma'} \in Q'(\db')$, and then (with a slight abuse of notation), use every $a_v \in\mL_{v}.\mathsf{Iterate}(\ba'_{\sigma'})$ to obtain an answer $(\ba'_{\sigma'}, a_v)$ to $Q(\db)$. The preprocessing phase guarantees that the iterator is nonempty for every answer $\ba'_{\sigma'}$.

\begin{algorithm}[t]
\caption{$\mathsf{EnumerationFullCQ}(\sigma)$}
\label{alg:full-scq-enumerate-iterative}
\KwIn{a signed-elimination sequence $\sigma = (v_1, v_2, \dots, v_n)$}
\KwOut{an enumeration of the query answers $Q(\db)$}
\ForEach{$a_{v_1} \in \mL_{v_1}.\mathsf{Iterate}\big(\emptyset\big)$}{
	\ForEach{$a_{v_2} \in \mL_{v_2}.\mathsf{Iterate}\big((a_{v_1})\big)$}{
		\ForEach{$a_{v_3} \in \mL_{v_3}.\mathsf{Iterate}\big((a_{v_1},a_{v_2})\big)$}{
			\dots \\
			\ForEach{$a_{v_n} \in \mL_{v_n}.\mathsf{Iterate}\big((a_{v_1}, \dots, a_{v_{n-1}})\big)$}{
				\textbf{emit} $(a_{v_1},a_{v_2},\dots, a_{v_n})$
			}
		}
	}
}

\end{algorithm}

\section{Enumeration of $\faqneg$} \label{sec:enum-faqneg}

In this section, we give a high-level description of the enumeration algorithm for $\faqneg$ queries as \rebut{Eq}~\ref{def:faqneg} that constructively proves \autoref{conj:main-upper}. The full version of the algorithm is presented in \autoref{sec:preprocessing}, and runs on a more general class of queries, called \emph{$\nestfaqneg$} queries, with similar notions of free-connex $\pn$-acyclicity.

Let $\varphi$ be a free-connex $\pn$-acyclic $\faqneg$ query \eqref{def:faqneg} with free variables $F = [f]$ and $\mH = ([n], \mE^+, \mE^-)$ be its associated $\pn$-acyclic signed hypergraph. Assume w.l.o.g that $\mH$ accepts a $\pn$-elimination sequence $\sigma = (1, 2, \dots, f \dots, n)$. Let $\db$ be a database instance of $\varphi$. In the preprocessing phase, the algorithm executes a sequence of \emph{$\pn$-elimination steps on the $\pn$-leaf $i+1$}, for each $i = n-1, \dots, f$ (in order). A $\pn$-elimination step returns an intermediate tuple $(\varphi_i, \db_i)$, where
\begin{itemize}
    \item $\varphi_i$ is a $\pn$-acyclic $\nestfaqneg$ query with free variables $[i]$ associated with $\mH_i = \removeleaf{\mH_{i+1}}{i+1}$
  ($\mH_{n-1} = \removeleaf{\mH}{n}$)
    \item $\db_{i}$ is an instance of $\varphi_i$ such that $\varphi_i(\db_i) = \varphi_{i+1}(\db_{i+1})$. 
\end{itemize}
The sequence ends up with a full $\pn$-acyclic $\nestfaqneg$ query $\varphi_f$ with free variables $F$ and $\db_f$ such that $\varphi(\db) = \varphi_f(\db_f)$. Next, for a full $\pn$-acyclic $\nestfaqneg$ query, we show in Appendix~\ref{sec:enum-full-faqneg} a reduction from the enumeration problem $\Enum(\varphi_f, \db_f)$  into $\Enum(Q^*_{f}, \db^*_{f})$, where $Q^*_{f}$ is a full $\pn$-acyclic $\cqneg$ and $\db^*_{f}$ is a database instance of $Q^*_{f}$. At this point, we simply apply the preprocessing algorithm for a full $\cqneg$. In the enumeration phase, as we emit each answer $\ba_{F}$ of $Q^*_{f}(\db_{f})$, we plug the emitted tuple $\ba_{F}$ into $\varphi_f$ to recover its weight $\varphi_f(\ba_{F}) = \varphi(\ba_{F})$.

We now take a closer look at a $\pn$-elimination step via a concrete running example. The formal (and rather technical) description is deferred to Appendix~\ref{sec:enum-free-connex-faqneg} up to Appendix~\ref{sec:putting_things_together}.

\begin{example} \label{ex:faqneg}
Consider the following $\pn$-acyclic $\faqneg$ query $\varphi(x_1, x_2)$, where 3 is a $\pn$-leaf of its associated hypergraph with a pivot hyperedge $\{3\}$ (corresponding to the positive factor $R_{3}$):
\begin{align*}
     \bigoplus_{x_3} R_{1}(x_1) \otimes R_{2}(x_2) \otimes R_{3}(x_3) \otimes \overline{R}_{123}(x_1, x_2, x_3) \otimes \overline{R}_{23}(x_2, x_3) 
\end{align*}
\end{example}

\begin{figure*}[t]
    \begin{minipage}{0.33\textwidth}
        \centering
        \begin{tikzpicture} [scale=0.47, every node/.style={scale=0.8}]
            \node[fill=black!40!white, circle,inner sep=1pt] (root) at (0, 0) {${\otimes}$}; 
        
            \node (oplus0) at (-4, -1.5) {$\vdash$};
            \node (oplus1) at (-0.5, -1.5) {$\vdash$};

            \node (dots3) at (5, -1.5) {$R_1$};
            \node (dots2) at (3, -1.5) {$R_2$};
            \node (dots1) at (1, -1.5) {$R_3$};

            \node (otimes00) at (-5, -3) {$R_{123}$};
            \node (otimes01) at (-3, -3) {$\bc_{123}$}; 
        
            \node (otimes10) at (-1.5, -3) {$R_{23}$};
            \node (otimes11) at (0.5, -3) {$\bc_{23}$};
        
            \draw (root) -- (oplus0);
            \draw (root) -- (oplus1);
            \draw (root) -- (dots1);
            \draw (root) -- (dots2);
            \draw (root) -- (dots3);
        
            \draw (oplus0) -- (otimes00);
            \draw (oplus0) -- (otimes01);
        
            \draw (oplus1) -- (otimes10);
            \draw (oplus1) -- (otimes11);
        \end{tikzpicture}
    \end{minipage}
    \begin{minipage}{0.33\textwidth}
        \centering
        \begin{tikzpicture} [scale=0.47, every node/.style={scale=0.8}]
            \node (root) at (0, 0) {$\otimes$}; 
        
            \node (oplus0) at (-3, -1) {$\vdash$};
            \node (oplus1) at (-2, -3) {$\vdash$};

            \node (dots3) at (4, -1) {$R_1$};
            \node (dots2) at (2, -1) {$R_2$};
            \node (dots1) at (1.5, -3) {$\bc_{123}$};
            \node (c123) at (0, -3) {$R_3$}; 

            \node (otimes00) at (-5, -1.75) {$R'_{123}$};
            \node[fill=black!40!white, circle,inner sep=1pt](otimes01) at (-0.5, -1.75) {${\otimes}$}; 
        
            \node (otimes10) at (-3.5, -4) {$R_{23}$};
            \node (otimes11) at (-0.5, -4) {$\bc_{23}$};

            \draw (root) -- (oplus0);
            \draw (otimes01) -- (oplus1);
            \draw (otimes01) -- (c123);
            \draw (otimes01) -- (dots1);
            \draw (root) -- (dots2);
            \draw (root) -- (dots3);
        
            \draw (oplus0) -- (otimes00);
            \draw (oplus0) -- (otimes01);
        
            \draw (oplus1) -- (otimes10);
            \draw (oplus1) -- (otimes11);
        \end{tikzpicture}
    \end{minipage}
    \begin{minipage}{0.33\textwidth}
        \centering
        \begin{tikzpicture} [scale=0.47, every node/.style={scale=0.8}]
            \node (root) at (0, 0) {$\otimes$}; 
        
            \node (oplus0) at (-3, -1) {$\vdash$};
            \node(oplus1) at (-2, -2.5) {$\vdash$};

            \node (dots3) at (4, -1) {$R_1$};
            \node (dots2) at (2, -1) {$R_2$};

            \node (otimes01) at (-0.5, -1.75) {$\otimes$}; 
        
            \node (otimes10) at (-3.5, -3.25) {$R'_{23}$};
            \node (otimes00) at (-5, -1.75) {$R'_{123}$};
            \node (otimes11) at (-0.5, -3.25) {$\otimes$};
            
            \node (otimes2) at (-2, -4.25) {$R'_3$};
            
            \draw (root) -- (oplus0);
            \draw (otimes01) -- (oplus1);
            \draw (root) -- (dots2);
            \draw (root) -- (dots3);
        
            \draw (oplus0) -- (otimes00);
            \draw (oplus0) -- (otimes01);
        
            \draw (oplus1) -- (otimes10);
            \draw (oplus1) -- (otimes11);
            \draw (otimes11) -- (otimes2);
        \end{tikzpicture}
    \end{minipage}
    \caption{The refactor steps on $\AST$ of $\varphi$ for the $\pn$-leaf 3, recursively from left to right. The shaded ${\otimes}$ nodes indicate the subtrees to be recursively refactored. The left is the original $\AST$ representation of $\varphi$, and the right is its refactored $\AST$.}
    \label{fig:faqneg-ast}
\end{figure*}
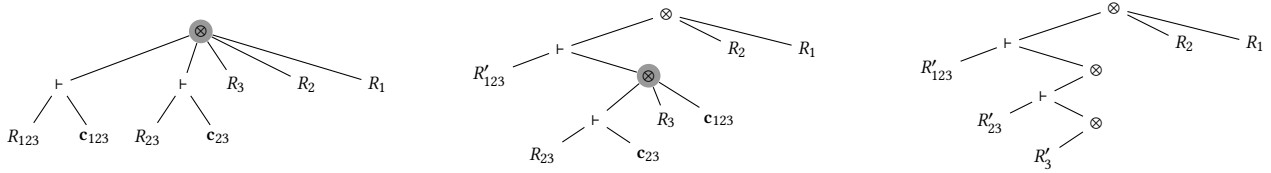

The $\pn$-elimination step for the $\pn$-leaf 3 is much more involved than the elimination step for a $\pn$-acyclic $\cqneg$ query. A naive attempt may aim for a reduction to the follwing $\varphi'$
\begin{align*}
    \varphi'(x_1, x_2) & = R_{1}(x_1) \otimes R_{2}(x_2)  \otimes \overline{R'}_{12}(x_1, x_2) \otimes \overline{R'}_{2}(x_2),
\end{align*}
asking for $\overline{R'}_{12}(x_1, x_2) \otimes \overline{R'}_{2}(x_2)  =  \bigoplus_{x_3} R_{3}(x_3) \otimes \overline{R}_{123}(x_1, x_2, x_3) \otimes \overline{R}_{23}(x_2, x_3)$. However, it is not attainable by the semantics of $\faqneg$ queries over a general semiring. Indeed, we examine the weight $\overline{R'}_{2}(a_2)$, for some $x_2 = a_2$ such that $\Pi_{x_3} \overline{R}_{23} (a_2, x_3) \neq \emptyset$: if $x_1 = a_1$ with $\Pi_{x_1, x_2} \overline{R}_{123} (a_1, a_2, x_3) = \emptyset$, then it should encode $\bigoplus_{x_3} R_{3}(x_3) \otimes \bc_{123} \otimes \overline{R}_{23}(a_2, x_3)$, where $\bc_{123}$ is the constant value for entries out of the table of $\overline{R}_{123}$; otherwise, we want $\overline{R'}_{2}(a_2) = \onebf$ and $\overline{R'}_{12}(a_1, a_2)$ to store $\bigoplus_{x_3} R_{3}(x_3) \otimes \overline{R}_{123}(a_1, a_2, x_3) \otimes \overline{R}_{23}(a_2, x_3)$. The two weights may not coincide over an arbitrary semiring.

This insufficiency calls for lifting $\faqneg$ queries to a more expressive class of queries $\nestfaqneg$, where we exhibit $\varphi$ as an \emph{abstract syntax tree} ($\AST$) depicted in Figure~\ref{fig:faqneg-ast} ({left}), in which we decompose the negative factors as 
\begin{align*}
    \overline{R}_{123} = R_{123} \oplus \indicator{\neg R_{123}} \otimes \bc_{123}, \qquad \overline{R}_{23} = R_{23} \oplus \indicator{\neg R_{23}} \otimes \bc_{23}
\end{align*}
where $\indicator{\neg R_K}$ is an indicator factor that maps to $\zerobf$ if $\bx_K \in R_K$ and $\onebf$ otherwise. As a shorthand, we write $x \vdash y := x \oplus \indicator{\neg x} \otimes y$. Now we illustrate the $\pn$-elimination step on the $\pn$-leaf 3 (corresponding to the variable $x_3$) in Example~\ref{ex:faqneg}. A $\pn$-elimination step can be further decomposed into three steps: $(i)$ refactor step, $(ii)$ oracle-construction step, and $(iii)$ aggregation step. 

\subsection{The Refactor Step}
We first illustrate the refactor step: it reorganizes the $\AST$ of $\varphi$ to make $x_3$ present only in one subtree rooted at a child of the root $\otimes$ node. First, it replaces $R_{123}$ with a new factor $R'_{123} := R_{123} \otimes (R_{23} \vdash \bc_{23}) \otimes R_{3}$, the entry table of which is exactly that of $R_{123}$, except for multiplying the extra weight $(R_{23} \vdash \bc_{23}) \otimes R_{3}$ to each table entry. Then, we get a new $\AST$ in Figure~\ref{fig:faqneg-ast} ({middle}) since
\begin{align*}
    (R_{123} \vdash \bc_{123}) \otimes (R_{23} \vdash \bc_{23}) \otimes R_{3} = R'_{123} \vdash \left((R_{23} \vdash \bc_{23}) \otimes R_{3} \otimes \bc_{123} \right).
\end{align*}
We keep recursing on the subtree for $(R_{23} \vdash \bc_{23}) \otimes R_{3} \otimes \bc_{123}$, thus leading to the desired $\AST$ in Figure~\ref{fig:faqneg-ast} ({right}) with new factors $R'_{23} := R_{23} \otimes (R_{3} \vdash \bc_{3})$ and $R'_3 := R_3 \otimes \bc_{123} \otimes \bc_{23}$ computed. Indeed,
\begin{align*}
    (R_{23} \vdash \bc_{23}) \otimes R_{3} \otimes \bc_{123}  = R'_{23} \vdash \left( R_{3} \otimes \bc_{123} \otimes \bc_{23} \right) = R'_{23} \vdash R'_3.
\end{align*}
As we go, $R_{3}$ sinks down and the linear inclusion $[3] \supseteq \{2, 3\} \supseteq \{3\}$ surfaces along the subtrees that contain $x_3$. In the end, $R_3$ absorbs the constants $\bc_{123}$ and $\bc_{23}$. The refactor step runs in linear time with no query (or database) size blow-up, as proven in Appendix~\ref{sec:rewriting}.

\subsection{The Oracle-construction Step}
Next, we illustrate the oracle-construction step. Chazelle and Rosenberg \cite{10.1145/73833.73848} showed that a semigroup \textsf{RangeSum} data structure (oracle) on an array of size $w$ can be constructed in time $O(w)$ to support a semigroup sum over any range in $O(\alpha(14 w, w))$ time, where $\alpha$ is the inverse Ackermann function. The oracle-construction step uses it as a black-box and builds a \textsf{RangeSum} oracle bottom-up for each subtree containing $x_3$, i.e. the subtree rooted at $\otimes$ nodes in Figure~\ref{fig:faqneg-ast} ({right}). We demonstrate via a simple database instance.

\begin{example}
    We assume the counting ring and on the refactored tree in Figure~\ref{fig:faqneg-ast} ({right}), let $R'_3$ store $\{\langle i, 1 \rangle, i \in [15]\}$, $R'_{23}$ store $\{\langle (\sfa_1, 3), 2 \rangle, \langle (\sfa_1, 9), 2 \rangle, \langle (\sfa_2, 6), 3 \rangle, \langle (\sfa_2, 11), 3 \rangle, \langle (\sfa_3, 8), 4 \rangle\}$ 
    and 
    $R'_{123}$ store $\langle (\sfb_1, \sfa_1, 4), 5 \rangle, \langle (\sfb_1, \sfa_1, 15), 5 \rangle,  \langle (\sfb_2, \sfa_1, 3), 6 \rangle$, $\langle (\sfb_2, \sfa_2, 12), 6 \rangle$. 
\end{example}

The oracle-construction step starts from an array representation of the database instance drawn in Figure~\ref{fig:range-sum}, indexed by all possible values of $x_3$. As an example, the tuple $\langle (\sfb_1, \sfa_1, 4), 5 \rangle$ of $R'_{123}$ can be accessed as the $4$-th element of the array $R'_{123}(\sfb_1, \sfa_1, x_3)$. The necessity of \textsf{RangeSum} becomes natural: after factoring out the term $R_1(\sfb_1) \otimes R_2(\sfa_1)$ (independent of $x_3$), the aggregation for $x_1 = \sfb_1, x_2 = \sfa_1$, from the lens of the array $R'_{123}(\sfb_1, \sfa_1, x_3)$, is
\begin{equation} \label{eq:range-sum-example}
    \begin{aligned}
        & \bigoplus_{x_3}  R'_{123}(\sfb_1, \sfa_1, x_3) \vdash ( R'_{23}(\sfa_1, x_3)  \vdash  R'_{3}(x_3) ) = \sum_{1 \leq x_3 \leq 2} 1 +  \\
    & \sum_{x_3 = 3} \textcolor{blue}{2}  + \sum_{x_3 = 4} \textcolor{olive}{5} + \sum_{5 \leq x_3 \leq 8} 1 + \sum_{x_3 = 9} \textcolor{blue}{2} + \sum_{10 \leq x_3 \leq 14} 1 + \sum_{x_3 = 15} \textcolor{olive}{5}
    \end{aligned}
\end{equation}
where each term is a sum over a range. To support fast aggregation, we construct a \textsf{RangeSum} oracle on the array of $R'_3(x_3)$. Then, we build a \textsf{RangeSum} oracle on the array of $\textcolor{blue}{R'_{23}(\sfa_1, x_3)}$ and $\textcolor{red}{R'_{23}(\sfa_2, x_3)}$. In particular, we query the oracle of $R'_3(x_3)$ to fill in the partial sums (as in the 3rd, 4th row in Figure~\ref{fig:range-sum}) and then construct \textsf{RangeSum} oracles over the partial sums. 
Take $\textcolor{blue}{R'_{23}(\sfa_1, x_3)}$ as the example, $[5, 8]:\textbf{4}$ denotes querying the oracle $R'_3(x_3)$ with $5 \leq x_3 \leq 8$ as range and getting back the value $\textbf{4}$. Then the \textsf{RangeSum} oracle for $\textcolor{blue}{R'_{23}(\sfa_1, x_3)}$ is constructed over the array (of partial sums):
\begin{align*}
    [1, 2]:\textbf{2} \quad \textcolor{blue}{\textbf{2}} \quad [4, 4]:\textbf{1} \quad [5, 8]:\textbf{4} \quad \textcolor{blue}{\textbf{2}} \quad [10, 14]:\textbf{5}. \quad [15, 15]:\textbf{1}
\end{align*}
where the ranges above split at $x_3 = 3, 9$ and $x_3 = 4, 15$ in account for $\textcolor{blue}{R'_{23}(\sfa_1, x_3)}$ and $\textcolor{olive}{R'_{123}(\sfb_1, \sfa_1, x_3)}$, respectively. The rationale here is to avoid mis-aligned ranges when querying the oracle of $\textcolor{blue}{R'_{23}(\sfa_1, x_3)}$ during the oracle-construction of $\textcolor{olive}{R'_{123}(\sfb_1, \sfa_1, x_3)}$. The details are in the 6th row of Figure~\ref{fig:range-sum}. The oracle-construction of $\textcolor{red}{R'_{23}(\sfa_2, x_3)}$ and $\textcolor{violet}{R'_{123}(\sfb_2, \sfa_2, x_3)}$ are similarly depicted in the 4th and 7th row of Figure~\ref{fig:range-sum}. It is worth noting that the splits of ranges should be choreographed for query range alignment, and yet can not be arbitrarily fine-grained due to possible blow-ups in time and space. Indeed, we formally show in Appendix~\ref{sec:oracle} that desired oracles can be carefully constructed in linear time and space.

\begin{figure*}[t] 
    \centering
    \begin{minipage}{1.1\textwidth}
        \begin{tikzpicture}
            \node[black] at (-1.6, 0) {\footnotesize $[1, 15]$};
            \node[black] at (0, 0) {$x_3$};
            \node[black] at (-0.1, -0.5) {$R'_3(x_3)$};
            \node[black] at (-1.45, -0.5) {\footnotesize $15$}; 

            \draw[thin, dotted] (-1.25, 0.2) -- (-1.25, -3.2);

            \foreach \x/\idx in {1, 2, 3, 4, 5, 6, 7, 8, 9, 10, 11, 12, 13, 14, 15}
                \node[inner sep=0.2pt] at (\x, -0.5) {\scriptsize \textbf{1}};

            \foreach \x/\idx in {1, 2, 3, 4, 5, 6, 7, 8, 9, 10, 11, 12, 13, 14, 15}
                \node[inner sep=0.2pt] at (\x, 0) {\footnotesize \idx};
            
            \node[blue] at (-0.2, -1) {$R'_{23}(\sfa_1, x_3)$};
            \foreach \x/\idx in {      3,                       9}
                \node[inner sep=0.2pt] at (\x, -1) {\scriptsize \textcolor{blue}{\textbf{2}}};
            
            \foreach \x/\idx in {    1.5                       }
                \node[inner sep=0.2pt] at (\x, -1) {\scriptsize \textcolor{black}{[1, 2]:\textbf{2}}};

            \draw[thin, black] (1, -1) -- (1.1, -1);
            \draw[thin, black] (1.9, -1) -- (2, -1);

            \foreach \x/\idx in {    4                       }
                \node[inner sep=0.2pt] at (\x, -1) {\scriptsize \textcolor{black}{[4, 4]:\textbf{1}}};
            
            \foreach \x/\idx in {    6.5                       }
                \node[inner sep=0.2pt] at (\x, -1) {\scriptsize \textcolor{black}{[5, 8]:\textbf{4}}};
            
            \draw[thin, black] (5, -1) -- (6.1, -1);
            \draw[thin, black] (6.9, -1) -- (8, -1);

            \foreach \x/\idx in {    12                       }
                \node[inner sep=0.2pt] at (\x, -1) {\scriptsize \textcolor{black}{[10, 14]:\textbf{5}}};

            \draw[thin, black] (10, -1) -- (11.5, -1);
            \draw[thin, black] (12.5, -1) -- (14, -1);
            
            \foreach \x/\idx in {    15                       }
                \node[inner sep=0.2pt] at (\x, -1) {\scriptsize \textcolor{black}{[15, 15]:\textbf{1}}};

            \node[red] at (-0.2, -1.5) {$R'_{23}(\sfa_2, x_3)$};
                \foreach \x/\idx in {      6,                       11}
                    \node[inner sep=0.2pt] at (\x, -1.5) {\scriptsize \textcolor{red}{\textbf{3}}};

            \foreach \x/\idx in {    1.5                       }
                \node[inner sep=0.2pt] at (\x, -1.5) {\scriptsize \textcolor{black}{[1, 2]:\textbf{2}}};

            \draw[thin, black] (1, -1.5) -- (1.1, -1.5);
            \draw[thin, black] (1.9, -1.5) -- (2, -1.5);

            \foreach \x/\idx in {    3                       }
                \node[inner sep=0.2pt] at (\x, -1.5) {\scriptsize \textcolor{black}{[3, 3]:\textbf{1}}};
            
            \foreach \x/\idx in {    4.5                       }
                \node[inner sep=0.2pt] at (\x, -1.5) {\scriptsize \textcolor{black}{[4, 5]:\textbf{2}}};
            
            \draw[thin, black] (4, -1.5) -- (4.1, -1.5);
            \draw[thin, black] (4.9, -1.5) -- (5, -1.5);

            \foreach \x/\idx in {    8.5                       }
                \node[inner sep=0.2pt] at (\x, -1.5) {\scriptsize \textcolor{black}{[7, 10]:\textbf{4}}};

            \draw[thin, black] (7, -1.5) -- (8.1, -1.5);
            \draw[thin, black] (8.9, -1.5) -- (10, -1.5);

            \foreach \x/\idx in {    12                       }
                \node[inner sep=0.2pt] at (\x, -1.5) {\scriptsize \textcolor{black}{[12, 12]:\textbf{1}}};

            \foreach \x/\idx in {    14                       }
                \node[inner sep=0.2pt] at (\x, -1.5) {\scriptsize \textcolor{black}{[13, 15]:\textbf{3}}};
            
            \draw[thin, black] (13, -1.5) -- (13.5, -1.5);
            \draw[thin, black] (14.5, -1.5) -- (15, -1.5);

            \node[purple] at (-0.2, -2) {$R'_{23}(\sfa_3, x_3)$};
            \node[purple] at (-1.5, -2) {\footnotesize $18$};
                \foreach \x/\idx in {      8}
                    \node[inner sep=0.2pt] at (\x, -2) {\scriptsize \textcolor{purple}{\textbf{4}}};

            \foreach \x/\idx in {    4                       }
                \node[inner sep=0.2pt] at (\x, -2) {\scriptsize \textcolor{black}{[1, 7]:\textbf{7}}};

            \draw[thin, black] (1, -2) -- (3.6, -2);
            \draw[thin, black] (4.4, -2) -- (7, -2);

            \foreach \x/\idx in {    12                       }
                \node[inner sep=0.2pt] at (\x, -2) {\scriptsize \textcolor{black}{[9, 15]:\textbf{7}}};

            \draw[thin, black] (9, -2) -- (11.6, -2);
            \draw[thin, black] (12.4, -2) -- (15, -2);

            \node[olive] at (-0.1, -2.5) {$R'_{123}(\sfb_1, \sfa_1, x_3)$};
            \node[olive] at (-1.5, -2.5) {\footnotesize $25$};
                    \foreach \x/\idx in {      4,                       15}
                        \node[inner sep=0.2pt] at (\x, -2.5) {\scriptsize \textcolor{olive}{\textbf{5}}};

            \foreach \x/\idx in {    2                       }
                \node[inner sep=0.2pt] at (\x, -2.5) {\scriptsize \textcolor{blue}{[1, 3]: \textbf{4}}};
            
            \draw[thin, blue] (1, -2.5) -- (1.6, -2.5);
            \draw[thin, blue] (2.4, -2.5) -- (3, -2.5);

            \foreach \x/\idx in {    9.5                       }
                \node[inner sep=0.2pt] at (\x, -2.5) {\scriptsize \textcolor{blue}{[5, 14]: \textbf{11}}};

            \draw[thin, blue] (5, -2.5) -- (9, -2.5);
            \draw[thin, blue] (10, -2.5) -- (14, -2.5);

            \node[violet] at (-0.1, -3) {$R'_{123}(\sfb_2, \sfa_2, x_3)$};
            \node[violet] at (-1.5, -3) {\footnotesize $29$};
                \foreach \x/\idx in {      3,              12}
                    \node[inner sep=0.2pt] at (\x, -3) {\scriptsize \textcolor{violet}{\textbf{6}}};

            \foreach \x/\idx in {    1.5                       }
                \node[inner sep=0.2pt] at (\x, -3) {\scriptsize \textcolor{red}{[1, 2]:\textbf{2}}};

            \draw[thin, red] (1, -3) -- (1.1, -3);
            \draw[thin, red] (1.9, -3) -- (2, -3);

            \foreach \x/\idx in {    7.5                       }
                \node[inner sep=0.2pt] at (\x, -3) {\scriptsize \textcolor{red}{[4, 11]:\textbf{12}}};

            \draw[thin, red] (4, -3) -- (7, -3);
            \draw[thin, red] (8, -3) -- (11, -3);

            \foreach \x/\idx in {    14                       }
                \node[inner sep=0.2pt] at (\x, -3) {\scriptsize \textcolor{red}{[13, 15]:\textbf{3}}};

            \draw[thin, red] (13, -3) -- (13.5, -3);
            \draw[thin, red] (14.5, -3) -- (15, -3);  
            
        \end{tikzpicture}
    \end{minipage}
    \caption{The oracle-construction steps on the refactored $\AST$. The oracle of \textcolor{blue}{$R'_{23}(\sfa_1, x_3)$}, \textcolor{red}{$R'_{23}(\sfa_2, x_3)$} and \textcolor{purple}{$R'_{23}(\sfa_3, x_3)$} are built atop the oracle of $R'_{3}(x_3)$; the oracle of \textcolor{olive}{$R'_{123}(\sfb_1, \sfa_1, x_3)$} and \textcolor{violet}{$R'_{123}(\sfb_2, \sfa_2, x_3)$} are built atop the oracle of \textcolor{blue}{$R'_{23}(\sfa_1, x_3)$} and \textcolor{red}{$R'_{23}(\sfa_2, x_3)$}, respectively. The leftmost column shows the requested range sums over $[1, 15]$ in the aggregation step.}
    \label{fig:range-sum}
\end{figure*}

\subsection{The Aggregation Step}
Finally, we go over the aggregation step that effectively peels off $x_3$ from the refactored $\AST$. To that end, it pushes the aggregation operator $\bigoplus_{x_3}$ downward the refactored $\AST$ from root to $R'_3(x_3)$, i.e. the opposite direction of the oracle-construction step. Start with the root $\otimes$ node, we will construct new factors $R''_{12}(x_1, x_2)$ and $R''_{2}(x_2)$ such that (let $\bigoplus_{x_3} R'_{3} = 15$):
\begin{align*}
    \bigoplus_{x_3}  R'_{123} & \vdash ( R'_{23}  \vdash  R'_{3} )  = R''_{12} \vdash \bigoplus_{x_3} ( R'_{23}  \vdash  R'_{3} ) \\
    & = R''_{12} \vdash ( R''_{2}  \vdash  \bigoplus_{x_3} R'_{3} ) = R''_{12} \vdash ( R''_{2}  \vdash  15 ).
\end{align*}
To realize the first equality, we ask the table of $R''_{12}(x_1, x_2)$ to store two entries, $\langle (\sfb_1, \sfa_1), 25\rangle$ and $\langle (\sfb_2, \sfa_2), 29\rangle$, where $25$ is the aggregation for $x_1 = \sfb_1, x_2 = \sfa_1$ in \eqref{eq:range-sum-example}, that can be obtained by a simple range query $[1, 15]$ on the oracle of \textcolor{olive}{$R'_{123}(\sfb_1, \sfa_1, x_3)$}. Similarly for $29$ from a $[1, 15]$ query on the oracle of \textcolor{violet}{$R'_{123}(\sfb_2, \sfa_2, x_3)$}. Indeed, the first equality holds because for all values of $(x_1, x_2)$ not encoded in the table of $R''_{12}$, we always have $R'_{123} = \zerobf$ and $\indicator{\neg R'_{123}} = \onebf$, therefore for those $(x_1, x_2)$s, $R'_{123} \vdash ( R'_{23}  \vdash  R'_{3} ) = \zerobf \oplus \onebf \otimes ( R'_{23}  \vdash  R'_{3} ) = R'_{23}  \vdash  R'_{3}$.

Now that we have pushed $\bigoplus_{x_3}$ to $( R'_{23}  \vdash  R'_{3} )$ (i.e. the $\otimes$ node on the 2nd level of the refactored $\AST$) and we insert a single tuple $\langle (\sfa_3), 18 \rangle$ into the new factor $R''_{2}(x_2)$, where $18$ is yet another $[1, 15]$ query on the oracle of \textcolor{purple}{$R'_{23}(\sfa_3, x_3)$}. The second equality holds because for all values of $x_2$ not encoded in the table of $R''_{2}$, we have $R'_{23} = \zerobf$ and $\indicator{\neg R'_{23}} = \onebf$. Thus, $R'_{23} \vdash R'_{3} = \zerobf \oplus \onebf \otimes R'_{3} = R'_{3}$. The resulting $\nestfaqneg$ query $\varphi''$ after the $\pn$-elimination on $3$ is
\begin{align*} \label{eq:full-faqneg-example}
    \varphi''(x_1, x_2) = R_{1}(x_1) \otimes R_{2}(x_2) \otimes \left( R''_{12}(x_1, x_2) \vdash ( R''_{2}(x_2)  \vdash  15 ) \right)
\end{align*}
where the $\AST$ of $\varphi''$ is exactly the refactored $\AST$ in Figure~\ref{fig:faqneg-ast} ({right}), except that $x_3$ is being eliminated. 
Now the $\pn$-elimination step is complete. We defer the formal analysis of the aggregation step to Appendix~\ref{sec:aggregation} and Appendix~\ref{sec:putting_things_together}.

\subsection{The Reduction to Enumeration of Full $\cqneg$}
Finally, we cast the enumeration of the (full) $\nestfaqneg$ query $\varphi''$ into the enumeration of the full $\pn$-acyclic $\cqneg$ query $Q^*$, by directly extracting factors from $\varphi''$ as relations:
\begin{align*}
    Q^*(x_1, x_2) \leftarrow R^*_{1}(x_1) \wedge R^*_{2}(x_2) \wedge \neg R^{**}_{12}(x_1, x_2)  \wedge \neg R^{**}_{2}(x_2) 
\end{align*}
where $R^*_{1}$ stores $a_1 \in R_{1}$ with $R_{1}(a_1) \neq \zerobf$ (similarly for $R^*_{2}$), and $R^{**}_{12}$ stores $(a_1, a_2) \in R''_{12}$ with $R''_{12}(a_1, a_2) = \zerobf$ (similarly for $R^{**}_{2}$). Intuitively, $Q^*$ forbids exactly the tuples $(a_1, a_2)$ such that $\varphi''(x_1, x_2) = \zerobf$ to be emitted. Thus we can safely enumerate answers of $Q^*$ using techniques in Section~\ref{sec:enumeration-full} and use $\varphi''$ to recover the non-$\zerobf$ weights. 
\section{Lower Bounds}
\label{sec:hardness}

In this section, we present conditional and unconditional lower bounds which complement our upper bounds. All lower bounds here refer to {\em self-join-free} queries, which means that each relation name appears at most once in the body of the query. Technical proofs of this section can be found in Appendix~\ref{appendix:hardness-proofs}.

\subsection{Lower Bounds for  $\cqneg$}
\label{subsec:hardness}

In this section, we show lower bound results for any $\cqneg$ that is not free-connex $\pn$-acyclic. We split this result into two theorems that use different conditional lower bounds.

\begin{description}
\item[Problem] $(k+1,k)$-Hyperclique
\item[Input] a $k$-uniform hypergraph\footnote{A hypergraph is said to be \emph{$k$-uniform} if every hyperedge contains exactly $k$ vertices.} (for $k > 2$) 
\item[Output] does it contain a hyperclique of size $k+1$, i.e. a set of $k+1$ vertices where every subset of size $k$ forms a hyperedge
\end{description}

\begin{conjecture}[$(k+1,k)$-Hyperclique]
\label{conj:hyperclique}
There is no algorithm that solves $(k+1,k)$-Hyperclique in $O(m)$ time, where $m$ is the number of edges in the input hypergraph.
\end{conjecture}

Readers are referred to~\cite{BerkholzGS20} for evidence why Conjecture~\ref{conj:hyperclique} is believable. When $k=2$, the (3,2)-Hyperclique problem is the problem of finding a triangle in a graph. 

\begin{conjecture}[Triangle]
\label{conj:triangle}
There is no algorithm that decides whether a graph with $n$ nodes contains a triangle in $O(n^2)$ time.
\end{conjecture}

The following is a combination of results from~\cite{BBThesis, BerkholzGS20}.

\begin{theorem}\label{thm:non-pn-acyclic boolean}
Let $Q$ be a $\cqneg$ that is not $\pn$-acyclic. Assuming Conjecture~\ref{conj:hyperclique} and Conjecture~\ref{conj:triangle}, then there is no algorithm for $Q$ that has linear preprocessing time and $O(1)$ delay.
\end{theorem}


To show a lower bound for queries that are not free-connex (but are $\pn$-acyclic), we will use a weaker lower bound conjecture that implies \textsc{Triangle}. 

\begin{conjecture}[$\bmm$]
\label{conj:bmm}
There is no algorithm that computes the product $A \times B$ of two $n \times n$ Boolean matrices $A$ and $B$ in $O(n^2)$ time.
\end{conjecture}

Evidence for Conjecture~\ref{conj:bmm} can be found in~\cite{Raz03}. Bagan et al.~\cite{BaganDG07} reduce the \textsf{BMM} problem to the non-free-connex acyclic query $Q(x, y) \leftarrow \bigvee_{z} A(x, z) \wedge B(z, y)$ and apply Conjecture~\ref{conj:bmm} to obtain a conditional lower bound. The {\em matrix multiplication exponent} $\omega$ is the smallest number such that for any $\epsilon >0$, there is an algorithm that multiplies two $n$-by-$n$ matrices with at most $O(n^{\omega+\epsilon})$ operations (assuming RAM model). The best bound known so-far on $\omega$ is (roughly) $\omega < 2.373$ in~\cite{Williams12, Gall14a}. We note that Conjecture~\ref{conj:bmm} does not violate the common belief that $\omega =2$, since that only implies that \textsf{BMM} can be computed in time $n^{2+o(1)}$. For non-free-connex CQs, a weaker lower bound conjecture was used, \textsf{sparse BMM} (the matrices have $m$ non-zero entries and no $O(m)$ algorithm exists). However, this conjecture cannot be applied in our case because we need to take the complement of the matrix to populate a negated atom, and that means that a sparse matrix becomes dense.

\begin{theorem}\label{thm:non-free-connex boolean}
 Let $Q$ be a $\cqneg$ that is $\pn$-acyclic and not free-connex. Assuming Conjecture~\ref{conj:bmm}, there is no algorithm with linear preprocessing time and $O(1)$ delay. 
\end{theorem}

\subsection{Lower Bounds for $\faqneg$}
\label{subsec:embedding}

We present next lower bounds for $\faqneg$ when restricted to queries with head $\varphi()$, which we denote as $\mathsf{SumProd}^{\neg}$. To show these bounds, we will use weaker conjectures than the ones used in the previous section.

\begin{conjecture}[Minimum-Weight $k$-Clique]
\label{conj:tropical_clique}
There is no algorithm that computes the minimum weight of a $k$-clique in a edge-weighted graph with $n$ nodes in $O(n^k)$ time.
\end{conjecture}


Our reduction from Minimum-Weight $k$-Clique~\cite{AbboudBDN18} is an application of the {clique embedding power} technique introduced in~\cite{ICALP}.

\begin{theorem}
Assuming Conjecture~\ref{conj:tropical_clique}, a $\mathsf{SumProd}^{\neg}$ $\varphi$ over the tropical semiring can be solved in linear time iff $\varphi$ is $\pn$-acyclic.
\end{theorem}

Over the counting ring, we can instead show that any lower bound for counting Boolean CQs transfers immediately to $\mathsf{SumProd}^{\neg}$.

\begin{theorem}
\label{lb:count}
Suppose that no linear-time algorithm can count the solutions of a non $\alpha$-acyclic Boolean CQ.
Then, there is no linear-time algorithm for a $\mathsf{SumProd}^{\neg}$ query over the counting ring that is not $\pn$-acyclic.
\end{theorem}

\subsection{An Unconditional Lower Bound for $\faqneg$}\label{subsec:unconitional}
Finally, we show an unconditional lower bound that provides some evidence on the necessity of the inverse Ackermann factor in the runtime of Theorem~\ref{conj:main-upper}.

The lower bound is based on the {additive structure} of the underlying semiring for $\faqneg$, i.e. a commutative semigroup with operator $\oplus$. It uses the {\em arithmetic model of computation}~\cite{Yao85a, ChazelleR91}, which charges one unit of computation for every $\oplus$ operation performed, while all other computation is free. Essentially, the computation can be viewed as a sequence of instructions of the form $z_i = a z_j \oplus b z_k$, where $\{z_i\}_i$ form an unbounded set of variables. Moreover, this sequence should be agnostic to the actual values of the semiring. The only thing we need is that the semigroup is {\em faithful}~\cite{Yao85a, ChazelleR91}, meaning that for every $T_1, T_2 \subseteq \{1,2,\dots, n\}$, and integers $\delta_i, \delta_j' > 0$,
$\bigoplus_{i \in T_1} \delta_i \cdot a_i = \bigoplus_{j \in T_2} \delta_j' \cdot a_j$
cannot be an identity for all $a_1, a_2, \dots, a_n \in S$ unless $T_1 = T_2$.
This is essentially saying that there is no ``magical shortcut'' to compute the sums. 

\begin{theorem}\label{thm:unconditional-lb}
Under the arithmetic model of computation, any constant delay enumeration algorithm (in data complexity) for
$$\varphi(x)=\bigoplus_{y} A(x) \otimes B(y) \otimes \indicator{\lnot R}(x, y)$$
on factors $A$, $B$ each of size $n$ and factor $R$ of size $2m$
must require $\Omega(m\cdot \alpha(m, m))$ preprocessing time for every $m \leq n$.
\end{theorem}



\section{Difference of CQs}
\label{sec:beyond}

As an application of our results, we consider the class of queries of the form $Q_1 - Q_2$, where $Q_1, Q_2$ are full CQs with the same set of variables. It is shown in a recent paper~\cite{CQDiff} that $Q_1 - Q_2$ can be computed in time $O(|\db| + \mathsf{OUT})$, where $\mathsf{OUT}$ is the output size of $Q_1 - Q_2$, if $Q_1$ is $\alpha$-acyclic and $Q_1 \wedge R_e$ is $\alpha$-acyclic for every $R_e$ in $Q_2$. We use our main theorem to strengthen this result by providing a constant-delay enumeration guarantee (Appendix~\ref{appendix:beyond}).

\begin{theorem}\label{thm:cq:diff}
Let $Q = Q_1 - Q_2$, where $Q_1, Q_2$ are full CQs over the same set of variables. If  $Q_1$ is $\alpha$-acyclic and $Q_1 \wedge R_e$ is $\alpha$-acyclic for every $R_e$ in $Q_2$, then the result $Q$ can be enumerated with constant delay after $O(|\db|)$ preprocessing time.
\end{theorem}

As a corollary, we obtain the following generalization to differences of non-full CQs (proven in Appendix~\ref{appendix:beyond}).

\begin{corollary} \label{cor:cq:diff}
Let $Q = Q_1 - Q_2$, where $Q_1, Q_2$ are CQs with the same set of free variables. If $Q$ is difference-linear (Def 2.3 in \cite{CQDiff}), then the output of $Q$ can be enumerated with constant delay after $O(|\db|)$ preprocessing time.
\end{corollary}

\section{Conclusion}

This paper has made an initial foray into a novel way of interpreting $\cqneg$ from the perspective of semiring and \textsf{FAQ}s \cite{DBLP:conf/pods/KhamisNR16}. We presented a constant-delay enumeration algorithm for the class of free-connex $\pn$-acyclic $\faqneg$ queries, after linear preprocessing (modulo an inverse Ackermann factor), and showed lower bounds for $\faqneg$ queries out of this class. 
We leave as an intriguing open question the parameterized complexity of general $\cqneg$ and $\faqneg$ queries (a brief discussion is in Appendix~\ref{appendix:width}).



\bibliographystyle{ACM-Reference-Format}

\bibliography{ref}

\onecolumn
\appendix

\section{Missing proofs in Section~\ref{sec:pn-acyclic}}
\label{appendix:pn-acyclic}

Let $\mH = (\mV, \mE^+, \mE^-)$ be a signed hypergraph. We define $\reduced{\mH}$ as the hypergraph obtained by iteratively removing a hyperedge $K \in \mE^+ \cup \mE^-$ if there exists some hyperedge $U \in \mE^+$ such that $K \subseteq U$.
We say a signed hypergraph $\mH$ is \emph{reduced} if $\mH = \reduced{\mH}$.
For a multiset of hyperedges $\mE$, we define $\rmvertex{\mE}{x} := \{K \setminus \{x\} \mid K \in \mE\}$.
Define $\rmvertex{\mH}{x}$ as the hypergraph obtained by removing $x$ from every hyperedge in $\mH$.

\subsection{Equivalence with~\cite{BBThesis}}

A vertex $x$ is said to be a \emph{bicolor-leaf} in $\mH$ in~\cite{BBThesis} if $x$ is a $\beta$-leaf in $\reduced{\mH}$. We show that this notion of bicolor-leaf in~\cite{BBThesis} is equivalent to our definition of $\pn$-leaf in Definition~\ref{defn:pn-leaf}.


\begin{lemma}
\label{lemma:pivot-edge}
Let $\mH = (\mV, \mE^+, \mE^-)$ be a signed-acyclic signed hypergraph. Then, $x$ is a $\pn$-leaf of $\mH$ if and only if $x$ is a $\beta$-leaf of $\reduced{\mH}$.
\end{lemma}

\begin{proof}
Let $\reduced{\mH} = (\mV, \mE'_+, \mE'_-)$.

\framebox{$\implies$} Assume that $x$ is a $\pn$-leaf in $\mH$. Let $U$ be the pivot edge of $x$ in $\mH$ and assume that $U \subseteq R_1 \subseteq R_2 \subseteq \dots \subseteq R_n$ such that $R_i \in \mE^-$ for $i \in [n]$.
Then $U, R_1, R_2,\dots, R_n \in \mE'_-$.
Let $K$ be an arbitrary hyperedge in $\mE^+ \cup \mE^- \setminus \{U, R_1,R_2,\dots, R_n\}$ that contains $x$. Then by definition of $\reduced{\mH}$, we must have that $K \subseteq U$, and therefore $K$ is not present in $\reduced{\mH}$.
Therefore, $U, R_1, R_2,\dots, R_n$ are all hyperedges in $\reduced{\mH}$ that contain $x$. Hence $x$ is a $\beta$-leaf of $\reduced{\mH}$.

\framebox{$\impliedby$} Assume that $x$ is a $\beta$-leaf in $\reduced{\mH}$. 
Let $R_1, R_2, \dots, R_n$ be all hyperedges in $\reduced{\mH}$ that contain $x$, and assume that and $R_i \subseteq R_{i+1}$ for $i \in [n-1]$.

We argue that $\{R_1\} = \mE^+ \cap \mE'_+$, i.e. $R_1 \in \mE'_+$ and for every $i \in \{2, 3, \dots, n-1\}$, $R_i \in \mE'_-$. 
Let $i$ be the smallest index such that $R_i \in \mE'_+$. 
Such an $i$ must exist, since otherwise otherwise we have $x \in \bigcup_{K \in \mE'_-} K$ but $x \notin \bigcup_{K \in \mE'_+} K$, a contradiction to $\bigcup_{K\in \mE'_-}K \subseteq \bigcup_{K \in \mE'_+}K$.
If $i > 1$, since $R_1 \subseteq R_i$, then $R_1$ should have been removed from $\reduced{\mH}$, a contradiction.
Finally, if there is any $j \in \{2, 3, \dots, n-1\}$ such that $R_j \in \mE'_+$, since $R_1 \subseteq R_j$, $R_1$ would have been removed from $\reduced{\mH}$, a contradiction.

Let $U = R_1$. We show that $U$ is a pivot edge in $\mH$.

For (1), let $R$ be an arbitrary hyperedge in $\mE^+$ that contains $x$. Consider the maximal sequence of hyperedges $\{R^{(i)}\}_{i\geq0}$ such that $R^{(0)} = R$, and for each $i \geq 0$, if $R^{(i)} \notin \mE'_+$, then by construction of $\reduced{\mH}$, let $R^{(i+1)}$ be the hyperedge in $\mE^+$ such that $R^{(i)} \subset R^{(i+1)}$. 
Since the hypergraph is finite, this sequence must terminate and assume that its last hyperedge is $R^{(t)}$.
We thus have that $R^{(t)} \in \mE'_+$, and by previous argument, $R^{(t)} = R_1$. 
Therefore, $R = R^{(0)} \subset R^{(1)} \subset \dots \subset R^{(t)} = R_1 = U$, as desired.

Consider (2). let $N$ be an arbitrary hyperedge in $\mE^-$ such that $x \in N$ and $N \not\subseteq U$.
We show that $N \in \mE'_-$.
Indeed, if not, we must have that there exists some hyperedge $K \in \mE^+$ such that $N \subseteq K$. Since $K \subseteq U$, we have $N \subseteq U$, a contradiction.
Therefore, we have $\{N \in \mE^- \mid x \in N, N \not\subseteq R_x\} \cup \{U\} = \{R_1, R_2, \dots, R_n\}$, which is linearly ordered by $\subseteq$ by construction and $U = R_1$ is the minimal element, as desired. 
\end{proof}

\subsection{Proof of Proposition~\ref{prop:pn-acyclic-leaf}}

Our proof for Proposition~\ref{prop:pn-acyclic-leaf} is inspired from the proof of Theorem~7 in~\cite{10.1145/2983573}.
Some additional definitions are required.
Two vertices $x$ and $y$ are said to be \emph{non-neighbors} in a signed hypergraph $\mH = (\mV, \mE^+, \mE^-)$ if there is no hyperedge $R \in \mE^+ \cup \mE^-$ such that $R \subset \mV$ and $x, y \in R$.

We need some additional structural properties on signed-acyclicity. 

\begin{lemma}
\label{lemma:remove-var-or-atom}
Let $\mH = (\mV, \mE)$ be an $\alpha$-acyclic hypergraph. Then we have 
\begin{enumerate}
\item the hypergraph $\rmvertex{\mH}{x}$ is $\alpha$-acyclic for every \rebut{vertex $x \in \mV$}; and
\label{it:rmvertex}
\item the hypergraph $\mH' = (\mV, \mE \setminus \{R\})$ is $\alpha$-acyclic for every two distinct hyperedges $R$ and $S$ in $\mH$ such that $R \subseteq S$. 
\label{it:rmatom}
\end{enumerate}
\end{lemma}
\begin{proof}
Let $\mT$ be the join tree of $\mH$ witnessed by the bijection mapping $\chi : \mE \rightarrow V(\mT)$.
Consider two items.

\framebox{(\ref{it:rmvertex})}
We have that $\rmvertex{\mH}{x} = (\mV\setminus \{x\}, \rmvertex{\mE}{x})$.
Consider the new mapping $\chi' : \rmvertex{\mE}{x} \rightarrow V(\mT)$ such that $\chi'(K \setminus \{x\}) = \chi(K)$ for every $K \in \mE$. 
It is easy to verify that for every vertex $v \in \mV\setminus\{x\}$, the set of nodes $\{\chi'(K \setminus \{x\}) \mid K \setminus \{x\} \in \rmvertex{\mE}{x}\}$ induces a connected component in $\mT$ since the set of nodes $\{\chi(K) \mid K \in \mE\}$ does.

\framebox{(\ref{it:rmatom})}
Let $R$ and $S$ be two distinct hyperedges in $\mH$ such that $R \subseteq S$. 
Let $\mT_1, \mT_2, \dots, \mT_k$ be the connected components of $\mT - \chi(R)$, the tree obtained by removing the node $\chi(R)$ and all edges incident to $\chi(R)$ from $\mT$. 
Assume that $\chi(R)$ is adjacent to $\chi({A_i})$ in $\mT_i$ for each $i\in[k]$ and without loss of generality that $\chi(S)$ is contained in $\mT_k$.
Consider the tree $\mT'$ obtained by removing $\chi(R)$ from $\mT$ and adding an edge between every $\chi({A_i})$ to $\chi(S)$ for each $i\in[k-1]$. 

We argue that $\mT'$ is a join tree of $\mH'$. Consider any vertex $x$ in $\mH'$ and let $K_x = \{\chi(K) \mid x \in K, K \in \mE \setminus \{R\}\}$. The goal is to show that $K_x$ induces a connected component in $\mT'$.

Consider $I = \{i \in [k] \mid V(\mT_i) \cap K_x \neq \emptyset\}$, the indices of all subtrees $\mT_i$ that contains some hyperedge containing $x$.
If $I = \{i\}$ for some index $i \in[k]$, then $K_x$ is properly contained in $\mT_i$, and since $\mT_i$ is a join tree, $\mT_x$ induces a connected component in $\mT_i$ and thus $\mT'$. 
Otherwise, $\card{I} \geq 2$. In this case, we must have $A_i \in K_x$ for every $i \in I$ and thus $x \in R \subseteq S$.
Therefore, $K_x \setminus V(\mT_k)$ induces a connected component in $\mT'$. By construction, every $\chi(A_i)$ is connected to $\chi(S)$ in $\mT'$ and $\chi(S) \in K_x$. Note that $K_x \cap V(\mT_k)$ also induces a connected component in $\mT_k$ and $\chi(S) \in K_x \cap V(\mT_k)$, $K_x$ induces a connected component in $\mT'$.
\end{proof}




\begin{lemma}
\label{lemma:two-beta-leaf}
Let $\mH = (\mV, \mE^+, \mE^-)$ be a reduced signed-acyclic signed hypergraph with at least two vertices. Then $\mH$ contains two non-neighbor $\beta$-leaves (and therefore signed-leaves).
\end{lemma}
\begin{proof}
We define the size of a signed hypergraph as the sum of the size of each hyperedge in it. 
We use an induction on the size $k$ of the reduced signed-acyclic signed hypergraph $\mH$.

\begin{description}
\item[Basis $k = 2$.] In this case, we can only have $\mH = (\{x, y\}, \{\{x\}, \{y\}\}, \emptyset)$, $\mH = (\{x, y\}, \{\{x\}, \{y\}\}, \{\{x,y\}\})$ or  $\mH = (\{x, y\}, \{\{x, y\}\}, \emptyset)$ and the claim follows.

\item[Inductive step.] Assume that the claim holds for any reduced signed-acyclic signed hypergraph with size between $2$ and $k-1$.
Let $\mH = (\mV, \mE^+, \mE^-)$ be a reduced signed-acyclic signed hypergraph with size $k$.

Assume first that there is some hyperedge $R \in \mE^+ \cup \mE^-$ in $\mH$ such that $R = \mV$.
If $R \in \mE^+$, since $\mH$ is reduced, we must have $\mH = (\mV, \{R\}, \emptyset)$, and the lemma follows since pair of vertices in $\mH$ are non-neighbor $\beta$-leaves. 
If $R \in \mE^-$, consider $\mH' = (\mV, \mE^+, \mE^- \setminus \{R\})$. Note that $\mH'$ is also signed-acyclic, reduced and of smaller size. 
Thus by the inductive hypothesis, $\mH'$ contains two non-neighbor $\beta$-leaves $x$ and $y$. 
Then $x$ and $y$ are also $\beta$-leaves in $\mH$, since $K \subseteq \mV = R$ for every hyperedge $K$ in $\mH$.
The vertices $x$ and $y$ are non-neighbors, because if not, there exists some hyperedge $K$ in $\mH$ such that $x, y \in K$ and $K \subset R$, and thus $K$ is in $\mH'$, a contradiction to that $x$ and $y$ are non-neighbors in $\mH'$

In what follows, we assume that $\mH$ contains no hyperedge $R$ with $R = \mV$ and has at least 3 vertices (by the inductive step).

\begin{claim}
\label{cl:alpha-to-pn}
If $x$ is a $\alpha$-leaf in a reduced signed hypergraph $\mH$, then there exists a $\beta$-leaf $y$ in $\mH$ such that $x$ and $y$ are non-neighbors in $\mH$. 
\end{claim}
\begin{proof}
Assume that $x$ is an $\alpha$-leaf in $\mH$. 
Then there exists a hyperedge $R_x \in \mE^+ \cup \mE^-$ such that for any hyperedge $S \in \mE^+ \cup \mE^-$ that contains $x$, $S \subseteq R_x$.
Note that $R_x \neq \mV$.

Consider the signed hypergraph $\mH' = \reduced{\rmvertex{\mH}{x}}$, i.e., $\mH'$ is obtained by first removing $x$ from $\mH$ and then taking its reduced hypergraph.

By Proposition~\ref{prop:pn-acyclic-preserved}, $\mH'$ is signed-acyclic with size less than $k$. 
By the inductive hypothesis, $\mH'$ contains two non-neighbor $\beta$-leaves $y$ and $z$.
Since $y$ and $x$ are non-neighbors and $R_x \neq \mV$, either $y$ or $z$ is not contained in $R_x\setminus\{x\}$ and we assume that $y\notin R_x \setminus \{x\}$.

We argue that there is no hyperedge $S$ in $\mH$ that can contain both vertices $x$ and $y$. 
Indeed, if not, assume that $S$ contains both $x$ and $y$ in $\mH$. 
By definition of $R_x$, we have that $S \subseteq R_x$, and we would have $y \in R_x$, a contradiction to our choice of $y$.

Let $R_1, R_2, \dots, R_n$ be all hyperedges in $\mH$ such that $y \in R_i$ for each $i \in [n]$. 
We argue that every $R_i$ also appear in $\mH' = \reduced{\rmvertex{\mH}{x}}$.
First, every $R_i$ appears in $\mH' = \rmvertex{\mH}{x}$ since we argued that there is no hyperedge in $\mH$ that contains both $x$ and $y$.
Suppose for contradiction that some hyperedge $R_i$ is not contained in $\reduced{\rmvertex{\mH}{x}}$. 
Then there exists some hyperedge $K\in \mE^+$ such that $R_i \subseteq K \setminus \{x\} \subseteq K$, a contradiction that $\mH$ is reduced.

By the inductive hypothesis, $y$ is a $\beta$-leaf in $\mH' = \reduced{\rmvertex{\mH}{x}}$, and thus $y$ is a $\beta$-leaf in $\mH$ since the chain $R_1, R_2, \dots, R_n$ in $\mH'$ remains in $\mH$.
\end{proof}

Since $\mH$ is signed-acyclic, $\mH$ is $\alpha$-acyclic and thus contains an $\alpha$-leaf $x$. 
By Claim~\ref{cl:alpha-to-pn}, there is a $\beta$-leaf $y$ in $\mH$ that is non-neighbor with $x$. Then $y$ is also an $\alpha$-leaf in $\mH$, and by Claim~\ref{cl:alpha-to-pn} again, there is a $\beta$-leaf $z$ in $\mH$ that is non-neighbor with $y$. Hence $y$ and $z$ are non-neighbor $\beta$-leaves in $\mH$, as desired.
\end{description}
This concludes the proof.
\end{proof}

\begin{lemma}
\label{lemma:two-signed-leaves}
Every signed-acyclic signed hypergraph with at least two-vertices have two non-neighbor signed-leaves.
\end{lemma}
\begin{proof}
When $\mH$ contains at least two vertices, by Lemma~\ref{lemma:remove-var-or-atom} and~\ref{lemma:two-beta-leaf}, $\reduced{\mH}$ is reduced, signed-acyclic and contains two non-neighbor $\beta$-leaves $x$ and $y$. 
By Lemma~\ref{lemma:pivot-edge}, $x$ and $y$ are both $\pn$-leaves of $\mH$.
\end{proof}

\begin{proof}[Proof of Proposition~\ref{prop:pn-acyclic-leaf}]
Let $\mH$ be a signed-acyclic signed hypergraph. If $\mH$ contains only one vertex $x$, then it is only possible that $\mH=(\{x\}, \{\{x\}\}, \mE^-)$ where $\mE^- \subseteq \{\{x\}\}$, and in all cases $x$ is a $\pn$-leaf of $\mH$. 
If $\mH$ contains at least two vertices, the claim follows by Lemma~\ref{lemma:two-signed-leaves}.
\end{proof}




\subsection{Proof of Proposition~\ref{prop:removal-sequence}}

\begin{corollary}
\label{coro:remove-var-or-atom-pn}
Let $\mH = (\mV, \mE^+, \mE^-)$ be a signed-acyclic signed hypergraph. Then we have 
\begin{enumerate}
\item the hypergraph $\rmvertex{\mH}{x}$ is signed-acyclic, for every \rebut{vertex $x \in \mV$}; and
\label{it:rmvertex-pn}
\item let $R\in \mE^+ \cup \mE^-$ and $S \in \mE^+$ be two distinct hyperedges with $R \subseteq S$, and we have $\mH' = (\mV, \mE^+ \setminus\{R\}, \mE^- \setminus \{R\})$ is signed-acyclic.
\label{it:rmatom-pn}
\end{enumerate}
\end{corollary}
\begin{proof}
We again consider two items.

\framebox{(\ref{it:rmvertex-pn})}
Let $\mE = \mE^+ \cup \mE^-$ and we have that $\rmvertex{\mH}{x} = (\mV \setminus \{x\}, \rmvertex{\mE^+}{x}, \rmvertex{\mE^-}{x})$.
Let $\mE_x' \subseteq \rmvertex{\mE^-}{x}$ and consider the hypergraph $\mH_x' = (\mV \setminus \{x\}, \rmvertex{\mE^+}{x} \cup \mE_x')$.
By construction, we must have some $\mE' \subseteq \mE^-$ such that $\mE_x' = \rmvertex{\mE^-}{x}$.
Since $\mH$ is signed-acyclic, we have that the hypergraph $\mH' = (\mV, \mE^+ \cup \mE')$ is $\alpha$-acyclic.
Then $\rmvertex{\mH'}{x} = (\mV\setminus\{x\}, \rmvertex{\mE^+}{x} \cup \rmvertex{\mE'}{x}) = (\mV\setminus\{x\}, \rmvertex{\mE^+}{x} \cup \mE'_x) = \mH'_x$ is also $\alpha$-acyclic by Lemma~\ref{lemma:remove-var-or-atom}, as desired.
 
\framebox{(\ref{it:rmatom-pn})}
The claim is straightforward if $R \in \mE^-$. 
Assume that $R \in \mE^+$.
Consider any hypergraph $\mH' = (\mV, (\mE^+  \cup \mE') \setminus \{R\})$ where $\mE' \subseteq \mE^-$.
Note that the hypergraph $\mH'' =  (\mV, \mE^+  \cup \mE')$ is $\alpha$-acyclic since $\mH$ is signed-acyclic. 
Since $\mH''$ is $\alpha$-acyclic, $R \subseteq S$ and $S \in \mE^+$, by Lemma~\ref{lemma:remove-var-or-atom}, $\mH'$ is $\alpha$-acyclic, as desired.
\end{proof}

\begin{proposition}
\label{prop:pn-acyclic-preserved}
If $\mH$ is a signed-acyclic signed hypergraph and $x$ is a $\pn$-leaf of $\mH$, then $\removeleaf{\mH}{x}$ is signed-acyclic. 
\end{proposition}

\begin{proof}
Immediate from Corollary~\ref{coro:remove-var-or-atom-pn}.
\end{proof}

\begin{proof}[Proof of Proposition~\ref{prop:removal-sequence}]
Let $\mH = ([n], \mE^+, \mE^-)$ be a signed-acyclic signed hypergraph. 
By Proposition~\ref{prop:pn-acyclic-preserved} and~\ref{prop:pn-acyclic-leaf}, $\mH$ always has a $\pn$-elimination sequence, and (1) follows.

For (2), let $R = \{u_1, u_2, \dots, u_k\}\in \mE^-$. The claim follows if $R = [n]$.
Assume that $\mH$ contains at least two vertices and $R \subset [n]$.

Construct the following sequence of hypergraphs $\mH_{n}$, $\mH_{n-1}$, $\dots$, $\mH_{k}$, where 
\begin{itemize}
\item $\mH_n = \mH$;
\item for every $j = n, n-1, \dots, k+1$, let $v_j$ be a signed-leaf in $\mH_{j}$ such that $v_j \notin R$, and let $\mH_{j-1} = \removeleaf{\mH_j}{v_j}$.
\end{itemize}

We argue that in the second step, such a $v_j$ must exist. For $j = n$, since $R \subset [n]$ and $\mH_n$ is signed-acyclic, by Lemma~\ref{lemma:two-signed-leaves}, $\mH_n$ contains two non-neighbor signed-leaves, and there must be a signed-leaf $v_n$ of $\mH_n$ such that $v_n \notin R$.
Hence, $R \subset [n] \setminus \{v_n\}$ and $R$ remain in $\removeleaf{\mH_n}{v_{n}}$ by definition.
This argument can thus continue inductively.

Note that $R$ would contain every vertex in $\mH_k$ since $\mH_k$ has exactly $k$ vertices. Then $\mH_k$ would admit a signed-elimination sequence $(u_1,u_2,\dots,u_k)$, and thus $\mH = \mH_n$ admits a signed elimination sequence 
$(u_1,u_2,\dots,u_k, v_{k+1}, \dots, v_n)$ as desired. 
\end{proof}

\section{Missing proofs in Section~\ref{sec:enumeration-full}} 
\label{appendix:enumeration-full-proof}

\rebuttal{The formal definitions for the operations $\mathsf{currNode}.\mathsf{nextM}(\ba_{W})$ and $\mathsf{currNode}.\mathsf{prevM}(\ba_{W})$ are defined in Algorithm~\ref{alg:nextm} and~\ref{alg:prevm}, respectively.}

\begin{figure*}
\scalebox{.84}{
\centering
\begin{minipage}{0.4\textwidth}
\subfloat[A database instance $\db(=\db_4)$.]{
\begin{tabular}{c c | c}
\begin{tabular}{l l l}
\multicolumn{3}{c}{$A$} \\
\hline
$x_1$ & $x_2$ & $x_3$ \\
\hline
\rowofthree{\sfa_1}{\sfb_1}{\sfc_1} \\
\rowofthree{\sfa_2}{\sfb_2}{\sfc_2} \\
\rowofthree{\sfa_3}{\sfb_3}{\sfc_3} \\
\rowofthree{\sfa_4}{\sfb_3}{\sfc_3} \\
& & \\
& & \\
& & \\
& & \\
& & \\
& & \\
\end{tabular}
&
\begin{tabular}{l l}
\multicolumn{2}{c}{$U$} \\
\hline
$x_3$ & $x_4$ \\
\hline
\rowcolor{Gray}
\rowoftwo{\sfc_1}{\sfd_1} \\
\rowoftwo{\sfc_1}{\sfd_2} \\
\rowoftwo{\sfc_2}{\sfd_1} \\
\rowcolor{white}
\rowoftwo{\sfc_2}{\sfd_3} \\
\rowcolor{Gray}
\rowoftwo{\sfc_2}{\sfd_4} \\
\rowoftwo{\sfc_2}{\sfd_5} \\
\rowoftwo{\sfc_3}{\sfd_2} \\
\rowcolor{white}
\rowoftwo{\sfc_3}{\sfd_3} \\
\rowcolor{Gray}
\rowoftwo{\sfc_3}{\sfd_4} \\
\rowoftwo{\sfc_3}{\sfd_5} \\
\end{tabular}
&
\begin{tabular}{l l l l}
& & & $V$ \\
\cline{4-4}
& & & $x_4$ \\
\cline{4-4}
& & & $\sfd_3$ \\
& \multicolumn{3}{c}{$R$} \\
\cline{2-4}
& $x_2$ & $x_3$ & $x_4$ \\
\cline{2-4}
\rowoffour{}{\textcolor{blue}{\sfb_3}}{\textcolor{blue}{\sfc_3}}{\textcolor{blue}{\sfd_2}} \\
\rowoffour{}{\textcolor{red}{\sfb_3}}{\textcolor{red}{\sfc_3}}{\textcolor{red}{\sfd_5}} \\
\multicolumn{4}{c}{$S$} \\
\hline
$x_1$ & $x_2$ & $x_3$ & $x_4$ \\
\hline
\rowoffour{\textcolor{violet}{\sfa_2}}{\textcolor{violet}{\sfb_2}}{\textcolor{violet}{\sfc_2}}{\textcolor{violet}{\sfd_4}} \\
\rowoffour{\sfa_4}{\sfb_3}{\sfc_3}{\sfd_4}
\end{tabular}
\end{tabular}
}
\quad
\subfloat[A new database instance $\db_3$.]{
\begin{tabular}{c c | c}
\begin{tabular}{l l l}
\multicolumn{3}{c}{$A$} \\
\hline
$x_1$ & $x_2$ & $x_3$ \\
\hline
\rowcolor{Gray}
\rowofthree{\sfa_1}{\sfb_1}{\sfc_1} \\
\rowofthree{\sfa_2}{\sfb_2}{\sfc_2} \\
\rowofthree{\sfa_3}{\sfb_3}{\sfc_3} \\
\rowcolor{white}
\rowofthree{\sfa_4}{\sfb_3}{\sfc_3} 
\end{tabular}
&
\begin{tabular}{l l l}
\multicolumn{1}{c}{$U$} & & \\
\cline{1-1}
$x_3$ & &\\
\cline{1-1}
$\sfc_1$ & &\\
$\sfc_2$ & &\\
$\sfc_3$ & &\\
 \\
\end{tabular}
&
\begin{tabular}{l l l}
& \multicolumn{2}{c}{$R$} \\
\cline{2-3}
& $x_2$ & $x_3$  \\
\cline{2-3}
\\
\multicolumn{3}{c}{$S$} \\
\hline
$x_1$ & $x_2$ & $x_3$ \\
\hline
\rowofthree{\sfa_4}{\sfb_3}{\sfc_3} \\
\end{tabular}
\end{tabular}	
}
\quad
\centering
\subfloat[$\db_2$.]{
\begin{tabular}{c}
\begin{tabular}{l l}
\multicolumn{2}{c}{$A$} \\
\hline
$x_1$ & $x_2$  \\
\hline
\rowoftwo{\sfa_1}{\sfb_1} \\
\rowoftwo{\sfa_2}{\sfb_2} \\
\rowoftwo{\sfa_3}{\sfb_3}
\end{tabular}
\end{tabular}	
}

\quad\centering

\subfloat[$\db_1$.]{
\begin{tabular}{c}
\begin{tabular}{l}
\multicolumn{1}{c}{$A$} \\
\hline
 $x_1$  \\
\hline
$\sfa_1$ \\
${\sfa_2}$ \\
${\sfa_3}$
\end{tabular}
\end{tabular}		
}
\end{minipage}%
\begin{minipage}{0.6\textwidth}
\subfloat[The list data structure $\mL_{4}$.]{
\begin{tikzpicture}[list/.style={rectangle,draw}, >=stealth, start chain,startnode/.style={
        draw,minimum width=0.75cm,minimum height=1.4cm}, bot/.style={}]   

\node[startnode] (c1) {$\sfc_1$};

\node[bot,left= of c1] (c1nullleft) {$\bot$};
\node[list,right=of c1] (d11) {$\sfd_1$};
\node[list,right=of d11] (d12) {$\sfd_2$};
\node[bot,right=of d12] (c1nullright) {$\bot$};

\path[<-] (c1nullleft.east) edge (c1.west);
\path[<->] (c1.east) edge (d11.west);
\path[<->] (d11.east) edge (d12.west);
\path[->] (d12.east) edge (c1nullright);

\node[startnode,below=0pt of c1] (c2) {$\sfc_2$};

\node[bot,left= of c2] (c2nullleft) {$\bot$};
\node[list,right=of c2] (d21) {$\sfd_1$};
\node[list,right=of d21] (d24) {\textcolor{violet}{$\sfd_4$}};
\node[list,right=of d24] (d25) {$\sfd_5$};
\node[bot,right=of d25] (c2nullright) {$\bot$};

\path[<-] (c2nullleft.east) edge (c2.west);
\path[<->] (c2.east) edge (d21.west);
\path[<->] (d21.east) edge (d24.west);
\path[<->] (d24.east) edge (d25.west);
\path[->] (d25.east) edge (c2nullright);

\node[startnode,below=0pt of c2] (c3) {$\sfc_3$};

\node[bot,left=of c3] (c3nullleft) {$\bot$};
\node[list,right=of c3] (d32) {\textcolor{blue}{$\sfd_2$}};
\node[list,right=of d32] (d34) {$\sfd_4$};
\node[list,right=of d34] (d35) {\textcolor{red}{$\sfd_5$}};
\node[bot,right=of d35] (c3nullright) {$\bot$};

\path[<-] (c3nullleft.east) edge (c3.west);
\path[<->] (c3.east) edge (d32.west);
\path[<->] (d32.east) edge (d34.west);
\path[<->] (d34.east) edge (d35.west);
\path[->] (d35.east) edge (c3nullright);


\path[dotted,<->] (d21.north east) edge[bend left] node[above] {\textcolor{violet}{$(\sfa_2, \sfb_2, \sfc_2)$}} (d25.north west);

\path[dotted,->] (c3.east) +(0,-0.2) edge[bend right] node[above] {\textcolor{black}{$(\sfa_4, \sfb_3, \sfc_3)$}} (c3nullright);

\path[densely dotted,<->] (c3.east) +(0,+0.2) edge[bend left] node[above] {$\textcolor{blue}{(\sfb_3, \sfc_3)}$} (d34.north west);
\path[densely dotted,->] (d34.north east) edge[bend left] node[above] {$\textcolor{red}{(\sfb_3, \sfc_3)}$} (c3nullright);



\end{tikzpicture}
} 
\quad \centering
\subfloat[The list data structure $\mL_{3}$.]{
\begin{tikzpicture}[list/.style={rectangle,draw}, >=stealth, start chain,startnode/.style={
        draw,minimum width=0.75cm,minimum height=1.4cm}, bot/.style={}]   

\node[startnode] (c1) {$(\sfa_1,\sfb_1)$};

\node[bot,left= of c1] (c1nullleft) {$\bot$};
\node[list,right=of c1] (d11) {$\sfc_1$};
\node[bot,right=of d11] (c1nullright) {$\bot$};

\path[<-] (c1nullleft.east) edge (c1.west);
\path[<->] (c1.east) edge (d11.west);
\path[<->] (d11.east) edge (c1nullright);

\node[startnode,below=0pt of c1] (c2) {$(\sfa_2,\sfb_2)$};

\node[bot,left= of c2] (c2nullleft) {$\bot$};
\node[list,right=of c2] (d21) {$\sfc_2$};
\node[bot,right=of d21] (c2nullright) {$\bot$};

\path[<-] (c2nullleft.east) edge (c2.west);
\path[<->] (c2.east) edge (d21.west);
\path[<->] (d21.east) edge (c2nullright);

\node[startnode,below=0pt of c2] (c3) {$(\sfa_3,\sfb_3)$};

\node[bot,left=of c3] (c3nullleft) {$\bot$};
\node[list,right=of c3] (d32) {$\sfc_3$};
\node[bot,right=of d32] (c3nullright) {$\bot$};

\path[<-] (c3nullleft.east) edge (c3.west);
\path[<->] (c3.east) edge (d32.west);
\path[<->] (d32.east) edge (c3nullright);







\end{tikzpicture}
} 
\centering
\subfloat[The list data structure $\mL_{2}$.]{
\begin{tikzpicture}[list/.style={rectangle,draw}, >=stealth, start chain,startnode/.style={
        draw,minimum width=0.75cm,minimum height=1.4cm}, bot/.style={}]   

\node[startnode] (c1) {$\sfa_1$};

\node[bot,left= of c1] (c1nullleft) {$\bot$};
\node[list,right=of c1] (d11) {$\sfb_1$};
\node[bot,right=of d11] (c1nullright) {$\bot$};

\path[<-] (c1nullleft.east) edge (c1.west);
\path[<->] (c1.east) edge (d11.west);
\path[<->] (d11.east) edge (c1nullright);

\node[startnode,below=0pt of c1] (c2) {$\sfa_2$};

\node[bot,left= of c2] (c2nullleft) {$\bot$};
\node[list,right=of c2] (d21) {$\sfb_2$};
\node[bot,right=of d21] (c2nullright) {$\bot$};

\path[<-] (c2nullleft.east) edge (c2.west);
\path[<->] (c2.east) edge (d21.west);
\path[<->] (d21.east) edge (c2nullright);

\node[startnode,below=0pt of c2] (c3) {$\sfa_3$};

\node[bot,left=of c3] (c3nullleft) {$\bot$};
\node[list,right=of c3] (d32) {$\sfb_3$};
\node[bot,right=of d32] (c3nullright) {$\bot$};

\path[<-] (c3nullleft.east) edge (c3.west);
\path[<->] (c3.east) edge (d32.west);
\path[<->] (d32.east) edge (c3nullright);







\end{tikzpicture}
} 
\quad \centering
\subfloat[The list data structure $\mL_{1}$.]{
\begin{tikzpicture}[list/.style={rectangle,draw}, >=stealth, start chain,startnode/.style={
        draw,minimum width=0.75cm,minimum height=1.4cm}, bot/.style={}]   

\node[startnode,below=0pt of c1] (c2) {$\emptyset$};
\node[above=of c2] {};

\node[bot,left= of c2] (c2nullleft) {$\bot$};
\node[list,right=of c2] (d21) {$\sfa_1$};
\node[list,right=of d21] (d24) {$\sfa_2$};
\node[list,right=of d24] (d25) {$\sfa_3$};
\node[bot,right=of d25] (c2nullright) {$\bot$};

\path[<-] (c2nullleft.east) edge (c2.west);
\path[<->] (c2.east) edge (d21.west);
\path[<->] (d21.east) edge (d24.west);
\path[<->] (d24.east) edge (d25.west);
\path[->] (d25.east) edge (c2nullright);

\end{tikzpicture}
} 
\end{minipage}%
}
\caption{\rebuttal{Intermediate databases and list data structures produced for Example~\ref{ex:preprocessing-full}.}}
\label{fig:preprocessing-appendix}
\end{figure*}
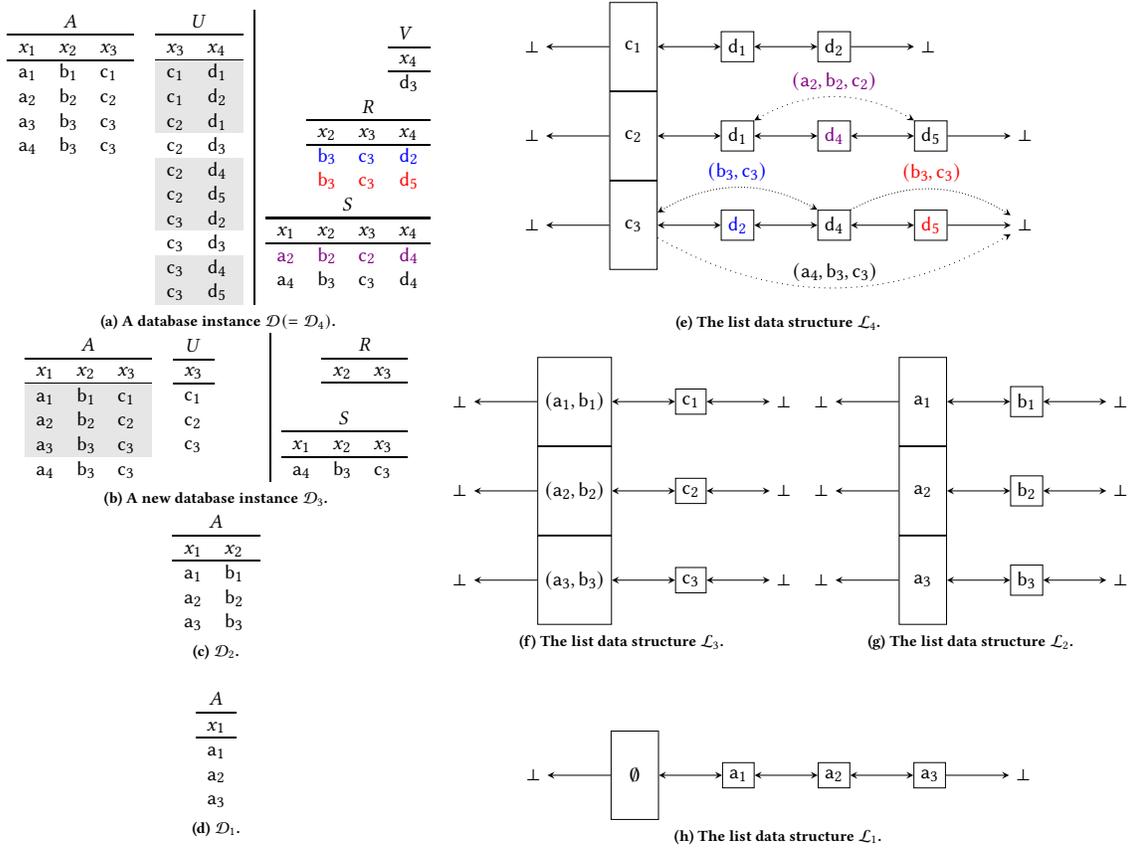

\rebuttal{

\begin{algorithm}[t]
\caption{$\mathsf{currNode}.\mathsf{nextM}(\ba_{W})$}
\label{alg:nextm}
\KwIn{a list node $\mathsf{currNode}$ from $\mL_v$, a tuple $\ba_{W}$ with $W \supseteq U \setminus \{v\}$}
\SetKwInOut{Output}{Global variables of $\mL_v$}
\Output{$U \in \mE^+$, $\{v\} = N_0 \subseteq N_1, N_2 \dots, N_i \in \mE^-$ s.t.~$v \in U$, $U \subseteq N_1 \subseteq N_2 \subseteq \dots \subseteq N_i$}
\KwOut{the next node of $\mathsf{currNode}$ following $\ba_{W}$}
$\ell \leftarrow $ the largest in $\{0,1,\dots,m\}$ such that $\mathsf{currNode}.\nextptr[\Pi_{N_i \setminus \{v\}} \ba_W]$ is defined \\
\Return{$\mathsf{currNode}.\nextptr[\Pi_{N_{\ell} \setminus \{v\}} \ba_W]$}
\end{algorithm}

\begin{algorithm}[t]
\caption{$\mathsf{currNode}.\mathsf{prevM}(\ba_{W})$}
\label{alg:prevm}
\KwIn{a list node $\mathsf{currNode}$ from $\mL_v$, a tuple $\ba_{W}$ with $W \supseteq U \setminus \{v\}$}
\SetKwInOut{Output}{Global variables of $\mL_v$}
\Output{$U \in \mE^+$, $\{v\} = N_0 \subseteq N_1, N_2 \dots, N_i \in \mE^-$ s.t.~$v \in U$, $U \subseteq N_1 \subseteq N_2 \subseteq \dots \subseteq N_i$}
\KwOut{the previous node of $\mathsf{currNode}$ following $\ba_{W}$}
$\ell \leftarrow $ the largest in $\{0,1,\dots,m\}$ such that $\mathsf{currNode}.\prevptr[\Pi_{N_i \setminus \{v\}} \ba_W]$ is defined \\
\Return{$\mathsf{currNode}.\prevptr[\Pi_{N_{\ell} \setminus \{v\}} \ba_W]$}
\end{algorithm}

\begin{example}
\label{ex:preprocessing-full}
Consider the query $Q(=Q_4)$ in Example~\ref{ex:pn:acyclic}, the database $\db$ in Figure~\ref{fig:preprocessing-appendix}(a) and a $\pn$-elimination sequence $\sigma = (1,2,3,4)$. 
\begin{align*}
Q_4(x_1,x_2,x_3,x_4) & \gets  A(x_1,x_2,x_3) \land U(x_3, x_4) \land \lnot V(x_4) \land \lnot R(x_2,x_3,x_4) \land \lnot S(x_1,x_2,x_3,x_4)
\end{align*}

\introparagraph{Preprocessing step}
We demonstrate the recursive steps of Algorithm~\ref{alg:full-scq-preprocessing} on inputs $(\mH_4, \db_4, (1, 2, 3, 4))$.

\textit{Step} (1). We remove the signed leaf $x_4$ from $\mH_4$, which yields the following query $Q_3(x_1, x_2, x_3)$ whose hypergraph corresponds to the hypergraph $\mH_3 = \removeleaf{\mH_4}{x_4}$ and emits a skipping list data structure $\mL_4$ as shown in Figure~\ref{fig:preprocessing-appendix}(e):
\begin{align*}
Q_3(x_1,x_2,x_3) & \gets  A(x_1,x_2,x_3) \land U(x_3) \land  \lnot R(x_2,x_3) \land \lnot S(x_1,x_2,x_3).
\end{align*}
The database is changed to $\db_3$ as in Figure~\ref{fig:preprocessing-appendix}(b).

\textit{Step} (2). We remove the signed leaf $x_3$ from $\mH_3$, yielding $Q_2(x_1, x_2)$ whose hypergraph corresponds to the hypergraph $\mH_2 = \removeleaf{\mH_3}{x_3}$ and emits $\mL_3$ as shown in Figure~\ref{fig:preprocessing-appendix}(f):
\begin{align*}
Q_2(x_1,x_2) & \gets  A(x_1,x_2).
\end{align*}
Note that in this step, the positive atom $U(x_3)$ and all negated atoms $\lnot R(x_2,x_3)$ and $\lnot S(x_1,x_2,x_3)$ are removed from $Q_3$ since their corresponding negative hyperedges are contained by the positive hyperedge corresponding to $A(x_1,x_2,x_3)$.
Further, the skipping list $\mL_3$ does not contain any skipping edge.
The database is changed to $\db_2$ as in Figure~\ref{fig:preprocessing-appendix}(c).

\textit{Step} (3). We further remove the signed leaf $x_2$ from $\mH_2$, finally yielding $Q_1(x_1)$ with a hypergraph corresponding to $\mH_1 = \removeleaf{\mH_2}{x_2}$ and emits $\mL_2$ shown in Figure~\ref{fig:preprocessing-appendix}(g):
\begin{align*}
Q_1(x_1) & \gets  A(x_1).
\end{align*}
Note that $\mL_2$ also does not contain any skipping edge.
The database is changed to $\db_1$ as in Figure~\ref{fig:preprocessing-appendix}(d).

\textit{Step} (4). We remove $x_1$ from $\mH_1$, and this step essentially creates a linked-list $\mL_1$ (shown in Figure~\ref{fig:preprocessing-appendix}(h)) on the remaining elements in the relation $A$ in Figure~\ref{fig:preprocessing-appendix}(d).

\medskip
\introparagraph{Enumeration step}
For the enumeration step, we first enumerate every element in $\mL_1$ (yielding $\sfa_1, \sfa_2$ and $\sfa_3$). 
Then we use that element enumerated in $\mL_1$ as a probing tuple in the enumeration process of every element in $\mL_2$. For example, $\mL_2.\mathsf{nextM}(\sfa_1)$ leads to $\sfb_1$, and $\mL_2.\mathsf{nextM}(\sfa_2)$ leads to $\sfb_2$.
We continue this step using the combined tuple enumerated from $\mL_1$ and $\mL_2$ (say, $(\sfa_1, \sfb_1)$), to enumerate the elements in $\mL_3$. For example, $\mL_3.\mathsf{nextM}((\sfa_1,\sfb_1))$ gives $c_1$ and $\mL_3.\mathsf{nextM}((\sfa_2,\sfb_2))$ gives $c_2$.

Finally, we use the combined tuple enumerated from $\mL_1$, $\mL_2$, $\mL_3$ (say, $(\sfa_2, \sfb_2, \sfc_2)$), to enumerate the elements in $\mL_4$. 
This step uses the skipping links: for example, $\mL_4.\mathsf{nextM}((\sfa_2,\sfb_2, \sfc_2))$ would locate the doubly linked-list stored at the key $\sfc_2$, and then we traverse that doubly linked-list using the tuple $(\sfa_2,\sfb_2, \sfc_2)$.
We first yield $\sfd_1$, but since the skipping links leaving $\sfd_1$ contain $(\sfa_2,\sfb_2, \sfc_2)$, we follow that skipping-link and reach $\sfd_5$, bypassing $\sfd_4$ and correctly enumerating $(\sfa_2, \sfb_2, \sfc_2, \sfd_1)$ and $(\sfa_2, \sfb_2, \sfc_2, \sfd_5)$ as answers.
\end{example}
}

\begin{algorithm}[t]
\caption{$\mathsf{BuildList}(v, U)$}
\label{alg:buildlist}
\KwIn{${U} \in \mE^+, v \in U$}
\KwOut{a data structure $\mL_v$}
$\mL_v \leftarrow $ an empty hashtable (returns $\bot$ always)\\
\ForEach{$\ba_{U} \in R_{U}$}{
	\If{$\mL_v[\Pi_{U \setminus \{v\}} \ba_{U}] = \bot$}{$\mathsf{newHead}.\prevptr[\emptyset] \gets \bot$ \\
	$\mathsf{newHead}.\nextptr[\emptyset] \gets \bot$ \\
	$\mL_v[\Pi_{U \setminus \{v\}} \ba_{U}]  \gets \mathsf{newHead}$
	}
	$head \gets \mL_v[\Pi_{U \setminus \{v\}} \ba_{U}] $ \\
	$\mathsf{Node}.\mathsf{val} \leftarrow \Pi_{\{v\}} \ba_{U}$ \\
	$\mathsf{Node}.\prevptr[\emptyset] \gets head $ \\
	$\mathsf{Node}.\nextptr[\emptyset] \gets head.\nextptr[\emptyset]$ \\
	$(head.\nextptr[\emptyset]).\prevptr[\emptyset] \gets \mathsf{Node}$ \\
}
\Return $\mL_v$
\end{algorithm}

Lemma~\ref{lemma:full-scq-correctness} establishes the correctness of our preprocessing/enumeration algorithms.

\begin{lemma}
\label{lemma:full-scq-correctness}
Let $Q$ be a $\pn$-acyclic full $\cqneg$ and $\db$ be a database instance.
Let $\sigma = \sigma' \cdot v$ be a $\pn$-elimination sequence of the signed hypergraph of $Q$. 
Let $Q'$, $\db'$, $\sigma'$ be the input to the recursive call of Algorithm~\ref{alg:full-scq-preprocessing}. 
Then the following statements hold: 
\begin{enumerate}
\item if $\ba_{\sigma}\in Q(\db)$, then $\ba_{\sigma'}\in Q'(\db')$; and 
\item if $\ba_{\sigma'}\in Q'(\db')$, then $\mL_v.\mathsf{Iterate}(\ba_{\sigma'})$ is nonempty and for every $a_v$ emitted by $\mL_v.\mathsf{Iterate}(\ba_{\sigma'})$, we have $(\ba_{\sigma'}, a_v)\in Q(\db)$.
\end{enumerate}
\end{lemma}

\begin{proof}
Consider two items.

\framebox{(1)} 
Assume that $\ba_{\sigma} \in Q(\db)$. 
Let $\ba_{\sigma'} = \Pi_{\sigma'} \ba_{\sigma}$.
Hence $\ba_{\sigma}$ satisfies all positive atoms of $Q$, and therefore $\ba_{\sigma'}$ must satisfy all positive atoms of $Q'$ by the $\alpha$-step.

Suppose for contradiction that $\Pi_{N_i \setminus \{v\}} \ba_{\sigma'} \in R_{N_i \setminus \{v\}}$ for some $i\in[m]$.
Therefore, we have that $\Pi_{U \setminus \{v\}} \ba_{\sigma'} \in \Pi_{U \setminus \{v\}} R_U$, and $\mL_v.\mathsf{nextM}(\Pi_{N_i \setminus \{v\}}\ba_{\sigma'})$ is empty.
This means that for every $a_v$ such that 
$(\Pi_{U \setminus \{v\}} \ba_{\sigma'}, a_v) \in R_U$ in $\db$, there is some $j\in[i]$ such that $(\Pi_{N_j\setminus \{v\}} \ba_{\sigma'}, a_v) \in R_{N_j}$. Then in particular, there is some $j\in[i]$ such that $(\Pi_{N_j\setminus \{v\}} \ba_{\sigma'}, \Pi_{\{v\}} \ba_{\sigma}) = \Pi_{N_j} \ba_{\sigma} \in R_{N_j}$.
However, since $\ba_{\sigma}\in Q(\db)$, $\Pi_{N_j} \ba_{\sigma} \notin R_{N_j}$, a contradiction. 
Therefore, $\Pi_{N_i \setminus \{v\}} \ba_{\sigma'} \notin \Pi_{N_i \setminus \{v\}} R_{N_i}$ in $\db'$ for every $i\in[m]$, and $\ba_{\sigma'}\in Q'(\db')$, as desired.

\framebox{(2)}
Assume that $\ba_{\sigma'}\in Q'(\db')$.
Then, we have $\Pi_{U \setminus \{v\}} \ba_{\sigma'} \in R_{U \setminus \{v\}}$, and thus $\Pi_{U \setminus \{v\}} \ba_{\sigma'} \in \mL_v$.
First, we argue that $\mL_v.\mathsf{Iterate}(\ba_{\sigma'})$ is nonempty. 
Indeed, if $\mL_v.\mathsf{Iterate}(\ba_{\sigma'})$ is empty, then $\mL_v.\mathsf{nextM}(\ba_{N_m})$ must also be empty since $N_m$ is the largest in the chain $U \subseteq N_1 \subseteq N_2 \subseteq \dots \subseteq N_m$.
Therefore, there exists some $i\in[m]$ such that $\Pi_{N_i \setminus \{v\}} \ba_{\sigma'} \in \Pi_{N_i \setminus \{v\}} R_{N_i}$ in $\db'$. 
However, since $\ba_{\sigma'}\in Q'(\db')$, $\Pi_{N_i \setminus \{v\}} \ba_{\sigma'} \notin \Pi_{N_i \setminus \{v\}} R_{N_i}$ in $\db'$, a contradiction.

Let $a_v$ be an arbitrary constant iterated by $\mL_v.\mathsf{Iterate}(\ba_{\sigma'})$.
We show that $(\ba_{\sigma'}, a_v) \in Q(\db)$. 
Since $\ba_{\sigma'} \in Q'(\db')$, $\Pi_{U \setminus \{v\}} \ba_{\sigma'} \in \Pi_{U \setminus \{v\}} R_{U}$ in $\db'$, and by construction, 
$(\Pi_{U \setminus \{v\}} \ba_{\sigma'}, a_v) \in R_U$ in $\db$ and thus satisfies every positive atom in $Q$ by the $\alpha$-step.  
It remains to show that $(\Pi_{N_i \setminus \{v\}} \ba_{\sigma'}, a_v) \notin R_{N_i}$ for every $i\in[m]$.
Indeed, if $(\Pi_{N_i \setminus \{v\}} \ba_{\sigma'}, a_v) \in R_{N_i}$ for some $i\in[m]$, then $a_v$ would have been skipped by a skipping link labeled $\Pi_{N_i \setminus \{v\}} \ba_{\sigma'}$, and thus will not be iterated by $\mL_v.\mathsf{Iterate}(\ba_{\sigma'})$, a contradiction.

The proof is now complete.
\end{proof}

\begin{proof}[Proof of Theorem~\ref{thm:boolean}]
\rebuttal{We first prove the theorem for full queries.}
Given a signed hypergraph $\mH = ([n],\mE^+, \mE^-)$, its signed-elimination sequence can be found in time $O(|Q|^3)$ using brute force: \rebuttal{we may find a signed leaf in $\mH$ in time $O(|Q|^2)$ by first iterating over every vertex and then checking whether it is a signed leaf by definition. Then we remove this signed leaf from $\mH$, and iteratively apply the previous process until the graph is empty within $O(|Q|)$ iterations. We remark that this step can potentially be improved.}

Let $\mH$ be the hypergraph of $Q$ before the preprocessing step. For every vertex $v$ in $\mH$, we denote $d(v)$ as the number of positive and negative hyperedges in $\mH$ that contains $v$. A key observation is that: for any data structure $\mL_v$, the procedures $\mathsf{prevM}$, $\mathsf{nextM}$ runs in time $O(d(v))$, since there are at most $d(v)$ projections to check. 
For implementation, we also need to add create a book keeping hash table from each tuple $\ba_{U}$ to the exact node in the hash table $\mL_v$ that contains the value $\Pi_{\{v\}} \ba_{U}$ for later use.
Therefore, the $\alpha$-step runs in $O(d(v) \cdot |R_U|)$ time, since for each tuple in $R_U$, we need to probe at most $d(v)$ relations to process it. For the $\beta$-step, the data structure $\mL_v$ can be initialized in $O(|R_U|)$ time for line $11$.
Note that line $4$ of $\mathsf{ExtendList}(\mL_v, N_i)$ only takes constant time using the book keeping hash table created in the $\alpha$-step.
For line $15$--$16$ and each $i \in \{1,2,\dots,m\}$, the running time is $O(d(v) \cdot |R_{N_i}|)$, since for each tuple in $R_{N_i}$, $\mathsf{nextM}$ and $\mathsf{prevM}$ are called, both requiring $O(d(v))$ time.

Hence one recursive step of the preprocessing phase runs in $O(d(v) \cdot (|R_U| + |R_{N_1}| + |R_{N_2}| + \dots |R_{N_m}|)) = O(d(v) \cdot |\db|)$ time.
Summing over all possible vertex $v$, we have that the preprocessing phase runs in time $\sum_{v \in \mV} O(d(v) \cdot |\db|) = O(|Q| \cdot |\db|)$.

For the enumeration phase, let $(c_1, c_2, \dots, c_n)$ and $(c_1', c_2', \dots, c_n')$ be two consecutive answers enumerated by Algorithm~\ref{alg:full-scq-enumerate-iterative}.
Let $i$ be the largest such that $c_j = c_j'$ for every $1 \leq j \leq i$.
For every $i + 1 \leq j \leq n$, $\mL_{v_j}.\mathsf{nextM}((c_1, c_2, \dots, c_j))$ is called and returns $\bot$, incurring a running time of $O(d({v_j}))$. 
Hence the overall delay is $\sum_{i+1 \leq j \leq n} O(d({v_j})) = O(|Q|)$.

\rebuttal{
Next, we consider the case where $Q$ is free-connex signed-acyclic with signed hypergraph $\mH = ([n], \mE^+, \mE^-)$ and free variables $F = [f]$ for an integer $0 \leq f < n$. 
Then the signed hypergraph $\mH' = ([n], \mE^+, \mE^- \cup \{F\})$ is signed-acyclic.
By Proposition~\ref{prop:removal-sequence}, there exists a signed-elimination sequence 
$$\sigma = (1, 2, \dots, f) \cdot \sigma'.$$

To enumerate $Q(\mathcal{D})$, the crux is to apply Algorithm~\ref{alg:full-scq-preprocessing} on inputs $(\mH', \db, \sigma)$ in the preprocessing phase, \emph{as if} $Q$ were a full query.
Subsequently in the enumeration step, we only uses the list data structures $\mathcal{L}_1$, $\mathcal{L}_2$, $\dots$, $\mathcal{L}_f$, ignoring every $\mathcal{L}_v$ where $v$ appears in the sequence $\sigma'$.
This is correct, because every answer to $Q(\mathcal{D})$ must also participate in the first $f$ positions of some full query answers, which can be recovered exactly by traversing only the first $f$ list data structures.
}

\end{proof}

\newpage
\section{Missing Details for Section~\ref{sec:enum-faqneg}: Enumeration Algorithms for $\faqneg$ and $\nestfaqneg$} \label{sec:preprocessing}

\myparagraph{Context-free Grammar ($\CFG$)}
If $u, v, w$ are strings of terminals and non-terminals, and $\nonterm ::= w$ is a rule of the $\CFG$, we say that $u \nonterm v$ \emph{yields} $uwv$ in the $\CFG$ (written as $u \nonterm v \yield uwv$). We say that $u$ \emph{derives} $v$ (written as $u \derive v$) in the $\CFG$ if $u = v$ or if there is a sequence $u_1, \dots, u_k$ for $k \geq 0$ such that 
$$ u \yield u_1 \yield u_2 \yield \cdots \yield u_k \yield v.
$$
A \emph{derivation} of a string for a grammar is a sequence of grammar rule applications that transform the start symbol into the string. A derivation proves that the string belongs to the grammar's language.

\medskip

\myparagraph{$\nestfaqneg$ Grammar \& Expression} 
We formally introduce \textit{nested functional aggregate queries with negation} ($\nestfaqneg$) over a single semiring $\bS = (\boldsymbol{D}, \oplus, \otimes, \mathbf{0}, \mathbf{1})$. As motivated in Section~\ref{sec:enum-faqneg}, $\nestfaqneg$ is a generalization of $\faqneg$ that is essential for our enumeration algorithms. 

For any $i \in [n]$, let $x_i$ denote a variable, and $a_i$ denote a value in the discrete domain $\dom(x_i)$ of the variable $x_i$. 
Overloading notation, we also refer to $[n]$ as the set of variables. 
For any subset $K \subseteq [n]$, we define $\bx_K = (x_i)_{i \in K}$ and $\ba_K = (a_i)_{i \in K} \in \prod_{i \in K} \dom(\bx_K)$, where $\dom(\bx_K) = \prod_{i \in K} \dom(x_i)$. In other words, $\bx_K$ is a tuple of variables, and $\ba_K$ is a tuple of specific values with support $K$.


\smallskip
Let $\mH = ([n], \mE^+, \mE^-)$ be a signed hypergraph. To each $K\in \mE^+$ (and $K \in \mE^-$), we associate a distinct function $R_K: \dom(\bx_K) \rightarrow \boldsymbol{D}$, called a \emph{factor}. We assume that all factors are represented via the \textit{listing representation}: each factor $R_K$ is a table of all tuples of the form $\langle \ba_K, R_K(\ba_K) \rangle$, where $R_K(\ba_K) \in \boldsymbol{D}$ is the \emph{weight} of the tuple $\ba_K$. For entries not in the table, the factor implicitly encodes their weights as $\zerobf$. Under set-theoretic context, we also use $R_K$ to denote the set of tuples of schema $\bx_K$ explicitly stored in the factor table and $\neg R_K = \dom (\bx_K) \setminus R_K$. Our definition of listing representation is slightly more general than \cite{ DBLP:conf/pods/KhamisNR16, FAQ-AI} in that we allow $R_K(\bx_K) = \mathbf{0}$ for some $\bx_K$ in the table (i.e. $\bx_K \in R_K$). A factor $R_K$ where $K \in \mE^+$ (resp. $R_N$ where $N \in \mE^-$) is called a \emph{positive} (resp. \emph{negative}) \emph{factor}. 

Over a single semiring $\bS = (\boldsymbol{D}, \oplus, \otimes, \mathbf{0}, \mathbf{1})$, a $\nestfaqneg$ expression $\formula(\bx_{[n]})$ associated with $\mH = ([n], \mE^+, \mE^-)$ is a string recognized by the following context-free grammar ($\CFG$):

\begin{equation}\label{def:cfg}
    \begin{aligned}
        \framebox{\CFG} \qquad \qquad \quad \; \\
        \text{ for all } S \subseteq [n], \\
        \nonterm(\bx_{\emptyset}) &::= \; e 
        && \text{ where } S = \emptyset \text{ and } e \in \boldsymbol{D} \setminus \{\zerobf\} \\
        \nonterm(\bx_{S}) &::= \; R_K(\bx_K) \otimes \nonterm(\bx_{S^-})
        && \text{ where } S = K \cup S^-, K \in \mE^+ \text{ and } S^- \subseteq S  \\
        & \; \; \mid \; (R_N(\bx_N) \vdash \nonterm(\bx_{N^-})) \otimes \nonterm(\bx_{S^-})
        && \text{ where } S = N \cup S^-, N^- \subseteq N \in \mE^- \text{ and } S^- \subseteq S
    \end{aligned}
\end{equation}

where
\begin{itemize}
    \item (terminals) $e, R_K(\bx_K)$ and $R_N(\bx_N)$, where $e \in \boldsymbol{D} \setminus \{\zerobf\}, K \in \mE^+$ and $N \in \mE^-$, the semiring operators $\otimes$ and $\vdash$, and the parenthesis are terminals of the $\CFG$. On its semantics, $R_N(\bx_N) \vdash \nonterm(\bx_{N^-})$ is a shorthand for $R_N(\bx_N) \oplus (\indicator{\neg R_N}(\bx_N) \otimes \nonterm(\bx_{N^-}))$, where $N^- \subseteq N$ and
    $\indicator{R_N}$ is an indicator factor defined as the following (also defined in \cite{FAQ-AI}):
    $$
    \indicator{R_N}(\bx_N)= 
    \begin{cases} \onebf & \text { if } \bx_K \in R_K \\ \zerobf & \text { otherwise. } \end{cases}
    $$
    Here, $\vdash$ is the non-commutative operator that takes a negative factor as its left operand and a $\nestfaqneg$ subexpression as its right operand.
    \item (non-terminals) $\nonterm(\bx_{S})$, where $S \subseteq [n]$, are the non-terminals of the $\CFG$, and in particular ($S = [n]$), $\nonterm(\bx_{[n]})$ is the start symbol of the $\CFG$. {A $\nestfaqneg$ subexpression is a string $\formula(\bx_S)$ derived from the $\nonterm(\bx_S)$ non-terminal, following a sequence of grammar rule applications of $\CFG$ as defined in \eqref{def:cfg}.}
    \item (size measure) the \emph{size} of a $\nestfaqneg$ subexpression $\formula(\bx_{[n]})$ is defined as $|\formula| = \sum_{i \in [n]} d(i)$, where $d(i)$ is the number of hyperedges in $\mH$ that contain $i$;
    \item (our assumptions) in our work, we assume that each factor $R_K, R_N$, where $K \in \mE^+, N \in \mE^-$, shows up precisely once in $\formula(\bx_{[n]})$, i.e., a one-to-one mapping from the hyperedges of $\mH$ to factors in $\formula(\bx_{[n]})$. Moreover, we assume that a $\nestfaqneg$ expression $\formula(\bx_{[n]})$ is {\em safe}, that is, every $i \in [n]$ appears in some $K \in \mE^+$ (i.e. $\bigcup_{{K \in \mE^+} } K = [n]$). Indeed, the ``unsafest'' case occurs when every branch takes the right child of $\vdash$ operator (thus more propensity to have un-guarded variables), because if takes the left child $R_N$, then the right sub-tree is immediately guarded by $R_N$; otherwise, the $\neg R_N$ can possibly be un-guarded. 
\end{itemize}


A $\nestfaqneg$ expression $\formula(\bx_{[n]})$ can be depicted pictorially as a \emph{abstract syntax tree} ($\AST$) drawn below, where
\begin{figure}[h]
    \centering
    \begin{tikzpicture} 
        \centering
        \node (root) at (0, 0) {$\bigotimes$}; 
    
        \node[draw,circle,inner sep=1.5pt] (oplus0) at (-6, -1) {$\vdash$};
        \node[draw,circle,inner sep=1.5pt] (oplus1) at (-2.5, -1) {$\vdash$};
    
        \node[draw,circle,inner sep=1.5pt] (oplus3) at (1.25, -1) {$\vdash$};

        \node (e) at (3, -1.25) {$e$};
        \node (rk1) at (5, -1.25) {$R_{K_1}$};
        \node (rkt) at (7, -1.25) {$R_{K_t}$};

        \node (dots1) at (-0.5, -1.15) {$\cdots$};
        \node (dots2) at (6, -1.25) {$\cdots$};
    
        \node (otimes00) at (-7, -2) {$R_{N_k}$};
        \node (otimes01) at (-5, -2) {$\bigotimes$}; 
    
        \node (otimes10) at (-3.5, -2) {$R_{N_{k-1}}$};
        \node (otimes11) at (-1.5, -2) {$\bigotimes$};

        \node (otimes20) at (0.25, -2) {$R_{N_{1}}$};
        \node (otimes21) at (2.25, -2) {$\bigotimes$};

        \node (dots2) at (-1.5, -4) {$\cdots$};

        \node (dots2) at (-5, -4) {$\cdots$};

        \node (dots3) at (2.25, -4) {$\cdots$};
    
        \draw (root) -- (oplus0);
        \draw (root) -- (oplus1);
        \draw (root) -- (oplus3);

        \draw (oplus0) -- (otimes00);
        \draw (oplus0) -- (otimes01);

        \draw (oplus1) -- (otimes10);
        \draw (oplus1) -- (otimes11);

        \draw (oplus3) -- (otimes20);
        \draw (oplus3) -- (otimes21);

        \draw[dashed] (otimes11) -- (-2.25, -4);
        \draw[dashed] (otimes11) -- (-0.75, -4);
        
        \draw[dashed] (otimes21) -- (1.5, -4);
        \draw[dashed] (otimes21) -- (3, -4);

        \draw[dashed] (otimes01) -- (-5.75, -4);
        \draw[dashed] (otimes01) -- (-4.25, -4);

        \draw (root) -- (e);
        \draw (root) -- (rk1);
        \draw (root) -- (rkt);

    \end{tikzpicture}
    \caption{The $\AST$ for a general $\nestfaqneg$ expression $\formula(\bx_{[n]})$ recognized by the $\CFG$ in \eqref{def:cfg}.}
\end{figure}
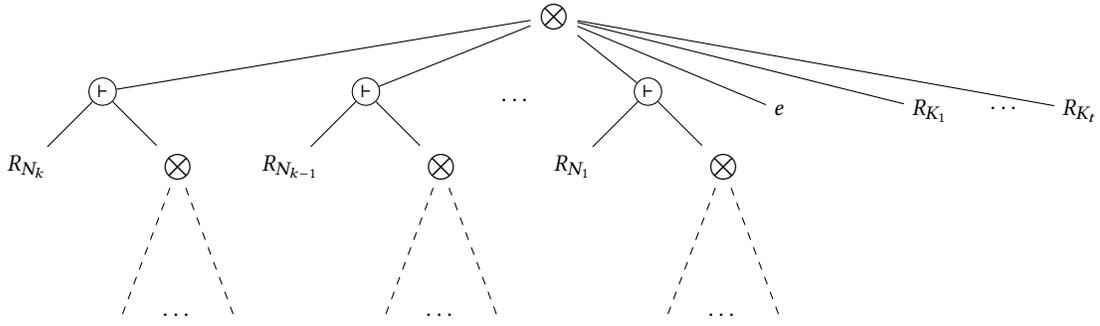

\begin{itemize}
    \item $\bigotimes$ and $\vdash$ are the intermediate nodes, and they are show up on alternating levels, starting from the root being a $\bigotimes$ node. The $\bigotimes$ node takes at least one child, but the  $\vdash$ node always takes two children. 
    \item $e, R_K(\bx_K)$ and $R_N(\bx_N)$, where $e \in \boldsymbol{D} \setminus \{\zerobf\}, K \in \mE^+$ and $N \in \mE^-$, are the leaves of the $\AST$.
\end{itemize}

\myparagraph{The $\nestfaqneg$ Queries.} 
A $\nestfaqneg$ query $\varphi$ (with free variables $F \subseteq [n]$) associated with a signed hypergraph $\mH = ([n], \mE^+, \mE^-)$ is defined as:
\begin{align}\label{def:nestfaqneg}
    \varphi(\bx_F) = \bigoplus_{\bx_{[n] \setminus F} \in \dom(\bx_{[n] \setminus F})} \; \formula(\bx_{[n]})
\end{align}
where
\begin{itemize}
    \item $\formula(\bx_{[n]})$ is a safe $\nestfaqneg$ expression recognized by the $\CFG$ in \eqref{def:cfg}. Sometimes we call $\formula(\bx_{[n]})$ a \emph{full} $\nestfaqneg$ query (i.e. $F = [n]$).
    \item $F = [f]\subseteq [n]$ is the set of \emph{free variables} for some integer $0 \leq f \leq n$.
    \item (query input and output) The input to a $\nestfaqneg$ query $\varphi$ is a database instance $\db$ that contains the constants $e \in \boldsymbol{D} \setminus \{\zerobf\}$ and a list representation for every factor in $\formula(\bx_{[n]})$. {The size of a database instance $|\db|$ is defined as the total number of rows to store the constants and the list representations.} An {\em answer} (or {\em output}) of $\varphi$ is a tuple $\langle \ba_F, \varphi(\ba_F)\rangle \in \dom(\bx_F)$ such that $\ba_F \in \dom(\bx_F)$ and $\varphi(\ba_F) \neq \zerobf$. The set of all answers of $\varphi$ is denoted by $\varphi(\db)$. 
    \item (complexity measure) {In this work, we measure the combined (and data) complexity of evaluating a $\nestfaqneg$ query $\varphi$ on a database instance $\db$ by the query size $|\varphi| = |\formula| = \sum_{i \in [n]} d(i)$ and the database size $|\db|$.} 
\end{itemize}

A fragement of $\nestfaqneg$ queries is the $\faqneg$ queries, where the corresponding $\AST$ is flat: the right child of the $\vdash$ nodes are $\bc_N \neq \zerobf$, that is, the default values of $R_N$. Indeed, the query can be compactly written as 
\begin{align*}
    \varphi(\bx_F) = \bigoplus_{\bx_{[n] \setminus F} \in \dom(\bx_{[n] \setminus F})} \; \bigotimes_{K \in \mE^+} R_K(\bx_K) \otimes \bigotimes_{N \in \mE^-} ( R_N(\bx_N) \vdash \bc_N ) 
\end{align*}
that coincides with the definition of $\faqneg$ queries in \eqref{def:faqneg} by letting $\overline{R}_N(\bx_N) = ( R_N(\bx_N) \vdash \bc_N )$. 


\begin{definition}[Free-connex $\pn$-acyclicity]
\label{defn:free-connex-pn-acyclic}
A $\nestfaqneg$ query $\varphi$ \eqref{def:nestfaqneg} is \emph{free-connex $\pn$-acyclic} if the signed hypergraph $\mH = ([n], \mE^+, \mE^- \cup \{F\})$ is $\pn$-acyclic. 
\end{definition}

In the following sections, we will show the following main theorem for free-connex $\pn$-acyclic $\nestfaqneg$ queries, which subsumes Theorem~\ref{conj:main-upper} as a special case.

\begin{theorem}\label{thm:faqneg-main}
    Let $\varphi$ be a free-connex $\pn$-acyclic $\nestfaqneg$ query with free variables $F$ over a commutative semiring $\bS$. Then, there is an algorithm that can enumerate the answers of $\varphi$ in $O(|\varphi|)$ delay, after a $O(|\varphi|^3 + |\varphi| \cdot |\db| \cdot \alpha(14 \cdot |\db|, |\db|))$ preprocessing time. 
\end{theorem}

Before jumping to the general case of free-connex $\pn$-acyclicity, we dedicate the next section to a special case, where $\varphi$ is a \emph{full} $\pn$-acyclic $\nestfaqneg$ query, i.e. $F = [n]$.

\subsection{Enumeration of full $\nestfaqneg$} \label{sec:enum-full-faqneg}

In this section, we study the enumeration problem $\Enum(\varphi, \db)$ for $\pn$-acyclic \emph{full} $\nestfaqneg$ queries. First, we recall the definition \eqref{def:nestfaqneg} and \eqref{def:cfg} that: a $\nestfaqneg$ query associated with a signed hypergraph $\mH = ([n], \mE^+, \mE^-)$ is full if $F = [n]$, that is, $\varphi(\bx_{[n]}) = \formula(\bx_{[n]})$, where $\formula(\bx_{[n]})$ is a safe $\nestfaqneg$ expression associated with $\mH$. The enumeration algorithm can be summarized as follows. 
\begin{enumerate}
    \item (pre-processing phase) we first reduce the $\Enum(\varphi, \db)$ problem into a $\Enum(Q^*, \db^*)$ problem, where $Q^*$ is a $\pn$-acyclic \emph{full} $\cqneg$ query that evaluates on a new database instance $\db^*$; then, we follow the pre-processing algorithm for $\pn$-acyclic \emph{full} $\cqneg$ queries presented in Section~\ref{sec:enumeration-full}.
    \item (enumeration phase) we enumerate the output tuples of $Q^*$ via the enumeration algorithm for $\pn$-acyclic \emph{full} $\cqneg$ queries presented in Section~\ref{sec:enumeration-full}; then we plug the tuples into $\varphi$ to get its corresponding weight.
\end{enumerate}

Now we present the construction of $Q^*$ and $\db^*$. First, the desired $Q^*$ is the following full $\cqneg$ query associated with $\mH$:
\begin{align} \label{eq:q-for-full-faqneg}
    Q^*(\bx_{[n]}) &= \bigwedge_{K \in \mE^+} R^*_K(\bx_{K}) \wedge \bigwedge_{N \in \mE^-} \neg R^*_N(\bx_{N}),
\end{align}
where for every $K \in \mE^+$, we add a positive atom $R^*_K(\bx_{K})$, and for every $N \in \mE^-$, we add a negative atom $R^*_N(\bx_{N})$ (and place a $\neg$ symbol in front) into the body of $Q^*$. It is easy to see that \eqref{eq:q-for-full-faqneg} is a full $\pn$-acyclic \emph{full} $\cqneg$ query associated with $\mH$. Next, we construct the corresponding database instance $\db^*$ for every atom in $Q^*$ defined as follows: 
\begin{equation}\label{eq:db-for-full-faqneg}
    \begin{aligned} 
        R^*_K &= \{\ba_K \in R_K \mid R_K(\ba_K) \neq \zerobf\},  && \text{ for every } K \in \mE^+, \\
        R^*_N &= \{\ba_N \in R_N \mid R_N(\ba_N) = \zerobf\}, && \text{ for every } N \in \mE^-.
    \end{aligned}
\end{equation}
Thus, $\db^*$ is of size $O(|\db|)$ and can be constructed in $O(|\db|)$ time by scanning over the list representation of every factor in $\db$. We reason about the such a construction of $\Enum(Q^*, \db^*)$ through the following theorem:

\begin{theorem}
\label{thm:enum-full-faqneg}
Let $\varphi$ be a $\pn$-acyclic full $\nestfaqneg$ query associated with a signed hypergraph $\mH = ([n], \mE^+, \mE^-)$ over a commutative semiring $\bS$, and $\db$ be a corresponding database instance. Then, a tuple $\ba_{[n]}$ is an answer of $\varphi$ (i.e. $\varphi(\ba_{[n]}) \neq \zerobf$) if and only if $\ba_{[n]}$ is an answer of $Q^*(\db^*)$ (i.e. $\ba_{[n]} \in Q^*(\db^*)$, or $Q^*(\ba_{[n]}) = \mathsf{true}$), where $Q^*$ and $\db^*$ are defined in \eqref{eq:q-for-full-faqneg} and \eqref{eq:db-for-full-faqneg}, respectively. 
\end{theorem}

\begin{proof}
    We first prove the ``if'' direction. Suppose $\ba = \ba_{[n]} \in Q^*(\db^*)$. Then, we have $\Pi_K \ba \in R^*_K$, for every $K \in \mE^+$, and $\Pi_N \ba \notin R^*_N$, for every $N \in \mE^-$. This implies that $R_K(\Pi_K \ba) \neq \zerobf$, for every $K \in \mE^+$. However, for every $N \in \mE^-$, either $\Pi_N \ba \notin R_N$, or $\Pi_N \ba \in R_N \wedge R_N(\Pi_N \ba) \neq \zerobf$. We prove that $\ba \in \varphi(\db)$, or in other words, $\formula(\ba) \neq \zerobf$, by induction on the $\nestfaqneg$ subexpression $\formula(\bx_{S})$ following the rules of the $\CFG$ \eqref{def:cfg} in a bottom-up fashion. The base case $S = \emptyset$ is trivial, because $\formula(\Pi_{\emptyset} \ba) = e$, where $ e \in \boldsymbol{D} \setminus \{\zerobf\}$. We argue the inductive step. If $\nonterm(\bx_S) \derive \formula(\bx_S)$ uses the production $\nonterm(\bx_{S}) ::= \; R_K(\bx_K) \otimes \nonterm(\bx_{S^-})$ next, and by induction hypothesis $\formula(\Pi_{S^-}\ba) \neq \zerobf$, then $\formula(\Pi_{S}\ba) = R_K(\Pi_{K}\ba) \otimes \formula(\Pi_{S^-}\ba) \neq \zerobf$. Otherwise, $\nonterm(\bx_S) \derive \formula(\bx_S)$ uses the production $(R_N(\bx_N) \vdash \nonterm(\bx_{N^-})) \otimes \nonterm(\bx_{S^-})$ as the next yield step, and by induction hypothesis $\formula(\Pi_{S^-}\ba) \neq \zerobf$ and $\formula(\Pi_{N^-}\ba) \neq \zerobf$. We distinguish the following two cases:
    \begin{itemize}
        \item If $\Pi_N \ba \notin R_N$, then 
        \begin{align*}
            R_N(\Pi_N \ba) \vdash \formula(\Pi_{N^-} \ba) = R_N(\Pi_N \ba) \oplus \indicator{\neg R_N} (\Pi_N \ba) \otimes \formula(\Pi_{N^-} \ba) = \zerobf \oplus \onebf \otimes \formula(\Pi_{N^-} \ba) = \formula(\Pi_{N^-} \ba) \neq \zerobf
        \end{align*}
        \item If $\Pi_N \ba \in R_N$ and $R_N(\Pi_N \ba) \neq \zerobf$, then 
        \begin{align*}
            R_N(\Pi_N \ba) \vdash \formula(\Pi_{N^-} \ba) = R_N(\Pi_N \ba) \oplus \indicator{\neg R_N} (\Pi_N \ba) \otimes \formula(\Pi_{N^-} \ba) = R_N(\Pi_N \ba) \oplus \zerobf \otimes \formula(\Pi_{N^-} \ba) = R_N(\Pi_N \ba) \neq \zerobf
        \end{align*}
    \end{itemize} 
    In both cases, we have $\formula(\Pi_{S}\ba) = (R_N(\Pi_N \ba) \vdash \formula(\bx_{N^-})) \otimes \formula(\Pi_{S^-}\ba) \neq \zerobf$. Therefore, $\varphi(\ba) \neq \zerobf$.

    \medskip
    We then prove the ``only if'' direction. Suppose $\varphi(\ba) \neq \zerobf$. We prove that $\ba \in Q^*(\db^*)$, or in other words, $\Pi_K \ba \in R^*_K$, for every $K \in \mE^+$, and $\Pi_N \ba \notin R^*_N$, for every $N \in \mE^-$. We prove by induction on the $\nestfaqneg$ subexpression $\formula(\bx_{S})$ following the rules of the $\CFG$ \eqref{def:cfg} in a top-down fashion, i.e. tracing the derivation from the start symbol $\nonterm(\bx_{[n]})$ of the $\CFG$ \eqref{def:cfg}. The base case is simply $\varphi(\ba)  = \formula(\ba) \neq \zerobf$. We now argue the inductive step and suppose $\nonterm(\bx_S)$ ($S \neq \emptyset$) is an intermediate non-terminal along the derivation of $\formula(\bx_{[n] })$. By inductive hypothesis, we have that $\formula(\Pi_{S} \ba) \neq \zerobf$. We then distinguish the following two cases:
    \begin{itemize}
        \item If $\nonterm(\bx_S) \derive \formula(\bx_S)$ uses the production $\nonterm(\bx_{S}) ::= R_K(\bx_K) \otimes \nonterm(\bx_{S^-})$ next, then $\formula(\ba_{S^-}) \neq \zerobf$ and $R_K(\Pi_K \ba) \neq \zerobf$. Therefore, $\Pi_K \ba \in R^*_K$.
        \item If $\nonterm(\bx_S) \derive \formula(\bx_S)$ uses the production $\nonterm(\bx_{S}) ::= (R_N(\bx_N) \vdash \nonterm(\bx_{N^-})) \otimes \nonterm(\bx_{S^-})$ next, then $\formula(\ba_{S^-}) \neq \zerobf$ and $(R_N(\Pi_N \ba) \vdash \formula(\Pi_{N^-} \ba)) \neq \zerobf$. The latter further implies that $\Pi_N \ba \notin R^*_N$ because if not, then 
        \begin{align*}
            R_N(\Pi_N \ba) \vdash \formula(\Pi_{N^-} \ba) = \zerobf \oplus \zerobf \otimes \formula(\Pi_{N^-} \ba) = \zerobf
        \end{align*}
        and we get a contradiction. Thus, we have proven that $\ba \in Q^*(\db^*)$ and close the proof.
    \end{itemize}
\end{proof}

An immediate corrollary from \autoref{thm:enum-full-faqneg} is the following, as a special case of \autoref{thm:faqneg-main}.
\begin{corollary} \label{cor:enum-full-faqneg}
    Let $\varphi$ be a \emph{full} $\pn$-acyclic $\nestfaqneg$ query over a commutative semiring $\bS$. Then, there is an algorithm that can enumerate the answers of $\varphi$ in $O(|\varphi|)$ delay, after a $O(|\varphi|^3 + |\varphi| \cdot |\db|)$ preprocessing time.
\end{corollary}

From the next section onwards, we study the enumeration problem of free-connex $\pn$-acyclic $\nestfaqneg$ queries.

\subsection{Enumeration of free-connex $\pn$-acyclic $\nestfaqneg$} \label{sec:enum-free-connex-faqneg}
In the following sections, we let $\varphi$ be a free-connex $\pn$-acyclic $\nestfaqneg$ query \eqref{def:nestfaqneg} with free variables $F = [f]$ and $\mH = ([n], \mE^+, \mE^-)$ be its associated signed hypergraph. We let $\formula(\bx_{[n]})$ be the $\nestfaqneg$ expression of $\varphi$, recognized by the $\CFG$ defined in \eqref{def:cfg}. As $\mH$ is $\pn$-acyclic, W.L.O.G, it accepts a $\pn$-elimination sequence $(1, 2, \dots, f \dots, n)$. 

On a high level, our preprocessing algorithm takes as input $(i)$ the $\nestfaqneg$ expression $\formula(\bx_{[n]})$, and $(ii)$ its database instance $\db$, and for each $i = n-1, n-2, \dots, f$ (following the $\pn$-elimination order), it executes a \textbf{$\pn$-elimination step on $i$}, that constructs an intermediate $\nestfaqneg$ expression $\formula_{i}(\bx_{[i]})$ and its database instance $\db_i$, where 
\begin{itemize}
    \item $\formula_{i}(\bx_{[i]})$ is associated with the signed hypergraph $\mH_i = \removeleaf{\mH_{i+1}}{i+1}$, i.e. the signed hypergraph after eliminating $n, n-1, \dots, i+1$ from $\mH$, (if $i = n-1$, then $\mH_{n-1} = \removeleaf{\mH}{n}$)
    \item $\db_{i}$ is an intermediate database instance of $\formula_{i}(\bx_{[i]})$ such that $\formula_{i}(\bx_{[i]}) = \bigoplus_{x_{i+1} \in \dom(x_{i+1})} \formula_{i+1}(\bx_{[i+1]}) $. In other words, for any $\ba_{[i]} \in \dom(\bx_{[i]})$, we have $\formula_{i}(\ba_{[i]}) = \bigoplus_{x_{i+1} \in \dom(x_{i+1})} \formula(x_{i+1}, \ba_{[i]})$, where $\formula(\bx_{[i+1]})$ is evaluated on $\db_{{i+1}}$ and $\formula_{i}(\bx_{[i]})$ is evaluated on $\db_{i}$.
\end{itemize}
The last $\nestfaqneg$ expression $\formula_f(\bx_{[f]})$, after the sequence of $\pn$-elimination steps on $n-1, n-2, \dots, f$, becomes a full $\nestfaqneg$ query whose associated signed hypergraph $\mH_f$ accepts a $\pn$-elimination sequence $(1, 2, \dots, f)$ and 
        \begin{align*}
            \varphi(\bx_F) = \bigoplus_{\bx_{[n] \setminus F} \in \dom(\bx_{[n] \setminus F})} \; \formula(\bx_{[n]}) = \bigoplus_{\bx_{[n-1] \setminus F} \in \dom(\bx_{[n-1] \setminus F})} \; \formula_{n-1}(\bx_{[n-1]}) = \cdots = \formula_f(\bx_{[f]}).
        \end{align*}

Thus, we degenerate to the case where $\varphi$ is a full $\pn$-acyclic $\nestfaqneg$ query and our algorithm simply follows our discussion in the last section (Appendix~\ref{sec:enum-full-faqneg}): reduce $\Enum(\formula_f(\bx_{[f]}), \db_{f})$ into $\Enum(Q^*_{f}(\bx_{[f]}), \db^*_{f})$, where $Q^*_{f}$ is a full $\pn$-acyclic $\cqneg$ and apply the preprocessing algorithm for full $\cqneg$. At the enumeration phase, we emit each answer $\ba_{F}$ of $Q^*_{f}(\db_{f})$ (applying the enumeration algorithm for full $\cqneg$) and plug the emitted tuple $\ba_{F}$ into $\formula_f$ to get recover its weight $\formula_f(\ba_{F}) = \varphi(\ba_{F})$.

\medskip
\introparagraph{$\pn$-elimination step}
W.L.O.G, we let $n$ be the given $\pn$-leaf of $\mH$ and appoint $U$ to be a pivot hyperedge, breaking ties arbitrarily. We call the factor $R_U(\bx_U)$ corresponding to $U$ the \emph{pivot factor}. A $\pn$-elimination step runs the following 3 algorithms consecutively:

\begin{enumerate}
    \item \textbf{the refactoring algorithm (Appendix~\ref{sec:rewriting})} The refactoring algorithm takes the given $\nestfaqneg$ expression $\formula(\bx_{[n]})$ and database instance $\db$ as input and returns $(\formula_n(\bx_{[n]}), \db_n)$ as output where 
    \begin{itemize}
        \item $\formula_n(\bx_{[n]})$ is a new $\nestfaqneg$ expression recognized by the following (more restrictive) $\CFG_n$ associated with the signed hypergraph $\mH_n = ([n], \mE^+_n, \mE^-_n)$, where $\mE^+_n = \{K \in \mE^+ \mid n \notin K\} \cup \{U \} \subseteq \mE^+$ and $\mE^-_n = \{N \in \mE^- \mid n \notin N \vee n \in U \subset N \} \subseteq \mE^-$,
            \begin{equation} \label{eq:cfg_n}
                \begin{aligned}
                    \framebox{$\CFG_n$} \qquad \qquad \; \\
                    \text{ for all } S \subseteq [n], \\
                    \nonterm_n(\bx_{\emptyset}) &::= \; e  
                    && \text{ where } S = \emptyset \text{ and } e \in \boldsymbol{D} \setminus \{\zerobf\} \\
                    \nonterm_n(\bx_{S}) &::= \; R_U(\bx_U) 
                    && \text{ where } S = U \\
                    & \; \; \mid \; R_K(\bx_K) \otimes \nonterm_n(\bx_{S^-})
                    && \text{ where } S = K \cup S^- \text{ and } K \in \{K \in \mE^+ \mid n \notin K\}   \\
                     & \; \; \mid \; (R_{N}(\bx_{N}) \vdash \nonterm_n(\bx_{N^-}))
                    && \text{ where } S = N \text{ and }  N^- \subseteq N \in \{N \in \mE^- \mid n \in U \subset N \}  \\
                    & \; \; \mid \; (R_N(\bx_N) \vdash \nonterm_n(\bx_{N^-})) \otimes \nonterm_n(\bx_{S^-})
                    && \text{ where } S = N \cup S^- \text{ and } N^- \subseteq N \in \{N \in \mE^- \mid n \notin N\} 
                \end{aligned}
            \end{equation}
        \item $\db_n$ is the corresponding database instance of $\formula_n(\bx_{[n]})$ such that $\formula_n(\bx_{[n]}) = \formula(\bx_{[n]})$. We observe that $\CFG_n$ is more restrictive than $\CFG$ \eqref{def:cfg} in that it has strictly less productions for the non-terminals $\nonterm_n(\bx_S)$ where $n \in S$. The refactoring algorithm uses the pivot factor $R_U$ to ``absorb'' the factors $R_K$ and $R_N$ where $n \in K \subseteq U$, $n \in N \subseteq U$. By the properties of a $\pn$-leaf $n$, the only hyperedges containing $n$ left in $\mH_n$ are $U$ and $N_1, \dots, N_k \in \mE^-$ for some $k$ such that $U \subseteq N_1 \subseteq \dots \subseteq N_k$.
    \end{itemize}
    \item \textbf{the oracle-construction algorithm (Appendix~\ref{sec:oracle})} The oracle-construction algorithm takes the $\nestfaqneg$ expression $\formula_n(\bx_{[n]})$ recognized by $\CFG_n$ \eqref{eq:cfg_n}, and the database instance $\db_n$ from the refactoring algorithm and constructs an oracle data structure for the following $\nestfaqneg$ subexpressions show up in $\formula_n(\bx_{[n]})$: $\formula_n(\bx_{U})$ and $\formula_n(\bx_{N_i})$, where $i \in [k]$. The oracle-construction algorithm is analogous to the $\beta$-step in the enumeration of full $\cqneg$ queries, but instead of linked lists, it builds arrays and \textsf{RangeSum} data structures that support fast aggregation in the upcoming aggregation algorithm.
    \item \textbf{the aggregation algorithm (Appendix~\ref{sec:aggregation})} The aggregation algorithm takes $(\formula_n(\bx_{[n]}), \db_n)$ and uses the constructed oracles 
    to finally constructs the desired $(\formula_{n-1}(\bx_{n-1}), \db_{n-1})$, where 
    \begin{itemize}
        \item $\formula_{n-1}(\bx_{n-1})$ is a new $\nestfaqneg$ expression recognized by the grammar $\CFG_{n-1}$ associated with $\mH_{n-1} = \removeleaf{\mH}{n}$, or equivalently, $\mH_{n-1} = ([n-1], \mE^+_{n-1}, \mE^-_{n-1})$, where $\mE^+_{n-1} = \{K \setminus \{n\} \mid K \in \mE^+_n \}$ and $\mE^-_{n-1} = \{N \setminus \{n\} \mid N \in \mE^-_n\}$. The grammar $\CFG_{n-1}$ is defined as follows:
        \begin{equation} \label{eq:cfg_n-1}
            \begin{aligned}
                \framebox{$\CFG_{n-1}$} \qquad \qquad \; \\
                \text{ for all } S \subseteq [n-1], \\
                    \nonterm_{n-1}(\bx_{\emptyset}) &::= \; e  
                    && \text{ where } S = \emptyset \text{ and } e \in \boldsymbol{D} \setminus \{\zerobf\}  \\
                    \nonterm_{n-1}(\bx_{S}) &::= \; R_K(\bx_K) \otimes \nonterm_{n-1}(\bx_{S^-}),  
                    && \text{ where } S = K \cup S^- \text{ and } K \in \mE^+_{n-1}, S^- \subseteq S  \\
                    & \; \; \mid \; (R_N(\bx_N) \vdash \nonterm_{n-1}(\bx_{N^-})) \otimes \nonterm_{n-1}(\bx_{S^-}), 
                    && \text{ where } S = N \cup S^-, N^- \subseteq N \in \mE^-_{n-1} \text{ and } S^- \subseteq S 
            \end{aligned}
        \end{equation}
        \item $\db_{n-1}$ is the corresponding database instance of $\formula_{n-1}(\bx_{n-1})$ such that $\formula_{n-1}(\db_{n-1}) = \bigoplus_{x_n \in \dom(x_n)} \formula_n(\bx_{[n]})$.
    \end{itemize}
\end{enumerate}

As a recap, a $\pn$-elimination step on $n$ aggregates out $x_n$ from the $\nestfaqneg$ expression $\formula(\bx_{[n]})$ and outputs a $\nestfaqneg$ expression $\formula_{n-1}(\bx_{[n-1]})$ associated with $\mH_{n-1}= \removeleaf{\mH}{n}$ and its database instance $\db_{n-1}$ such that $\formula_{n-1}(\bx_{[n-1]}) = \bigoplus_{x_n \in \dom(x_n)} \formula(\bx_{[n]})$. Now, we can keep applying the $\pn$-elimination step on the next $\pn$-leaf $n-1$ and so on, untill all variables in $[n] \setminus F$ have been removed. 

\subsection{The refactoring algorithm} \label{sec:rewriting}
In this section, we present the refactoring algorithm that refactors $\formula(\bx_{[n]})$ into $\formula_n(\bx_{[n]})$, a $\nestfaqneg$ expression associated with $\mH_n$. We start from $\nonterm(\bx_{[n]}) \derive \formula(\bx_{[n]})$, the derivation in $\CFG$ \eqref{def:cfg}, and refactor it step-by-step into a derivation of $\formula_n(\bx_{[n]})$ in $\CFG_n$ \eqref{eq:cfg_n}, i.e. $\nonterm_n(\bx_{[n]}) \derive \formula_n(\bx_{[n]})$. Along the way, we will also update the database instance $\db$ accordingly, leading to $\db_n$ at the end. In particular, we let 
$$
\formula_n(\bx_{[n]}) = \Rewrite \left(\nonterm(\bx_{\emptyset}) \yield \onebf, \nonterm(\bx_{[n]}) \derive \formula(\bx_{[n]}) \right),
$$
where $\Rewrite \left(\nonterm(\bx_{M}) \derive \mu(\bx_{M}) , \nonterm(\bx_{S}) \derive \formula(\bx_{S}) \right)$, for some $M, S \subseteq [n]$, is an invokation of a recursive algorithm $\Rewrite$ that takes the following inputs: 
\begin{enumerate}
    \item a stashed derivation of a $\nestfaqneg$ subexpression $\mu(\bx_{M})$, i.e. $\nonterm(\bx_{M}) \derive \mu(\bx_{M})$, such that its first yield step will be maintained to be one of the followings:
    \begin{itemize}
        \item if $n \notin M$, then it follows that $M = \emptyset$ and $\mu(\bx_{M}) \yield \onebf$; or 
        \item otherwise $n \in M$, then it applies the production of $\CFG$ that uses the largest hyperedge among all productions that can be used as the next yield step, if there is any (an arbitrary choice otherwise)
    \end{itemize}
    \item a derivation of a $\nestfaqneg$ subexpression $\formula(\bx_{S})$, i.e. $\nonterm(\bx_{S}) \derive \formula(\bx_{S})$, for some $S \subseteq [n]$;
    \item a database instance $\db$ of $\formula(\bx_{[n]})$ as a global input,
\end{enumerate}
and returns a $\nestfaqneg$ subexpression $\formula_n(\bx_{M \cup S})$ recognized by $\CFG_n$ and a database instance $\db_n$ corresponding to $\formula_n(\bx_{[n]})$, such that $\nonterm_n(\bx_{M \cup S}) \derive \formula_{n}(\bx_{M  \cup S}) = \mu(\bx_{M}) \otimes \formula(\bx_{S})$. The refactoring algorithm $\Rewrite$ is illustrated in Algorithm~\ref{alg:rewrite} and formally, we prove the following lemma.

\begin{algorithm}[t]
    \KwIn{two derivations in $\CFG$, $\nonterm(\bx_{M}) \derive \mu(\bx_{M})$ and $\nonterm(\bx_{S}) \derive \formula(\bx_{S})$, for some $M, S \subseteq [n]$ }
    \KwIn{a global database instance $\db$ of $\formula(\bx_{[n]})$}
    \KwOut{an expression $\formula_n(\bx_{M \cup S})$ recognized by $\CFG_n$ such that $\nonterm_n(\bx_{M \cup S}) \derive \formula_n(\bx_{M \cup S}) = \mu(\bx_{M}) \otimes \formula(\bx_{S})$.}
    \BlankLine
    \If{$\nonterm(\bx_{S}) \yield \onebf$}{
        \If(\Comment{base case} \label{alg:rewrite:base-case}){$\nonterm(\bx_{M}) \yield \onebf$}{
            \Return $\onebf$   \;
        }
        \ElseIf(\Comment{(2.2)} \label{alg:rewrite:2.2}){$\nonterm(\bx_{M}) \yield R_U(\bx_U) \otimes \nonterm(\bx_{M^-})$}{
            $\absorb{R_U}(\bx_U) \leftarrow R_U(\bx_U) \otimes \mu(\bx_{M^-})$  \;
            \Return $\absorb{R_U}(\bx_U)$ \label{alg:rewrite:2.2} \;
        }
        \ElseIf(\Comment{(2.1)} \label{alg:rewrite:2.1}){$\nonterm(\bx_{M}) \yield (R_N(\bx_N) \vdash \nonterm(\bx_{N^-})) \otimes \nonterm(\bx_{M^-})$}{
            $\absorb{R_N}(\bx_N) \leftarrow R_N(\bx_N) \otimes \mu(\bx_{M^-})$  \label{alg:rewrite:2.1} \;
            \Return $\left( \absorb{R_N}(\bx_N) \vdash \Rewrite \left(\nonterm(\bx_{M^-}) \derive \mu(\bx_{M^-}), \nonterm(\bx_{N^-}) \derive \formula(\bx_{N^-}) \right) \right)$ \label{alg:rewrite:case-pos-base} \;
        }
    }
    \ElseIf(\Comment{(3.3)} \label{alg:rewrite:3.3}){$\nonterm(\bx_{S}) \yield e$}{
            \Return $e \otimes e_{\emptyset} \otimes \Rewrite(\widetilde{\nonterm}(\bx_{M}) \derive \widetilde{\mu}(\bx_{M}), \nonterm(\bx_{S}) \derive \onebf)$  \label{alg:rewrite:case-e} \;
    } 

    \ElseIf(\Comment{$K \in \mE^+, S^- \subseteq [n]$ such that $S = K \cup S^-$} \label{alg:rewrite:3.1} ){$\nonterm(\bx_{S}) \yield R_K(\bx_K) \otimes \nonterm(\bx_{S^-})$}{    
        \If(\Comment{(3.1)}){$n \notin K$}{
            \Return $R_K(\bx_K) \otimes \Rewrite(\nonterm(\bx_{M}) \derive \mu(\bx_{M}), \nonterm(\bx_{S^-}) \derive \formula(\bx_{S^-}))$   \label{alg:rewrite:case-pos1} \;
        }
        \Else(\Comment{(4.1)} \label{alg:rewrite:4.1}){
            \textbf{augment} the stash derivation $\nonterm(\bx_{M}) \derive \mu(\bx_{M})$ into 
            $\nonterm(\bx_{M \cup K}) \yield R_K(\bx_K) \otimes \nonterm(\bx_{M}) \derive \mu(\bx_{M \cup K})$  \;
            \Return $\Rewrite(\nonterm(\bx_{M \cup K}) \derive \mu(\bx_{M \cup K}), \nonterm(\bx_{S^-}) \derive \formula(\bx_{S^-}))$ \label{alg:rewrite:case-pos2} \;
        }
    }
    \ElseIf(\Comment{$N \in \mE^-, N^- \subseteq N, S^- \subseteq [n]$ such that $S = N \cup S^-$}  \label{alg:rewrite:case-neg}) {$\nonterm(\bx_{S}) \yield (R_N(\bx_N) \vdash \nonterm(\bx_{N^-})) \otimes \nonterm(\bx_{S^-})$ }  
    {  
        \If(\Comment{(3.2)} \label{alg:rewrite:3.2} ){$n \notin N$}{
            \Return $(R_N(\bx_N) \vdash \mu(\bx_{N^-})) \otimes \Rewrite(R_U(\bx_U), \nonterm(\bx_{M}) \derive \mu(\bx_{M}), \nonterm(\bx_{S^-}) \derive \formula(\bx_{S^-}))$  \label{alg:rewrite:case-neg1} \;
        }
        \Else(\Comment{(4.2)} \label{alg:rewrite:4.2} ){
            \textbf{augment} the stash derivation $\nonterm(\bx_{M}) \derive \mu(\bx_{M})$ into 
            $\nonterm(\bx_{M \cup N}) \yield  \left(R_N(\bx_N) \vdash \nonterm(\bx_{N^-}) \right) \otimes \nonterm(\bx_{M}) \derive \mu(\bx_{M \cup N})$  \;
            \Return $\Rewrite(\nonterm(\bx_{M \cup N}) \derive \mu(\bx_{M \cup N}), \nonterm(\bx_{S^-}) \derive \formula(\bx_{S^-}))$ \label{alg:rewrite:case-neg2.2} \;
        }
    }
    \caption{The refactoring algorithm $\Rewrite(\nonterm(\bx_{M}) \derive \mu(\bx_{M}), \nonterm(\bx_{S}) \derive \formula(\bx_{S}))$}
    \label{alg:rewrite}
\end{algorithm}

\begin{lemma} \label{lem:rewrite}
    Let $\mH = ([n], \mE^+, \mE^-)$ be a signed hypergraph, $n$ be a $\pn$-leaf of $\mH$ and $U$ be a pivot hyperedge. Let $\varphi$ be a free-connex $\pn$-acyclic $\nestfaqneg$ query \eqref{def:nestfaqneg} associated with $\mH$ and $\db$ be a (global) database instance of $\varphi$. There is an algorithm that takes the following as input
    \begin{enumerate}
        \item a derivation of a $\nestfaqneg$ subexpression $\mu(\bx_{M})$, i.e. $\nonterm(\bx_{M}) \derive \mu(\bx_{M})$, for some $M \subseteq [n]$ such that if $n \notin M$, then $M = \emptyset$ and $\mu(\bx_{M}) \yield \onebf$;
        \item a derivation of a $\nestfaqneg$ subexpression $\formula(\bx_{S})$, i.e. $\nonterm(\bx_{S}) \derive \formula(\bx_{S})$, for some $S \subseteq [n]$;
        \item a global database instance $\db$ of $\formula(\bx_{[n]})$
    \end{enumerate}
    and then returns a subexpression $\formula_n(\bx_{M \cup S})$ of $\CFG_n$ \eqref{eq:cfg_n} such that $\nonterm_n(\bx_{M \cup S}) \derive \formula_n(\bx_{M \cup S})$ and a corresponding (global) database instance $\db_n$ such that
    \begin{enumerate}
        \item $\nonterm_n(\bx_{M \cup S}) \derive \formula_n(\bx_{M \cup S}) = \mu(\bx_{M}) \otimes \formula(\bx_{S})$,
        \item $|\db_n| = O(\dbsize)$,
        \item the length of the derivation $\nonterm_n(\bx_{M \cup S}) \derive \formula_n(\bx_{M \cup S})$ is at most $|\eta|$, where $|\eta|$ is the length of the derivation $\nonterm(\bx_{S}) \derive \formula(\bx_{S})$ plus the length of the derivation $\nonterm(\bx_{M}) \derive \mu(\bx_{M})$.
        \item runs in time $O(|\eta|^2  + d(n) \cdot \dbsize)$, where $d(n)$ is the number of hyperedges in $\mH$ that contains $n$.
    \end{enumerate}
\end{lemma}

\begin{proof}
    We prove the lemma by induction on $|\eta| \geq 2$ for every if-else branch in $\Rewrite$. We first examine the case where \textcolor{black}{the next yield step of $\nonterm(\bx_{S})$ is $\nonterm(\bx_{\emptyset}) \yield \formula(\bx_{\emptyset})  = \onebf$, so $S = \emptyset$}. We divide into $2$ main cases (case (1) and (2)), one for each next yield step of $\nonterm(\bx_{M})$: 
    \begin{enumerate}
        \item (\autoref{alg:rewrite:base-case}) the base case is when $M = \emptyset$ and $\nonterm(\bx_{M}) \yield \mu(\bx_{\emptyset}) = \onebf$, in which case ($M = S = \emptyset, |\mu| = |\formula|  = 1$) we simply return {$\onebf$} and the lemma is trivially true. Indeed, it uses the production \textcolor{black}{$\nonterm_n(\bx_{\emptyset}) ::= \onebf$ in $\CFG_n$}.
        \item otherwise, ($n \in M)$, {recall $S = \emptyset$ now, so there must be one largest hyperedge in $\mu(\bx_M)$, by the $\beta$-property of the $\pn$-leaf $n$} and we have the following cases:
        \begin{enumerate}
            \item [(2.1)] (\autoref{alg:rewrite:2.1}) let the next yield step of $\nonterm(\bx_{M})$ be 
               \begin{align*}
                    \nonterm(\bx_{M}) \yield (R_{N}(\bx_{N}) \vdash \nonterm(\bx_{N^-})) \otimes \nonterm(\bx_{M^-}) \derive (R_{N}(\bx_{N}) \vdash \mu(\bx_{N^-})) \otimes \mu(\bx_{M^-})
                \end{align*}
                where $N \in \mE^-$, $N^- \subseteq N$ and $M^- \subseteq M$ such that $M = N \cup M^-$. As $N$ is the largest hyperedge used among the next yield steps of $\nonterm(\bx_{M})$, we have $M^- \subseteq N$. Then we have
                \begin{align*}
                    \nonterm(\bx_{M}) \otimes \nonterm(\bx_{S}) & \derive (R_{N}(\bx_{N}) \vdash \nonterm(\bx_{N^-})) \otimes \nonterm(\bx_{M^-})  \otimes \onebf\\
                    & \derive (R_{N}(\bx_{N}) \vdash \mu(\bx_{N^-})) \otimes \mu(\bx_{M^-}) \otimes \onebf \\
                    & = \; \absorb{R_{N}}(\bx_{N}) \vdash \left(\mu(\bx_{N^-}) \otimes \mu(\bx_{M^-}) \right)
                \end{align*}
                where $\absorb{R_{N}}(\bx_{N}) = R_{N}(\bx_{N}) \otimes \mu(\bx_{M^-})$ is an updated negative factor that can be computed in $O(|R_N|)$ time. Recall that we return in this case
                \begin{align*}
                    \textcolor{black}{\left(\absorb{R_{N}}(\bx_{N}) \vdash \Rewrite \left(\nonterm(\bx_{M^-}) \derive \mu(\bx_{M^-}), \nonterm(\bx_{N^-}) \derive \mu(\bx_{N^-}) \right)\right)}.
                \end{align*}
                Now, we observe that the first yield step in the stashed derivation $\nonterm(\bx_{M}) \yield (R_{N}(\bx_{N}) \vdash \nonterm(\bx_{N^-})) \otimes \nonterm(\bx_{M^-})  \derive \formula(\bx_{M}) $ is trimmed off in this inner recursive call. That is, $|\eta|$ decrements by 1 in the recursion. By the induction hypothesis,
                \begin{align*}
                    \nonterm_n(\bx_{M^- \cup N^-}) & \derive \formula_n(\bx_{M^- \cup N^-}) = \mu(\bx_{M^-}) \otimes \mu(\bx_{N^-})  = \Rewrite \left(\nonterm(\bx_{M^-}) \derive \mu(\bx_{M^-}), \nonterm(\bx_{N^-}) \derive \mu(\bx_{N^-}) \right).
                \end{align*}
                Thus, we get
            \begin{align*}
                \mu(\bx_{M}) \otimes \formula(\bx_{S}) & = (R_{N}(\bx_{N}) \vdash \mu(\bx_{N^-})) \otimes \mu(\bx_{M^-}) \otimes \onebf \\
                & = \absorb{R_{N}}(\bx_{N}) \vdash \left(\mu(\bx_{M^-}) \otimes \mu(\bx_{N^-})\right) \\
                & = \left(\absorb{R_{N}}(\bx_{N}) \vdash \Rewrite \left(\nonterm(\bx_{M^-}) \derive \mu(\bx_{M^-}), \nonterm(\bx_{N^-}) \derive \mu(\bx_{N^-}) \right)\right).
            \end{align*}
            and $\nonterm_n(\bx_{M \cup S})$ indeed derives this subexpression in $\CFG_n$ because (now $S = \emptyset$ and $M = N$, so $M \cup S = N$)
            \begin{align*}
                \nonterm_n(\bx_{M \cup S}) & \; \yield \absorb{R_{N}}(\bx_{N}) \vdash \nonterm_n(\bx_{M^- \cup N^- })  \\
                & \; \yield \absorb{R_{N}}(\bx_{N}) \vdash \nonterm_n(\bx_{M^- \cup N^- })  \\
                & \; \derive \absorb{R_{N}}(\bx_{N}) \vdash \formula_n(\bx_{M^- \cup N^-})  && \text{ (induction hypothesis) }\\
                & \; = \absorb{R_{N}}(\bx_{N}) \vdash \left(\mu(\bx_{M^-}) \otimes \mu(\bx_{N^-})\right) \\
                & \; =  \mu(\bx_{M}) \otimes \formula(\bx_{S})  
            \end{align*}
            where \textcolor{black}{the first yield step here uses the production $\nonterm_n(\bx_{N}) ::= R_{N}(\bx_{N}) \vdash \nonterm_n(\bx_{N^-})$, where $S = N$ and $N^- \subseteq N \in \{N \in \mE^- \mid n \in U \subset N\}$}.
            \item [(2.2)] (\autoref{alg:rewrite:2.2}) {if there is no $N \supseteq U$ (only $N \subset U$ lower), then in the children of the root, there must be $R_U$, because $R_U$ can no longer hide lower in any subtrees of the AST}, thus let the next yield step of $\nonterm(\bx_{M})$ be 
             \begin{align*}
                \nonterm(\bx_{M}) \yield R_U(\bx_U) \otimes \nonterm(\bx_{M^-}) \derive R_U(\bx_U) \otimes \mu(\bx_{M^-})
             \end{align*}
             where $M^- \subseteq M$. As $U$ is the largest hyperedge used among the next yield steps of $\nonterm(\bx_{M})$, we have $M^- \subseteq U$. Then we have
                \begin{align*}
                    \nonterm(\bx_{M}) \otimes \nonterm(\bx_{S}) & \derive R_U(\bx_U) \otimes \nonterm(\bx_{M^-}) \otimes \onebf \\
                    & \derive R_U(\bx_U) \otimes \mu(\bx_{M^-})  \otimes \onebf  \\
                    & = \; \absorb{R_U}(\bx_U) 
                \end{align*}
                where $\absorb{R_U}(\bx_U) = R_U(\bx_U) \otimes \mu(\bx_{M^-})$ is an updated positive factor that can be computed in $O(|R_U|)$ time.Recall that we simply return in this case \textcolor{black}{$\absorb{R_U}(\bx_U)$} and the lemma is trivially true. Indeed, the next production of $\CFG_n$ to be used is \textcolor{black}{$\nonterm_n(\bx_{U}) ::= R_U(\bx_U)$}.
        \end{enumerate}    
    \end{enumerate}
    Next, we examine the case where \textcolor{black}{the next yield step of $\nonterm(\bx_{S})$ is not $\nonterm(\bx_{S}) \yield \onebf$}. Now, $\Rewrite$ prioritizes processing the next yield steps of $\nonterm(\bx_{S})$ over those of $\nonterm(\bx_{M})$. We distinguish the following two cases: one for $n$ not being contained in the hyperedge used in the next yield step, and the other for the opposite.
    
    \begin{enumerate}
        \item [(3)] First, we examine the case where $n$ is not contained in the hyperedge in the next yield step of $\nonterm(\bx_{S})$. There are 3 subcases, one for each next yield step of $\nonterm(\bx_{M})$:
            \begin{enumerate}
                \item [(3.1)] (\autoref{alg:rewrite:3.1}) $\nonterm(\bx_{S}) \yield  R_K(\bx_K) \otimes \nonterm(\bx_{S^-}) \derive R_K(\bx_K) \otimes \formula(\bx_{S^-})$, for some $K \in \mE^+, S^- \subseteq [n]$ such that $S = K \cup S^-$. If $n \notin K$, then by the induction hypothesis, 
                \begin{align*}
                    \nonterm_n(\bx_{M \cup S^-}) \derive \formula_n(\bx_{M \cup S^-}) = \Rewrite(\nonterm(\bx_{M}) \derive \mu(\bx_{M}), \nonterm(\bx_{S^-}) \derive \formula(\bx_{S^-}))
                \end{align*}
                since $|\eta|$ decrements by 1 (the first yield step $\nonterm(\bx_{S}) \yield  R_K(\bx_K) \otimes \nonterm(\bx_{S^-})$ is trimmed off in the recursive call). In the $\CFG$, we have
                \begin{align*}
                    \nonterm(\bx_{M}) \otimes \nonterm(\bx_{S}) & \yield \; \nonterm(\bx_{M}) \otimes \left( R_K(\bx_K) \otimes \nonterm(\bx_{S^-}) \right) \\
                    & \derive \; R_K(\bx_K) \otimes \left( \mu(\bx_{M}) \otimes \formula(\bx_{S^-}) \right) \\
                    &= \; R_K(\bx_K) \otimes \Rewrite(\nonterm(\bx_{M}) \derive  \mu(\bx_{M}), \nonterm(\bx_{S^-}) \derive  \formula(\bx_{S^-}))
                \end{align*}
                (recall that we return {$R_K(\bx_K) \otimes \Rewrite(\nonterm(\bx_{M}) \derive \mu(\bx_{M}), \nonterm(\bx_{S^-}) \derive \formula(\bx_{S^-}))$}). Moreover, $\nonterm_n(\bx_{M \cup S})$ derives this expression in $\CFG_n$ because 
                \begin{align*}
                    \nonterm_n(\bx_{ M \cup S}) \yield R_K(\bx_K) \otimes \nonterm_n(\bx_{M \cup S^-}) \derive R_K(\bx_K) \otimes \formula_n(\bx_{M \cup S^-}) 
                \end{align*}
                where \textcolor{black}{the first yield step here uses the production $\nonterm_n(\bx_{S}) ::= R_K(\bx_K) \otimes \nonterm_n(\bx_{S^-})$, where $S = K \cup S^-$ and $K \in \{K \in \mE^+ \mid n \notin K\} $}.
                \item [(3.2)] (\autoref{alg:rewrite:3.2}) $\nonterm(\bx_{S}) \yield (R_N(\bx_N) \vdash \nonterm(\bx_{N^-})) \otimes \nonterm(\bx_{S^-}) \derive (R_N(\bx_N) \vdash \mu(\bx_{N^-})) \otimes \formula(\bx_{S^-}) $, for some $N \in \mE^-, N^- \subseteq N$, and $S^- \subseteq [n]$ such that $S = N \cup S^-$. If $n \notin N$, then by the induction hypothesis,
                \begin{align*}
                    \nonterm_n(\bx_{M \cup S^-}) \derive \formula_n(\bx_{M \cup S^-}) = \Rewrite(\nonterm(\bx_{M}) \derive \mu(\bx_{M}), \nonterm(\bx_{S^-}) \derive \formula(\bx_{S^-}))
                \end{align*}
                since $|\eta|$ decrements by 1 (the first yield step $\nonterm(\bx_{S}) \yield (R_N(\bx_N) \vdash \nonterm(\bx_{N^-})) \otimes \nonterm(\bx_{S^-})$ is trimmed off). Indeed, in the $\CFG$, we have
                \begin{align*}
                    \nonterm(\bx_{M}) \otimes \nonterm(\bx_{S}) & \yield \; \nonterm(\bx_{M}) \otimes (R_N(\bx_N) \vdash \nonterm(\bx_{N^-})) \otimes \nonterm(\bx_{S^-}) \\
                    & = \; (R_N(\bx_N) \vdash \nonterm(\bx_{N^-})) \otimes \nonterm(\bx_{M}) \otimes  \nonterm(\bx_{S^-})   \\
                    & \derive \; (R_N(\bx_N) \vdash \mu(\bx_{N^-})) \otimes \left( \mu(\bx_{M}) \otimes \formula(\bx_{S^-}) \right) \\
                    & = \; (R_N(\bx_N) \vdash \mu(\bx_{N^-})) \otimes \Rewrite(\nonterm(\bx_{M}) \derive  \mu(\bx_{M}), \nonterm(\bx_{S^-}) \derive \formula(\bx_{S^-}))
                \end{align*}
                (recall that we return {$(R_N(\bx_N) \vdash \mu(\bx_{N^-})) \otimes \Rewrite(\nonterm(\bx_{M}) \derive \mu(\bx_{M}) , \nonterm(\bx_{S^-}) \derive \formula(\bx_{S^-}))$}). Moreover, $\nonterm_n(\bx_{M \cup S})$ derives this expression in $\CFG_n$ because 
                \begin{align*}
                    \nonterm_n(\bx_{ M \cup S}) \yield (R_N(\bx_N) \vdash \nonterm_n(\bx_{N^-}) ) \otimes \nonterm_n(\bx_{M \cup S^-}) \derive (R_N(\bx_N) \vdash \mu_n(\bx_{N^-}) ) \otimes \nonterm_n(\bx_{M \cup S^-}) \derive (R_N(\bx_N) \vdash \mu_n(\bx_{N^-}) ) \otimes \formula_n(\bx_{M \cup S^-}) .
                \end{align*}
                where \textcolor{black}{(i) the first yield step here uses the production $\nonterm_n(\bx_{M \cup S}) ::= (R_N(\bx_N) \vdash \nonterm_n(\bx_{N^-})) \otimes \nonterm_n(\bx_{M \cup S^-})$, where $S = N \cup S^-$ and $N^- \subseteq N \in \{N \in \mE^- \mid n \notin N\}$}, (ii) the second step is valid because $\nonterm_n(\bx_{N^-}) \derive \mu_n(\bx_{N^-}) $ in $\CFG_n$ as $n \notin N^-$ and (iii) the last derivation follows from the induction hypothesis. 
            \item [(3.3)] (\autoref{alg:rewrite:3.3}) Lastly, if $\nonterm(\bx_{S}) \yield e$ (so $S = \emptyset$). In the $\CFG$, let $\nonterm(\bx_{M}) \yield \widetilde{\nonterm}(\bx_{M}) \otimes \nonterm(\bx_{\emptyset})$, where $\widetilde{\nonterm}(\bx_{M}) \derive \widetilde{\mu}(\bx_{M})$ and $\nonterm(\bx_{\emptyset}) \yield e_{\emptyset} \in \boldsymbol{D} \setminus \{\zerobf\}$, we have
            \begin{align*}
                \nonterm(\bx_{M}) \otimes \nonterm(\bx_{S}) & \yield \; \left(\widetilde{\nonterm}(\bx_{M}) \otimes \nonterm(\bx_{\emptyset})\right) \otimes e \\
                & \yield \; \left(\widetilde{\nonterm}(\bx_{M}) \otimes e_{\emptyset} \right) \otimes e \\
                & \derive \; (e \otimes  e_{\emptyset}) \otimes \widetilde{\mu}(\bx_{M})  && \text{ where } \mu(\bx_{M}) = \widetilde{\mu}(\bx_{M}) \otimes e_{\emptyset} \\
                & = \; (e \otimes  e_{\emptyset}) \otimes \Rewrite(\widetilde{\nonterm}(\bx_{M}) \derive \widetilde{\mu}(\bx_{M}), \nonterm(\bx_{\emptyset}) \yield \onebf)
            \end{align*}
            Recall that we return in this case {$e \otimes e_{\emptyset} \otimes \Rewrite(\widetilde{\nonterm}(\bx_{M}) \derive \widetilde{\mu}(\bx_{M}), \nonterm(\bx_{\emptyset}) \yield \onebf)$}. The yield step $\nonterm(\bx_{S}) \yield e$ is trimmed off, so $|\eta|$ decrements by 1 in the recursive call. 
            Moreover, $\nonterm_n(\bx_{M \cup S})$ derives this expression in $\CFG_n$ because
            \begin{align*}
                \nonterm_n(\bx_{M \cup S}) = \widetilde{\nonterm_n}(\bx_{M \cup S}) \otimes \nonterm_n(\bx_{\emptyset}) \yield \widetilde{\nonterm_n}(\bx_{M \cup S}) \otimes (e \otimes  e_{\emptyset})  \derive \widetilde{\mu_n}(\bx_{M}) \otimes (e \otimes  e_{\emptyset})  = \mu(\bx_{M}) \otimes e
            \end{align*}
            where \textcolor{black}{the first yield step here uses the production $ \nonterm_n(\bx_{\emptyset}) ::= e \in \boldsymbol{D} \setminus \{\zerobf\}$ in $\CFG_n$}.
            \end{enumerate}
        \item [(4)] otherwise, $n$ is contained in the hyperedge in the next yield step of $\nonterm(\bx_{S})$. In this case, we apply a `stash' step that temporarily `stashes' this yield step into $\nonterm(\bx_{M})$ and then keep processing the next yield step of $\nonterm(\bx_{S})$.
            \begin{enumerate}
                \item [(4.1)] (\autoref{alg:rewrite:4.1}) $\nonterm(\bx_{S}) \yield  R_K(\bx_K) \otimes \nonterm(\bx_{S^-}) \derive R_K(\bx_K) \otimes \formula(\bx_{S^-})$, for some $K \in \mE^+, S^- \subseteq [n]$ such that $S = K \cup S^-$. Now $n \in K$. Let $\mu(\bx_{M \cup K})  = R_K(\bx_K) \otimes \mu(\bx_{M})$, we have
                \begin{align*}
                    \nonterm(\bx_{M}) \otimes \nonterm(\bx_{S}) & \yield \; \nonterm(\bx_{M}) \otimes R_K(\bx_K) \otimes \nonterm(\bx_{S^-}) \\
                    & \derive \; \mu(\bx_{M}) \otimes R_K(\bx_K) \otimes \formula(\bx_{S^-}) \\
                    &= \;  \mu(\bx_{M \cup K}) \otimes \formula(\bx_{S^-})
                \end{align*}
                We insert a new yield step for $R_K$ into the stashed derivation $\nonterm(\bx_{M}) \derive \mu(\bx_{M})$ and keep track of the largest hyperedge to be used as the next yield step of $\nonterm(\bx_{M})$ in the augmented derivation. W.L.O.G, the agumented derivation becomes
                $$
                \nonterm(\bx_{M \cup K}) \yield R_K(\bx_K) \otimes \nonterm(\bx_{M}) \derive R_K(\bx_K) \otimes \mu(\bx_{M}) = \mu(\bx_{M \cup K})
                $$
                and then we return {$\Rewrite \left(\nonterm(\bx_{M \cup K}) \derive \mu(\bx_{M \cup K}), \nonterm(\bx_{S^-}) \derive \formula(\bx_{S^-}) \right)$}. 
                
                \item [(4.2)] (\autoref{alg:rewrite:4.2}) $\nonterm(\bx_{S}) \yield (R_N(\bx_N) \vdash \nonterm(\bx_{N^-})) \otimes \nonterm(\bx_{S^-}) \derive (R_N(\bx_N) \vdash \mu(\bx_{N^-})) \otimes \formula(\bx_{S^-}) $, for some $N \in \mE^-, N^- \subseteq N$, and $S^- \subseteq [n]$ such that $S = N \cup S^-$. Now $n \in N$. Let $\mu(\bx_{M \cup N})  = \left(R_N(\bx_N) \vdash \mu(\bx_{N^-}) \right) \otimes \mu(\bx_{M})$, we have
                    \begin{align*}
                        \nonterm(\bx_{M}) \otimes \nonterm(\bx_{S}) & \yield \; \nonterm(\bx_{M}) \otimes \left(R_N(\bx_N) \vdash \nonterm(\bx_{N^-}) \right) \otimes \nonterm(\bx_{S^-}) \\
                        & \derive \; \mu(\bx_{M}) \otimes \left(R_N(\bx_N) \vdash \mu(\bx_{N^-}) \right) \otimes \formula(\bx_{S^-}) \\
                        &= \;  \left[\left(R_N(\bx_N) \vdash \mu(\bx_{N^-}) \right) \otimes \mu(\bx_{M})\right] \otimes \formula(\bx_{S^-})  \\
                        &= \; \mu(\bx_{M \cup N}) \otimes \mu(\bx_{S^-})  
                    \end{align*}
                    $$
                    \nonterm(\bx_{M \cup N}) \yield \left(R_N(\bx_N) \vdash \nonterm(\bx_{N^-}) \right) \otimes \nonterm(\bx_{M}) \derive \left(R_N(\bx_N) \vdash \mu(\bx_{N^-}) \right) \otimes \mu(\bx_{M}) = \mu(\bx_{M \cup N}) 
                    $$
                    and then we return {$\Rewrite \left(\nonterm(\bx_{M \cup N}) \derive \mu(\bx_{M \cup N}), \nonterm(\bx_{S^-}) \derive \formula(\bx_{S^-}) \right)$}. 
            \end{enumerate}
            Case (4) is the only if-else branch where $|\eta|$ stays the same in the recursive call, because this step essentially `stashes' the next yield step of $\nonterm(\bx_{S})$ into the  derivation for $\nonterm(\bx_{M})$. Therefore, the length of the derivation for $\formula(\bx_S)$ decrements by 1 but that of $\mu(\bx_M)$ increments by 1 in the recursive call. However, in the worst-case, this if-else branch can be visited at most $O(|\eta|)$ times, each visit incurs an $O(|\eta|)$ time overhead to place the yield step of $R_N$ into the stashed derivation $\nonterm(\bx_{M \cup N}) \derive \mu(\bx_{M \cup N})$ so that the largest hyperedge can be directly accessed upon request. Eventually, $\Rewrite$ will recurse back to one of the previous cases and $|\eta|$ will decrement by 1 in that subsequent recursive call. As a result, the correctness of the lemma in this case simply follows through from the correctness of $\Rewrite \left(\nonterm(\bx_{M \cup N}) \derive \mu(\bx_{M \cup N}), \nonterm(\bx_{S^-}) \derive \formula(\bx_{S^-}) \right)$. 
    \end{enumerate}
    Lastly, we justify that $\Rewrite$ runs in time $O(|\eta|^2 + d(n) \cdot \dbsize)$. The $\Rewrite$ algorithm recursively visits each yield step in both derivations $\nonterm(\bx_{M}) \derive \mu(\bx_{M})$ and $\nonterm(\bx_{S}) \derive \formula(\bx_{S})$ at most twice (accounting for ``stashes''). For each visit, if it does not fall into the last stash case (4), $|\eta|$ decrements by 1, and $\Rewrite$ takes $O(\dbsize)$ time to update the global database instance $\db$. Furthermore, every such update (on the database instance of $\db$) leverages the $\alpha$ or $\beta$ properties of the $\pn$-leaf $n$ to guarantee that $|\db_n| \leq |\db|$ and the update can only happen once for each hyperedge containing $n$ (since that corresponding factor does not appear in the subsequent recursive call). Thus, all updates leading to $\db_n$ take $O(d(n) \cdot \dbsize)$ time in total.
    
    For the last case (4), as discussed in paragraph of case (4), $\Rewrite$ can fall into it for at most $O(|\eta|)$ times, each time with a $O(|\eta|)$ time overhead before the next recursive call (for the `stashes'). Therefore, the total time complexity of $\Rewrite$ is $O(|\eta|^2 + d(n) \cdot \dbsize)$.
\end{proof}
    
The following theorem is a direct consequence of Lemma~\ref{lem:rewrite}.
\begin{theorem}
    \label{thm:rewrite}
    There is an algorithm that takes as input an $\nestfaqneg$ expression $\formula(\bx_{[n]})$ associated with $\mH$, thus recognized by $\CFG$ \eqref{def:cfg}, and a database instance $\db$, and then returns a $\nestfaqneg$ expression $\formula_n(\bx_{[n]})$ associated with $\mH_n$, thus recognized by $\CFG_n$ \eqref{eq:cfg_n}, and a new database instance $\db_n$ such that $\formula(\bx_{[n]}) = \formula_n(\bx_{[n]})$, $|\formula_n| \leq |\formula|$ and $|\db_n| \leq |\db|$. Moreover, the algorithm runs in time $O(|\formula|^2 + d(n) \cdot \dbsize)$, where $d(n)$ is the number of hyperedges containing $n$ in $\mH$.
\end{theorem}

\begin{proof}
    We simply call $\formula(\bx_{[n]}) = \formula_n(\bx_{[n]}) = \Rewrite \left(\nonterm(\bx_{\emptyset}) \yield \onebf, \nonterm(\bx_{[n]}) \derive \formula(\bx_{[n]}) \right)$ and trace the recursive call steps of $\Rewrite$ to construct the derivation $\nonterm_n(\bx_{[n]}) \derive \formula_n(\bx_{[n]})$ using production rules of $\CFG_n$ \eqref{eq:cfg_n}. As $\Rewrite$ (Algorithm~\ref{alg:rewrite}) only recurses on $\nonterm_n(\bx_S)$ where $n \in S$, and if not, it just follows from the original derivations in $\CFG$ (those derivations can be attached to the end of the derivation for every $\nestfaqneg$ subexpression $\nonterm_n(\bx_S) \derive \formula_n(\bx_S)$ where $n \notin S$). We therefore close the proof by invoking Lemma~\ref{lem:rewrite}. 
\end{proof}

\subsection{The aggregation algorithm} \label{sec:aggregation}
We assume in this section that we have obtained $(\formula_n(\bx_{[n]}), \db_n)$ from the call of $\Rewrite$ as in Theorem~\ref{thm:rewrite}. We now shift our focus directly to the aggregation algorithm that eliminates $x_n$, where $n$ is the $\pn$-leaf of the signed hypergraph $\mH$. We defer the oracle-construction algorithm to Appendix~\ref{sec:oracle} for now because the aggregation algorithm will provide an intuitive motivation that necessitates the oracles to be constructed.

The aggregation algorithm takes as input $\formula_n(\bx_{[n]})$ recognized by $\CFG_n$ \eqref{eq:cfg_n} and a database instance $\db_n$ and returns a new $\nestfaqneg$ expression $\formula_{n-1}(\bx_{[n-1]})$ recognized by $\CFG_{n-1}$ \eqref{eq:cfg_n-1}, and a corresponding database instance $\db_{n-1}$ such that $\bigoplus_{x_n \in \dom(x_n)}\formula_n(\bx_{[n]}) = \formula_{n-1}(\bx_{[n-1]})$. As a recap, the production rules of $\CFG_n$ are: for all $S \subseteq [n]$,
\begin{align}
    \qquad \qquad \nonterm_n(\bx_{\emptyset}) &::= \; e  
    && \text{ where } S = \emptyset \text{ and } e \in \boldsymbol{D} \setminus \{\zerobf\}\label{eq:cfg_n:rule1} \\
    \qquad \qquad \nonterm_n(\bx_{S}) &::= \; R_U(\bx_U)
    && \text{ where } S = U \label{eq:cfg_n:rule2} \\
    & \; \; \mid \; R_K(\bx_K) \otimes \nonterm_n(\bx_{S^-})
    && \text{ where } S = K \cup S^- \text{ and } K \in \{K \in \mE^+ \mid n \notin K\}  \label{eq:cfg_n:rule3} \\
    \qquad \qquad & \; \; \mid \; (R_{N}(\bx_{N}) \vdash \nonterm_n(\bx_{N^-}))
    && \text{ where } S = N \text{ and }  N^- \subseteq N \in \{N \in \mE^- \mid n \in U \subset N \} \label{eq:cfg_n:rule4} \\
    \qquad \qquad & \; \; \mid \; (R_N(\bx_N) \vdash \nonterm_n(\bx_{N^-})) \otimes \nonterm_n(\bx_{S^-})
    && \text{ where } S = N \cup S^- \text{ and } N^- \subseteq N \in \{N \in \mE^- \mid n \notin N\} \label{eq:cfg_n:rule5} 
\end{align}

The main upside of turning the $\nestfaqneg$ expression $\formula(\bx_{[n]})$ (recognized by $\CFG$) into $\formula_n(\bx_{[n]})$ (recognized by $\CFG_n$) is that the productions of $\CFG_n$ are layed out in a principled way: $x_n$ is only contained in one terminal (or non-terminal). From an $\AST$ standpoint, after refactoring, $x_n$ only resides in at most one subtree at every intermediate $\bigotimes$ node.

We safely assume that $R_{N_k}, R_{N_{k-1}}, \cdots, R_{N_1}, R_U$ are the only factors of $\formula_n(\bx_{[n]})$ that contain $x_n$ in $\formula_n(\bx_{[n]})$ and $N_k \supseteq N_{k-1} \supseteq \cdots \supseteq N_1 \supseteq U$ (we let $N_0 = U$ for convenience), gifted by the $\beta$-property of the $\pn$-leaf $n$. A convenient way to derive $\formula_n(\bx_{[n]})$ in $\CFG_n$ is as follows (let $N_{i} \supseteq N^-_{i} = S_{i-1} \cup N_{i-1}$ for $i \in [k]$)
\begin{equation} \label{eq:derivation}
    \begin{aligned}
        \nonterm_n(\bx_{[n]}) & \derive \formula_n(\bx_{S_k}) \otimes \nonterm_n(\bx_{N_k})  \\
    & \yield \formula_n(\bx_{S_k}) \otimes \left( R_{N_k}(\bx_{N_k}) \vdash \nonterm_n(\bx_{N^-_k}) \right) \\
    & \derive \formula_n(\bx_{S_k}) \otimes \left(R_{N_k}(\bx_{N_k}) \vdash \formula_n(\bx_{S_{k-1}}) \otimes \nonterm_n(\bx_{N_{k-1}}) \right) \\
    & \yield \formula_n(\bx_{S_k}) \otimes \left(R_{N_k}(\bx_{N_k}) \vdash \formula_n(\bx_{S_{k-1}}) \otimes \left(R_{N_{k-1}}(\bx_{N_{k-1}}) \vdash \nonterm_n(\bx_{N^-_{k-1}}) \right) \right) \\
    & \; \vdots \\
    & \yield \formula_n(\bx_{S_k}) \otimes \left(R_{N_k}(\bx_{N_k}) \vdash \formula_n(\bx_{S_{k-1}}) \otimes \left(R_{N_{k-1}}(\bx_{N_{k-1}}) \vdash  \cdots R_{N_{1}}(\bx_{N_{1}}) \vdash\left(\formula_n(\bx_{S_1}) \otimes (R_{N_{1}}(\bx_{N_{1}}) \vdash \nonterm_n(\bx_{N^-_1})) \right) \cdots  \right) \right)   \\
    & \derive \formula_n(\bx_{S_k}) \otimes \left(R_{N_k}(\bx_{N_k}) \vdash \formula_n(\bx_{S_{k-1}}) \otimes \left(R_{N_{k-1}}(\bx_{N_{k-1}}) \vdash \cdots R_{N_{1}}(\bx_{N_{1}}) \vdash \left(\formula_n(\bx_{S_1}) \otimes (R_{N_{1}}(\bx_{N_{1}}) \vdash \formula_n(\bx_{S_{0}}) \otimes \nonterm_n(\bx_U)) \right) \cdots  \right) \right) \\
    & \yield \formula_n(\bx_{S_k}) \otimes \left(R_{N_k}(\bx_{N_k}) \vdash \formula_n(\bx_{S_{k-1}}) \otimes \left(R_{N_{k-1}}(\bx_{N_{k-1}}) \vdash \cdots R_{N_{1}}(\bx_{N_{1}}) \vdash \left(  \formula_n(\bx_{S_1}) \otimes (R_{N_{1}}(\bx_{N_{1}}) \vdash \formula_n(\bx_{S_{0}}) \otimes R_U(\bx_U)) \right)  \cdots   \right) \right) \\
    & = \; \formula_n(\bx_{[n]})
    \end{aligned}
\end{equation}
where every $\derive$ (a derive step) in \eqref{eq:derivation} applies the productions \eqref{eq:cfg_n:rule1}, \eqref{eq:cfg_n:rule3} and \eqref{eq:cfg_n:rule5} to factor out all the terms ($e, R_K(\bx_K)$ or $R_N(\bx_N) \vdash \nonterm_n(\bx_{N^-})$) that does not contain $x_n$ and wrap them in some $\formula_n(\bx_{S_i})$ such that $\nonterm_n(\bx_{S_i}) \derive \formula_n(\bx_{S_i})$, where $i \in [k]$ and $n \notin S_i$. Then it is followed by $\yield$ (a yield step) that applies the production \eqref{eq:cfg_n:rule4} that produces the negative factors $R_{N_k}, R_{N_{k-1}}, \cdots, R_{N_1}$ from $\nonterm_n(\bx_{N_k}), \nonterm_n(\bx_{N_{k-1}}), \cdots, \nonterm_n(\bx_{N_1})$, except for the last yield step where we use the production \eqref{eq:cfg_n:rule2} to produce the pivot factor $R_U(\bx_U)$ from $\nonterm_n(\bx_U)$.

On a high level, our aggregation algorithm will follow the derivation \eqref{eq:derivation} of $\formula_n(\bx_{[n]})$ line-by-line and ``push-in'' the aggregation operator $\bigoplus_{x_n \in \dom(x_n)}$ into the more and more nested subexpressions. For succinctness, we use $\bigoplus_{x_n}$ as an abbreviation of $\bigoplus_{x_n \in \dom(x_n)}$ in the following. This ``push-in'' step for every derive step is immediate: we can simply push the aggregation operator $\bigoplus_{x_n}$ by factoring out the term $\formula_n(\bx_{S_i})$ that does not contain $x_n$. That is, for $i = 0, 1, \ldots, k$,
\begin{align*}
    \bigoplus_{x_n} \formula_n(\bx_{S_i}) \otimes \left(R_{N_i}(\bx_{N_i}) \vdash \formula_n(\bx_{N^-_{i}})  \right) & = \formula_n(\bx_{S_i}) \otimes \bigoplus_{x_n}  \left(R_{N_i}(\bx_{N_i}) \vdash \formula_n(\bx_{N^-_{i}})  \right) \\
    \bigoplus_{x_n} \formula_n(\bx_{S_0}) \otimes R_U(\bx_U) & = \formula_n(\bx_{S_0}) \otimes \bigoplus_{x_n} R_U(\bx_U) \\
    & = \formula_n(\bx_{S_0}) \otimes R_{U \setminus \{n\}}(\bx_{U \setminus \{n\}}) 
\end{align*}
Here, in the last yield step (applying the production \eqref{eq:cfg_n:rule2}), $\bigoplus_{x_n} R_U(\bx_U)$ is straightforward: we can scan the list representation of the pivot factor $R_U$ once and get a new factor $R_{U \setminus \{n\}}(\bx_{U \setminus \{n\}})$ in time $O(|R_U|)$. 

Now we turn to the tricky yield steps in the derivation \eqref{eq:derivation} that applies the production \eqref{eq:cfg_n:rule4} of the $\CFG_n$. What we want here is to ``push-in'' the aggregation operator $\bigoplus_{x_n}$ after the $\vdash$ operator so that the aggregation algorithm can proceed with the subexpression $\formula_n(\bx_{N^-})$. To that end, first we recall the semantics of the $\vdash$ operator: 
\begin{equation*}
    \bigoplus_{x_n} R_N(\bx_N) \vdash \formula_n(\bx_{N^-})  = \bigoplus_{x_n} \left( R_N(\bx_N) \oplus \indicator{\neg R_N}(\bx_N) \otimes \formula_n(\bx_{N^-}) \right).
\end{equation*}
We distinguish the following two disjoint cases:
\begin{enumerate}
    \item if $\ba_{N \setminus \{n\}} \notin \Pi_{N \setminus \{n\}} R_N$, then the aggreation simply becomes $\bigoplus_{x_n} \formula_n(\bx_{N^-})$, because $R_N(x_n, \ba_{N \setminus \{n\}}) = \zerobf$ and $\indicator{\neg R_N}(x_n, \ba_{N \setminus \{n\}}) = \onebf$, for all $x_n \in \dom(x_n)$. Thus, for such tuples $\ba_{N \setminus \{n\}}$, the push-in step is trivial.
    \item otherwise, we have $\ba_{N \setminus \{n\}} \in \Pi_{N \setminus \{n\}} R_N(\bx_N)$. In this case, we define a new factor $R_{N \setminus \{n\}}$ as follows:
    \begin{align*}
        R_{N \setminus \{n\}}(\ba_{N \setminus \{n\}})  = 
        \begin{cases} \bigoplus_{x_n} R_N({x_n, \ba_{N \setminus \{n\}}}) \oplus \indicator{\neg R_N}({x_n, \ba_{N \setminus \{n\}}}) \otimes \formula_n({x_n, \ba_{N^- \setminus \{n\}}}) & \text { if } \ba_{N \setminus \{n\}} \in \Pi_{N \setminus \{n\}} R_N(\bx_N) \\ \zerobf & \text { otherwise. } \end{cases}
    \end{align*}
    It is easy to see that the list representation of $R_{N \setminus \{n\}}$ is of size $O(|R_N|)$.
\end{enumerate}
Therefore, if we can obtain (the list representation of) this new factor $R_{N \setminus \{n\}}$ efficiently, the ``push-in'' of the aggregation operator becomes immediate because
\begin{align*}
    \bigoplus_{x_n} \formula_n(\bx_{N}) & = \bigoplus_{x_n} \left( R_N(\bx_N) \vdash \formula_n(\bx_{N^-}) \right) \\
    & = \bigoplus_{x_n} \left( R_N(\bx_N) \oplus \indicator{\neg R_N}(\bx_N) \otimes \formula_n(\bx_{N^-}) \right) \\
    & = R_{N \setminus \{n\}}(\bx_{N \setminus \{n\}}) \oplus \indicator{\neg R_{N \setminus \{n\}}}(\bx_{N \setminus \{n\}}) \otimes \bigoplus_{x_n}  \formula_n(\bx_{N^-}) \\
    & = R_{N \setminus \{n\}}(\bx_{N \setminus \{n\}}) \vdash \bigoplus_{x_n}  \formula_n(\bx_{N^-}) 
\end{align*}

Back to our big picture, the aggregation step ``pushes-in'' the aggregation operator $\bigoplus_{x_n}$ one line at a time following the derivation \eqref{eq:derivation}:
\begin{equation} \label{eq:derivation-aggregate}
    \begin{aligned}
        \bigoplus_{x_n} \formula_n(\bx_{[n]}) & = \formula_n(\bx_{S_k}) \otimes \bigoplus_{x_n} \formula_n(\bx_{N_k})  \\
        & = \formula_n(\bx_{S_k}) \otimes \left( R_{N_k \setminus \{n\}}(\bx_{N_k \setminus \{n\}}) \vdash \bigoplus_{x_n} \formula_n(\bx_{N^-_k}) \right) \\
        & = \formula_n(\bx_{S_k}) \otimes \left(R_{N_k \setminus \{n\}}(\bx_{N_k \setminus \{n\} } ) \vdash \formula_n(\bx_{S_{k-1}}) \otimes \bigoplus_{x_n} \formula_n(\bx_{N_{k-1}}) \right) \\
        & = \formula_n(\bx_{S_k}) \otimes \left(R_{N_k\setminus \{n\}}(\bx_{N_k \setminus \{n\} }) \vdash \formula_n(\bx_{S_{k-1}}) \otimes \left(R_{N_{k-1} \setminus \{n\} }(\bx_{N_{k-1} \setminus \{n\} }) \vdash \bigoplus_{x_n}  \formula_n(\bx_{N^-_{k-1}}) \right) \right) \\
        & \; \vdots \\
        & = \formula_n(\bx_{S_k}) \otimes \left(R_{N_k}(\bx_{N_k}) \vdash \formula_n(\bx_{S_{k-1}}) \otimes \left(  \cdots R_{N_{1}}(\bx_{N_{1}}) \vdash\left(\formula_n(\bx_{S_1}) \otimes (R_{N_{1}}(\bx_{N_{1}}) \vdash \bigoplus_{x_n} \formula_n(\bx_{N^-_1})) \right) \cdots  \right) \right)   \\
        & = \formula_n(\bx_{S_k}) \otimes \left(R_{N_k}(\bx_{N_k}) \vdash \formula_n(\bx_{S_{k-1}}) \otimes \left( \cdots R_{N_{1}}(\bx_{N_{1}}) \vdash \left(\formula_n(\bx_{S_1}) \otimes (R_{N_{1}}(\bx_{N_{1}}) \vdash \formula_n(\bx_{S_{0}}) \otimes \bigoplus_{x_n} \formula_n(\bx_U)) \right) \cdots  \right) \right) \\
        & = \formula_n(\bx_{S_k}) \otimes \left(R_{N_k}(\bx_{N_k}) \vdash \formula_n(\bx_{S_{k-1}}) \otimes \left( \cdots R_{N_{1}}(\bx_{N_{1}}) \vdash \left(  \formula_n(\bx_{S_1}) \otimes (R_{N_{1}}(\bx_{N_{1}}) \vdash \formula_n(\bx_{S_{0}}) \otimes \bigoplus_{x_n} R_U(\bx_U)) \right)  \cdots   \right) \right) \\
        & =: \formula_{n-1}(\bx_{[n-1]})
    \end{aligned}
\end{equation}

As one can see from the comparison with \eqref{eq:derivation}, the derivation of $\formula_{n-1}(\bx_{[n-1]})$ exactly follows the derivation of $\formula_{n}(\bx_{[n]})$ except that $x_n$ has been peeled off from the set of variables and every $\nestfaqneg$ subexpression show up in $\formula_{n}(\bx_{[n]})$. So, $\formula_{n-1}(\bx_{[n-1]})$ is a $\nestfaqneg$ expression recognized by the following intermediate context-free grammar, obtained by directly removing $x_n$ from every non-terminal of the grammar $\CFG_n$: for all $S \subseteq [n-1]$,
\begin{align*}
        \nonterm_{n-1}(\bx_{\emptyset}) &::= \; e  
        && \text{ where }  S = \emptyset \text{ and } e \in \boldsymbol{D} \setminus \{\zerobf\}   \\
        \nonterm_{n-1}(\bx_{S \setminus \{n\}}) &::= \; R_{U \setminus \{n\}}(\bx_{U \setminus \{n\}}) 
        && \text{ where } S = U \\
        & \; \; \mid \; R_{K}(\bx_{K}) \otimes \nonterm_{n-1}(\bx_{S^- \setminus \{n\}})  \label{eq:nonterm-1}
        && \text{ where } S = K \cup S^- \text{ and } K \in \{K \in \mE^+ \mid n \notin K\}   \\
        & \; \; \mid \; (R_{N \setminus \{n\}}(\bx_{N \setminus \{n\}}) \vdash \nonterm_{n-1}(\bx_{N^- \setminus \{n\}}))
        && \text{ where } S = N \setminus \{n\} \text{ and } N^-  \subseteq N \in \{N \in \mE^-  \mid n \in U \subset N \}  \\
        & \; \; \mid \; (R_N(\bx_N) \vdash \nonterm_{n-1}(\bx_{N^-})) \otimes \nonterm_{n-1}(\bx_{S^- \setminus \{n\}}) 
        && \text{ where } S = N \cup S^- \text{ and } N^- \subseteq N \in \{N \in \mE^- \mid n \notin N\} 
\end{align*}


In spite of its discrepencies to $\CFG_{n-1}$ \eqref{eq:cfg_n-1}, every production of the above grammar can be derived from $\CFG_{n-1}$. Indeed, recall that $\CFG_{n-1}$ is associated with the signed hypergraph $\mH_{n-1} = \removeleaf{\mH}{n} = ([n-1], \mE_{n-1}^+, \mE_{n-1}^-)$. Then, $U \setminus \{n\} \in \mE_{n-1}^+$, $\{K \in \mE^+ \mid n \notin K\} \subseteq \mE_{n-1}^+$ and $\{N \setminus \{n\} \mid N \in \mE^-, n \notin N \vee n \in U \subset N  \}  \subseteq \mE_{n-1}^-$. Therefore, the $\nestfaqneg$ expression $\formula_{n-1}(\bx_{[n-1]})$ is also recognized by $\CFG_{n-1}$ at the end of the elimination step for the $\pn$-leaf $n$ and the $\pn$-elimination step for $n$ now completes.

The only missing (and tricky) piece is how to compute the list representation of the new factor $R_{N \setminus \{n\}}$ efficiently. In the next section, we will introduce an efficient algorithm for this task using \textsf{RangeSum} data structures and it then to the following main theorem. Its proof is deferred to Appendix~\ref{sec:putting_things_together}, after the introduction of \textsf{RangeSum} data structures.

\begin{theorem} \label{thm:aggregation-main}
    Let $\mH_n$ be a signed hypergraph with a $\pn$-leaf $n$. Let $\formula_n(\bx_{[n]})$ be a $\nestfaqneg$ expression associated with $\mM_n$ (thus recognized by $\CFG_n$ \eqref{eq:cfg_n}) and $\db_n$ be its database instance. There is an (aggregation) algorithm that takes $(\formula_n(\bx_{[n]}), \db_n)$ as input and outputs $(\formula_{n-1}(\bx_{[n-1]}), \db_{n-1})$ where
    \begin{enumerate}
        \item $\formula_{n-1}(\bx_{[n-1]})$ is the $\nestfaqneg$ expression associated with $\mH_{n-1} = \removeleaf{\mH}{n}$ obtained by directly peeling off in $x_n$ from $\formula_n(\bx_{[n]})$, thus it is recognized by $\CFG_{n-1}$ \eqref{eq:cfg_n-1} and $|\formula_{n-1}| = O(|\formula_n|)$,
        \item $\db_{n-1}$ is a database instance of $\formula_{n-1}(\bx_{[n-1]})$ such that $\bigoplus_{x_n \in \dom(x_n)}\formula_n(\bx_{[n]}) = \formula_{n-1}(\bx_{[n-1]})$ and $|\db_{n-1}| = O(|\db_n|)$.
    \end{enumerate}
    and the algorithm runs in $O(|\formula_n| + d(n) \cdot |\db_n| \cdot \ackfunc{\db_n})$ time, where $d(n)$ is the number of hyperedges containing $n$ in $\mH_n$.
\end{theorem}

{The inverse Ackermann factor $\ackfunc{\db_n}$ in the aggregation algorithm is inherited from the algorithm supporting \textsf{RangeSum} queries. Thus, the guarantee of the \textsf{RangeSum} problem carries over to our aggregation algorithm: for semigroups that accepts additive inverse, or for the semigroup with minimum (or maximum) as the operation (\textsf{RMQ} as in \cite{GBT84}), the aggregation algorithm runs in $O(|\formula_n| \cdot |\db_n|)$ time, without the additional inverse Ackermann overhead.}

\subsection{The oracle-construction algorithm} \label{sec:oracle}
\myparagraph{\textsf{RangeSum}} Chazelle and Rosenberg \cite{10.1145/73833.73848} studied the range query problem \textsf{RangeSum} in the semigroup model defined as follows. Preprocess a array $A$ of $\arraysize$ elements from a semigroup $\bS = (\boldsymbol{D}, \oplus)$, and then support the following query: given a query range $[\open, \close]$, return the range sum $\bigoplus_{\interval \in [\open, \close]} A[\interval]$, i.e. $\bigoplus_{\open \leq \interval \leq \close} A[\interval]$. In particular, they proved the following theorem.

\begin{theorem}[\cite{10.1145/73833.73848}] \label{thm:range-sum}
	There is a data structure of size $O(\arraysize)$ that works in the word RAM, and support a \textsf{RangeSum} query in $O(\alpha(14 \arraysize, \arraysize))$ time, where $\alpha$ is the inverse Ackermann function, after spending $O(\arraysize)$ preprocessing time. 
\end{theorem}
It is the absence of additive inverse for general semirings that results in the hardness of $\mathsf{RangeSum}$ and an unconditional hard instance can be recursively constructed that mimics the definition of the inverse Ackermann function $\alpha(\cdot, \cdot)$ as in~\cite{10.1145/73833.73848}. It is easy to see that \textsf{RangeSum} queries can be answered in $O(1)$ time after $O(\arraysize)$ preprocessing, if the underlying semigroup $\bS$ allows for additive inverse: one candidate algorithm can precompute the partial sums $\bigoplus_{\interval \in [1, \close]} A[\interval]$ and represent each \textsf{RangeSum} query $[\open, \close]$ as the semigroup sum of $\bigoplus_{ \interval \in [1, \close]} A[\interval]$ and the additive inverse of $\bigoplus_{\interval \in [1, \open-1]} A[\interval]$.  

The \textsf{RangeSum} problem with minimum as the semigroup operation, is studied intensively and typically known as \textsf{RangeMinimumQuery} (\textsf{RMQ}). In their influential paper \cite{GBT84}, Gabow, Bentley and Tarjan shown that there is an algorithm that works in the word RAM and supports a \textsf{RMQ} query in $O(1)$ time, after spending $O(\arraysize)$ preprocessing time and space. They recognized the Cartesian tree as the instrumental data structure that was introduced by Vuillemin \cite{Vuillemin80} in the context of average time analysis of searching.


In this section, we introduce a key oracle (called \textsf{RangeSumOrcale}) that supports efficient computation of the new negative factors $R_{N \setminus \{n\}}$ in the aggregation algorithm introduced in \eqref{eq:derivation-aggregate} of the last section. In particular, \textsf{RangeSumOrcale} uses the \textsf{RangeSum} data structures as a black-box, thus inherits the inverse Ackermann factor of Theorem~\ref{thm:range-sum} under general semirings. We first introduce the following definition.

\medskip
\myparagraph{\textsf{RangeSumOracle}}
Let $A$ be an an array of size $\arraysize$ over a semigroup $\bS = (\boldsymbol{D}, \oplus)$, where we implicitly assume that $A[\interval] = \zerobf$ if $\interval \notin [1, \arraysize]$. An \emph{array decomposition} of $A$ is an array of pairs as follows:
\begin{align*}
    \left([1, \interval_1], \bigoplus_{\interval \in [1, \interval_1]} A[\interval] \right), \;
    \left([\interval_1 + 1, \interval_2], \bigoplus_{\interval \in [\interval_1 + 1, \interval_2]} A[\interval] \right), \;
    \ldots, 
    \left([\interval_r + 1, \arraysize], \bigoplus_{\interval \in [\interval_p + 1, \arraysize]} A[\interval] \right), \;
    \left([w+1, \bot], \zerobf \right)
\end{align*}
where $1 \leq \interval_1 \leq \interval_2 \leq \cdots \leq \interval_{p} = \arraysize$ ($\interval_i$, where $i \in [p]$, and $0, \bot$, are called the \emph{break points} of the array decomposition) and  
each pair contains a range and a semigroup sum over that range. For convenience, we add a dummy pair $([w + 1, \bot], \zerobf)$ indicating the end of the array decomposition. A \emph{trivial} array decomposition of $A$ is the array of pairs 
\begin{align*}
    \left([1, 1], A[1] \right), \;
    \left([2, 2], A[2] \right), \;
    \ldots,
    \left([\arraysize, \arraysize], A[\arraysize] \right), \;
    \left([w+1, \bot], \zerobf \right).
\end{align*}
whose set of break points is $\{1, 2, \ldots, \arraysize\}$.

An $\RangeSumOracle$ of the array $A$ is a $\RangeSum$ data structure built on an array decomposition of $A$ such that: given a query range $[\interval_i + 1, \interval_j]$, where $\interval_i < \interval_j$ are two break points of the array decomposition (so that the query range aligns with the smaller range sums stored in the array decomposition), it returns the range sum over the query range
$$
\bigoplus_{i \leq k < j }  \bigoplus_{\interval \in [\interval_k + 1, \interval_{k+1}]} A[\interval] = \bigoplus_{\interval \in [\interval_i + 1, \interval_j]} A[\interval]
$$
It is easy to see from \autoref{thm:range-sum} that there is a data structure of size $O(\arraysize)$ and support a $\RangeSumOracle$ query in $O(\alpha(14 \arraysize, \arraysize))$ time, where $\alpha(\cdot, \cdot)$ is the inverse Ackermann function, after spending $O(\arraysize)$ preprocessing time.
Indeed, for a trivial array decomposition, the $\RangeSumOracle$ is essentially a $\RangeSum$ data structure of $A$.

\medskip
Now we formally describe our $\RangeSumOracle$(s). To better identify $\nestfaqneg$ subexpressions show up in $\formula_n(\bx_{[n]})$, we let $\nonterm_n(\bx_{N_i}) \derive \formula_n(\bx_{N_i})$ for $n \in U = N_0 \subseteq N^-_1 \subseteq N_1 \subseteq \cdots \subseteq N^-_k \subseteq N_k$ and $\nonterm_n(\bx_S) \derive \mu_n(\bx_S)$ for all $n \notin S \subseteq [n]$. Then, \eqref{eq:derivation} can be equivalently written as:

\begin{equation} \label{eq:derivation-oracle}
    \begin{aligned}
    \formula_n(\bx_{[n]}) & = \mu_n(\bx_{S_k}) \otimes \formula_n(\bx_{N_k})  \\
    & = \mu_n(\bx_{S_k}) \otimes \left( R_{N_k}(\bx_{N_k}) \vdash \formula_n(\bx_{N^-_k}) \right) \\
    & = \mu_n(\bx_{S_k}) \otimes \left(R_{N_k}(\bx_{N_k}) \vdash \mu_n(\bx_{S_{k-1}}) \otimes \formula_n(\bx_{N_{k-1}}) \right) \\
    & = \mu_n(\bx_{S_k}) \otimes \left(R_{N_k}(\bx_{N_k}) \vdash \mu_n(\bx_{S_{k-1}}) \otimes \left(R_{N_{k-1}}(\bx_{N_{k-1}}) \vdash \formula_n(\bx_{N^-_{k-1}}) \right) \right) \\
    & \; \vdots \\
    & = \mu_n(\bx_{S_k}) \otimes \left(R_{N_k}(\bx_{N_k}) \vdash \mu_n(\bx_{S_{k-1}}) \otimes \left(R_{N_{k-1}}(\bx_{N_{k-1}}) \vdash  \cdots R_{N_{1}}(\bx_{N_{1}}) \vdash\left(\mu_n(\bx_{S_1}) \otimes (R_{N_{1}}(\bx_{N_{1}}) \vdash \formula_n(\bx_{N^-_1})) \right) \cdots  \right) \right)   \\
    & = \mu_n(\bx_{S_k}) \otimes \left(R_{N_k}(\bx_{N_k}) \vdash \mu_n(\bx_{S_{k-1}}) \otimes \left(R_{N_{k-1}}(\bx_{N_{k-1}}) \vdash \cdots R_{N_{1}}(\bx_{N_{1}}) \vdash \left(\mu_n(\bx_{S_1}) \otimes (R_{N_{1}}(\bx_{N_{1}}) \vdash \mu_n(\bx_{S_{0}}) \otimes \formula_n(\bx_U)) \right) \cdots  \right) \right) \\
    & = \mu_n(\bx_{S_k}) \otimes \left(R_{N_k}(\bx_{N_k}) \vdash \mu_n(\bx_{S_{k-1}}) \otimes \left(R_{N_{k-1}}(\bx_{N_{k-1}}) \vdash \cdots R_{N_{1}}(\bx_{N_{1}}) \vdash \left(  \mu_n(\bx_{S_1}) \otimes (R_{N_{1}}(\bx_{N_{1}}) \vdash \mu_n(\bx_{S_{0}}) \otimes R_U(\bx_U)) \right)  \cdots   \right) \right) 
    \end{aligned}
\end{equation}
We will construct a $\RangeSumOracle$ for the $\nestfaqneg$ subexpression $\formula_n(\bx_{U})$ (let $U = N_0$) and then $\formula_n(\bx_{N_1}), \formula_n(\bx_{N_2}), \ldots, \formula_n(\bx_{N_k})$, in a bottom-up order as layed out in \eqref{eq:derivation-oracle}. Indeed, the $\BuildOracle$ algorithm described in Algorithm~\ref{alg:oracle} takes as input the $\nestfaqneg$ expression $\formula_n(\bx_{[n]})$ recognized by $\CFG_n$ and the corresponding database instance $\db_n$ and returns a $\RangeSumOracle$ for the every $\nestfaqneg$ subexpression $\formula_n(\bx_{N_i})$, where $i = 0, 1, \ldots, k$. Now we walk through its steps, starting from the inner-most $\formula_n(\bx_{U}) = R_U(\bx_U)$ in \eqref{eq:derivation-oracle}, $R_U$ being the pivot factor.

\begin{algorithm}
    \caption{$\BuildOracle(\formula_n(\bx_{[n]}), \db_n)$} \label{alg:oracle}
    \KwIn{a $\nestfaqneg$ expression $\formula_n(\bx_{[n]})$ recognized by $\CFG_n$ and a global database instance $\db_n$}
    \KwOut{An $\RangeSumOracle$ $\Oracle_i(\bx_{N_i \setminus \{n\}})$ for each $\formula_n(\bx_{N_i})$, where $i = 0, 1, \ldots, k$ and $U = N_0 \subseteq N_1 \subseteq \cdots \subseteq N_k$ are the only hyperedges containing $n$}

    \ForEach(\Comment{base case for $\formula_n(\bx_U)$} \label{line:oracle-base-start}
    ){$\ba_{U \setminus \{n\}} \in \bigcup_{0 \leq i \leq k} \Pi_{U \setminus \{n\}} R_{N_i}$}{
         \textbf{init} $A_0(\ba_{U \setminus \{n\}}) \gets []$, $\Oracle_0(\ba_{U \setminus \{n\}}) \gets []$, $\interval_{\ba_{U \setminus \{n\}}} \gets 1$ \\
    }
    
    \ForEach{$\ba_{U} \in \bigcup_{0 \leq i \leq k} \Pi_{U} R_{N_i}$}{
        $A_0(\ba_{U \setminus \{n\}})[\interval_{\ba_{U \setminus \{n\}}}] \gets \Pi_{\{n\}} \ba_{U}$ \\
        \textbf{append} $\left([\interval_{\ba_{U \setminus \{n\}}}, \interval_{\ba_{U \setminus \{n\}}}], R_U(\ba_{U}) \right)$ \textbf{to} $\Oracle_0(\ba_{U \setminus \{n\}})$ \\
        $\interval_{\ba_{U \setminus \{n\}}} \gets \interval_{\ba_{U \setminus \{n\}}} + 1$ \\
    }
    \ForEach{$\ba_{U \setminus \{n\}} \in \bigcup_{0 \leq i \leq k} \Pi_{U \setminus \{n\}} R_{N_i}$}{
        \textbf{append} $\left([\interval_{\ba_{U \setminus \{n\}}} + 1, \bot], \; \zerobf \right)$ \textbf{to} $\Oracle_0(\ba_{U \setminus \{n\}})$ \label{line:oracle-base-end} \\
        \textbf{construct} $\RangeSum$ data structures \textbf{on} $\Oracle_0(\ba_{U \setminus \{n\}})$ \\
    }
    
    \ForEach(\Comment{inductive case for $\formula_n(\bx_{N_i})$}){$i = 1, 2, \ldots k$ (in order) \label{line:oracle-inductive}}{
        \ForEach{$\ba_{N_i \setminus \{n\}} \in \bigcup_{i \leq s \leq k} \Pi_{N_i \setminus \{n\}} R_{N_s}$}{ 
            \textbf{init} $\Oracle_i(\ba_{N_i \setminus \{n\}}) \gets []$, $\open_{\ba_{N_i \setminus \{n\}}} \gets 0$  \\
        }
        
        \ForEach{$\ba_{N_{i-1} \setminus \{n\}} \in \bigcup_{i-1 \leq s \leq k} \Pi_{N_{i-1} \setminus \{n\}} R_{N_s} $}{
            $\close \gets 1$ \\
            \While{$\close \neq \bot$ }{
                $a_n \gets A_0(\Pi_{U \setminus \{n\}}\ba_{N_{i-1} \setminus \{n\}})[\close]$ \\
                \ForEach{$\ba_{N_i \setminus \{n\}} \in \bigcup_{i \leq s \leq k} \Pi_{N_{i} \setminus \{n\}} R_{N_s}(a_n, \ba_{N_{i-1} \setminus \{n\}}, \bx_{N_s \setminus N_{i-1}})  $}{
                    $M_1 \gets \bigoplus_{\interval \in [\open_{\ba_{N_i \setminus \{n\}}} + 1, \close -1]}\formula_n \left(A_0(\Pi_{U \setminus \{n\}} \ba_{N_{i-1} \setminus\{n\}})[\interval],  \ba_{N_{i-1} \setminus\{n\}} \right)$ \label{line:oracle-set-breakpt-true-start} \\
                    $M_2 \gets \formula_n(a_n, \Pi_{N_{i-1} \setminus \{n\}} \ba_{N_{i} \setminus \{n\}})$ \\
                    \textbf{append} 
                    $\left( [\open_{\ba_{N_i \setminus \{n\}}} + 1, \close -1 ], \; \mu_n(\Pi_{S_{i-1} \setminus \{n\}} \ba_{N_i \setminus\{n\}}) \otimes M_1 \right)$ \textbf{to} $\Oracle_i(\ba_{N_i \setminus \{n\}})$ \\
                    \textbf{append} 
                    $\left( [\close, \close], \; R_{N_i}(a_n, \ba_{N_i \setminus \{n\}}) \oplus \indicator{\neg R_{N_i} } (a_n, \ba_{N_i \setminus \{n\}}) \otimes \mu_n(\Pi_{S_{i-1} \setminus \{n\}} \ba_{N_i \setminus\{n\}}) \otimes M_2 \right)$ \textbf{to} $\Oracle_i(\ba_{N_i \setminus \{n\}})$   \label{line:oracle-set-breakpt-true-end}  \\
                    $\open_{\ba_{N_i \setminus \{n\}}} \gets \close + 1$ \\
                }
                \textbf{move} $\close$ to the next break point of $\Oracle_{i-1}(\ba_{N_{i-1} \setminus \{n\}})$ \\
            }
        }
        \ForEach{$\ba_{N_i \setminus \{n\}} \in \bigcup_{i \leq s \leq k} \Pi_{N_i \setminus \{n\}} R_{N_s}$}{
            \textbf{append} $([\open_{\ba_{N_i \setminus \{n\}}} + 1, \bot], \; \zerobf)$ into $\Oracle_i(\ba_{N_i \setminus \{n\}} )$ \\
            \textbf{construct} $\RangeSum$ data structures \textbf{on} $\Oracle_i(\ba_{N_i \setminus \{n\}} )$ \\
        }
    }
    \Return $\left(\Oracle_i(\bx_{N_i \setminus \{n\}})\right)_{i = 0, 1, \ldots, k}$
\end{algorithm}

\medskip
\introparagraph{\textcolor{black}{the base case of $\formula_n(\bx_{U}) = R_U(\bx_U)$}}
The base case first sets up the indices for accessing all $x_n$ values for $\RangeSum$ data structures constructed later. More precisely, we first build a hashtable $A_0(\bx_{U \setminus \{n\}})$ from $\bigcup_{0 \leq i \leq k} \Pi_{U} R_{N_i}$, where $\bx_{U \setminus \{n\}}$ is its key and each entry $\bx_{U \setminus \{n\}} = \ba_{U \setminus \{n\}} \in \Oracle_0(\bx_{U \setminus \{n\}})$ stores an array of $x_n$ values where $(x_n, \ba_{U \setminus \{n\}}) \in \bigcup_{0 \leq i \leq k} \Pi_{U} R_{N_i}$. The array is identified as $A_0(\ba_{U \setminus \{n\}})$ and the ordering of $x_n$ values in the array can be posited arbitrarily but fixed afterwards. Thus, $A_0(\bx_{U \setminus \{n\}})$ is of size $|\bigcup_{0 \leq i \leq k} \Pi_{U} R_{N_i}| = O(|\db_n|)$ and allows for every tuple $\ba_{U \setminus \{n\}} \in \bigcup_{0 \leq i \leq k} \Pi_{U \setminus \{n\}} R_{N_i}$, a direct (array) access to all possible $x_n$ values that could appear alongside $\ba_{U \setminus \{n\}}$ in the database instance $\db_n$. It is easy to see that this construction costs $O(|\db_n|)$ time and space.

Now we start constructing the base-case oracle $\Oracle_0(\bx_{U \setminus \{n\}})$ for $\formula_n(\bx_{U}) = R_U(\bx_U)$. It follows exactly as $A_0(\bx_{U \setminus \{n\}})$ except that each entry $\bx_{U \setminus \{n\}} = \ba_{U \setminus \{n\}} \in \Oracle_0(\bx_{U \setminus \{n\}})$ stores instead a trivial array decomposition of $A_0(\ba_{U \setminus \{n\}})$, replacing each value of $x_n$ in $A_0(\ba_{U \setminus \{n\}})$ by its weight $R_U(x_n, \ba_{U \setminus \{n\}}) \in \boldsymbol{D}$. So the $\interval$-th element in $A_0(\ba_{U \setminus \{n\}})$, say $a_n = A_0(\ba_{U \setminus \{n\}})[\interval]$, corresponds to 
the $\interval$-th entry in the array decomposition $\Oracle_0(\ba_{U \setminus \{n\}})$, i.e. $\left([\interval, \interval], R_U(a_n , \ba_{U \setminus \{n\}}) \right) = \left([\interval, \interval], \formula_n(a_n, \ba_{U \setminus \{n\}}) \right)$. Thus, $\Oracle_0(\bx_{U \setminus \{n\}})$ is of size $|A_0(\bx_{U \setminus \{n\}})| = O(|\db_n|)$. We recall that for $(a_n, \ba_{U \setminus \{n\}}) \notin R_U$, we have $R_U(a_n, \ba_{U \setminus \{n\}}) = \zerobf$. Applying \autoref{thm:range-sum}, we construct a $\RangeSumOracle$ on each array decomposition $\Oracle_0(\ba_{U \setminus \{n\}})$, after spending $O(|\db_n|)$ preprocessing time building the $\RangeSum$ data structures. Abusing notations, we denote the entire oracle as 
$\Oracle_0(\bx_{U \setminus \{n\}})$ to implicitly indicate that it is a hashtable of array decompositions equipped with the $\RangeSumOracle$ that supports the following query in $O(\ackfunc{\db_n})$ time:
\begin{description}
    \item[oracle] $\Oracle_0(\bx_{U \setminus \{n\}})$
    \item[input] a tuple $\ba_{U \setminus \{n\}} \in \bigcup_{0 \leq i \leq k} \Pi_{U \setminus \{n\}} R_{N_i}$ (to identity the array decomposition $\Oracle_i(\ba_{N_i \setminus \{n\}})$ to be queried) and a query range $[\open+1, \close]$, where $ 0 \leq \open < \close$ are two break points of the array decomposition.
    \item[output] a range sum $\bigoplus_{\interval \in [\open+1, \close]} \formula_n \left(A_0(\ba_{U \setminus \{n\}})[\interval], \; \ba_{U \setminus \{n\}} \right)$.
\end{description}

\introparagraph{\textcolor{black}{the inductive case of $\formula_n(\bx_{N_i})$}}
For $\formula_n(\bx_{N_i})$, where $i \in [k]$, following \eqref{eq:derivation-oracle}, we have that $\formula_n(\bx_{N_i}) = R_{N_i}(\bx_{N_i }) \vdash  \formula_n(\bx_{N^-_i}) = R_{N_i}(x_n, \ba_{N_i \setminus\{n\} }) \vdash  \mu_n(\bx_{S_{i-1}}) \otimes \formula_n(\bx_{N_{i-1}})$. Thus, for a fixing of $\bx_{N_i \setminus \{n\}} = \ba_{N_i \setminus \{n\}}$, we have

\begin{equation} \label{eq:oracle-derivation}
    \begin{aligned}
    \formula_n(x_n, \ba_{N_i \setminus\{n\}}) & = R_{N_i}(x_n, \ba_{N_i \setminus\{n\} }) \vdash  \formula_n(x_n, \Pi_{N^-_{i} \setminus \{n\}} \ba_{N_i \setminus\{n\}}) \\
    & = R_{N_i}(x_n, \ba_{N_i \setminus\{n\} }) \vdash  \mu_n(\Pi_{S_{i-1} \setminus \{n\}} \ba_{N_i \setminus\{n\}}) \otimes \formula_n(x_n, \Pi_{N_{i-1} \setminus \{n\}} \ba_{N_i \setminus\{n\}}) 
    \end{aligned}
\end{equation}

Following the bottom-up construction, for the $i$-th $(i > 0)$ time entering in the for-loop on \autoref{line:oracle-inductive}, we have constructed the $(i-1)$-th $\RangeSumOracle$(s) denoted as $\Oracle_{i-1}(\bx_{N_{i-1} \setminus \{n\}})$ for $\formula_n(\bx_{N_{i-1}})$. Specifically, the $\RangeSumOracle$(s) $\Oracle_{i-1}(\bx_{N_{i-1} \setminus \{n\}})$ is a hashtable of size $O(|\db_n|)$ containing array decompositions as entries that supports the following range query in $O(\ackfunc{\db_n})$ time:

\begin{description}
    \item[oracle] $\Oracle_{i-1}(\bx_{N_{i-1} \setminus \{n\}})$
    \item[input] a tuple $\ba_{N_{i-1} \setminus \{n\}} \in \bigcup_{i-1 \leq s \leq k} \Pi_{N_{i-1} \setminus \{n\}} R_{N_s}$ (i.e. a query key to identity the array decomposition $\Oracle_{i-1}(\ba_{N_{i-1} \setminus \{n\}})$ to be queried) and a query range $[\open+1, \close]$, where $0 \leq \open \leq \close$ are two break points of the array decomposition $\Oracle_{i-1}(\ba_{N_{i-1} \setminus \{n\}})$. In particular, the break points of the array decomposition $\Oracle_i(\ba_{N_i \setminus \{n\}})$ are $\interval$ and $\interval-1$, where
    \begin{align*}
        \interval \in \left\{\interval \mid \left( A_0(\Pi_{U \setminus \{n\}} \ba_{N_{i-1} \setminus \{n\}})[\interval], \ba_{N_{i-1} \setminus \{n\}}\right) \in \bigcup_{i-1 \leq s\leq k} \Pi_{N_{i-1}} R_{N_s} \right\}
    \end{align*}
    \item[output] a range sum $\bigoplus_{\interval \in [\open+1, \close]} \formula_n\left(A_0(\Pi_{U \setminus \{n\}} \ba_{N_{i-1} \setminus \{n\}})[\interval], \; \ba_{N_{i-1} \setminus \{n\}} \right)$, (or $\zerobf$ if $\open = \close$).
\end{description}

We now start constructing a new $\RangeSumOracle$ $\Oracle_{i}(\bx_{N_{i} \setminus \{n\}})$ for $\formula_n(\bx_{N_i})$, using the $\RangeSumOracle$ $\Oracle_{i-1}(\bx_{N_{i-1} \setminus \{n\}})$. We initialize $\Oracle_i(\bx_{N_i \setminus \{n\}})$ as a hashtable with $\bx_{N_i \setminus \{n\}}$ as its key, where each entry $\bx_{N_i \setminus \{n\}} = \ba_{N_i \setminus \{n\}} \in \bigcup_{i \leq s \leq k} \Pi_{N_t \setminus \{n\}} R_{N_s}$ is intialized as an empty array decomposition, denoted as $\Oracle_i( \ba_{N_i \setminus \{n\}})$. Furthermore, we initialize a set of scanning iterators $\open_{\ba_{N_i \setminus \{n\}}} = 0$, one for each $\ba_{N_i \setminus \{n\}} \in \bigcup_{i \leq s \leq k} \Pi_{N_t \setminus \{n\}} R_{N_s}$ that pinpoints the last constructed break point throughout its ongoing construction process. 

Next, we will make one scan over every entry (array decomposition) stored in $\Oracle_{i-1}(\bx_{N_{i-1} \setminus \{n\}})$ by iterating over its break points, and for each array decomposition $\Oracle_{i-1}(\ba_{N_i \setminus \{n\}})$, we construct the array decompositions $\Oracle_i(\ba_{N_i \setminus \{n\}})$ for all $\ba_{N_i \setminus \{n\}} \in \bigcup_{i \leq s \leq k} \Pi_{N_t \setminus \{n\}} R_{N_s}$ such that $\Pi_{N_{i-1} \setminus \{n\}} \ba_{N_i \setminus \{n\}} = \ba_{N_{i-1} \setminus \{n\}}$ simultaneously. Let us now describe the exact steps when scanning an array decomposition $\Oracle_{i-1}(\ba_{N_{i-1} \setminus \{n\}})$ in the oracle $\Oracle_{i-1}(\bx_{N_{i-1} \setminus \{n\}})$. First, we initialize a scanning iterator $\close = 1$ to scan the array decomposition $\Oracle_{i-1}(\ba_{N_{i-1} \setminus \{n\}})$ by moving $\close$ forward to the next break point. 

As we are scanning at the break point $\close$ of the array decomposition $\Oracle_{i-1}(\bx_{N_{i-1} \setminus \{n\}})$, we retrieve its corresponding $x_n$ value via $A_0(\Pi_{U \setminus \{n\}} \ba_{N_{i-1} \setminus \{n\}})[\close] = a_n$. For every ongoing construction of $\Oracle_i(\ba_{N_i \setminus \{n\}})$ (i.e. $\Pi_{N_{i-1} \setminus \{n\}} \ba_{N_i \setminus \{n\}} = \ba_{N_{i-1} \setminus \{n\}}$), if $(a_n, \ba_{N_{i} \setminus \{n\}}) \in \bigcup_{i \leq s \leq k} \Pi_{N_{i}} R_{N_s}$, we set both $\close-1$ and $\close$ as it new break points by appending the following two pairs to the array decomposition $\Oracle_i(\ba_{N_i \setminus \{n\}})$:
\begin{enumerate}
    \item the first pair to be appended is
    \begin{align*}
        \left( [\open_{\ba_{N_i \setminus \{n\}}} + 1, \close -1 ], \; \mu_n(\Pi_{S_{i-1} \setminus \{n\}} \ba_{N_i \setminus\{n\}}) \otimes M_1 \right)
    \end{align*}
    where $M_1$ is a call to the $\RangeSumOracle$ $\Oracle_{i-1}(\ba_{N_{i-1} \setminus \{n\}})$ with the input parameters $\ba_{N_{i-1} \setminus\{n\}} = \Pi_{N_{i-1} \setminus\{n\}} \ba_{N_i \setminus \{n\}}$ (the query key) and $[\open_{\ba_{N_i \setminus \{n\}}} + 1, \close -1]$ (the query range). Therefore, by induction hypothesis, 
    \begin{align*}
        M_1 = \bigoplus_{\interval \in [\open_{\ba_{N_i \setminus \{n\}}} + 1, \close -1]}\formula_n \left(A_0(\Pi_{U \setminus \{n\}} \ba_{N_{i-1} \setminus\{n\}})[\interval],  \ba_{N_{i-1} \setminus\{n\}} \right)
    \end{align*}
    and $M_1$ is returned by $\Oracle_{i-1}(\bx_{N_{i-1} \setminus \{n\}})$ in time $O(\ackfunc{\db_n})$.
    \item the second pair to be appended is
    \begin{align*}
        \left( [\close, \close], \; R_{N_i}(a_n, \ba_{N_i \setminus \{n\}}) \oplus \indicator{\neg R_{N_i} } (a_n, \ba_{N_i \setminus \{n\}}) \otimes \mu_n(\Pi_{S_{i-1} \setminus \{n\}} \ba_{N_i \setminus\{n\}}) \otimes M_2 \right)
    \end{align*}
    where $M_2$ is a call to the $\RangeSumOracle$ $\Oracle_{i-1}(\ba_{N_{i-1} \setminus \{n\}})$ with the input parameters $\ba_{N_{i-1} \setminus\{n\}} = \Pi_{N_{i-1} \setminus\{n\}} \ba_{N_i \setminus \{n\}}$ (the query key) and $[\close, \close]$ (the query range). Therefore, by induction hypothesis,
    \begin{align*}
        M_2 = \bigoplus_{\interval \in [\close, \close]}\formula_n \left(A_0(\Pi_{U \setminus \{n\}} \ba_{N_{i-1} \setminus\{n\}})[\interval], \ba_{N_{i-1} \setminus \{n\}} \right) = \formula_n(a_n, \ba_{N_{i-1} \setminus \{n\}}) = \formula_n(a_n, \Pi_{N_{i-1} \setminus \{n\}} \ba_{N_{i} \setminus \{n\}})
    \end{align*}
    and $M_2$ is returned by $\Oracle_{i-1}(\bx_{N_{i-1} \setminus \{n\}})$ in time $O(\ackfunc{\db_n})$.
\end{enumerate}

The correctness of the second pair simply follows from \eqref{eq:oracle-derivation}, where we fix the value of $x_n$ to be $a_n = A_0(\Pi_{U \setminus \{n\}} \ba_{N_{i-1} \setminus\{n\}})[\interval]$. We now reason the correctness of the first pair as follows: by construction of break points, $\open_{\ba_{N_i \setminus \{n\}}}$ and $\close$ are two consecutive break points for $\Oracle_{i}(\bx_{N_{i} \setminus \{n\}})$, thus 
$(A_0(\Pi_{U \setminus \{n\}} \ba_{N_{i-1} \setminus\{n\}})[\interval], \ba_{N_i \setminus\{n\} }) \notin R_{N_i}$, for all $\interval \in [\open_{\ba_{N_i \setminus \{n\}}} + 1, \close -1]$. Therefore, $R_{N_i} (A_0(\Pi_{U \setminus \{n\}} \ba_{N_{i-1} \setminus\{n\}})[\interval]) = \zerobf$ and 
$\indicator{R_{N_i}} (A_0(\Pi_{U \setminus \{n\}} \ba_{N_{i-1} \setminus\{n\}})[\interval]) = \onebf$ for all $\interval \in [\open_{\ba_{N_i \setminus \{n\}}} + 1, \close -1]$. Still following \eqref{eq:oracle-derivation}, we have that
\begin{equation*} 
    \begin{aligned}
    \formula_n(A_0(\Pi_{U \setminus \{n\}} \ba_{N_{i-1} \setminus\{n\}})[\interval], \ba_{N_i \setminus\{n\}}) & = R_{N_i} (A_0(\Pi_{U \setminus \{n\}} \ba_{N_{i-1} \setminus\{n\}})[\interval]) \oplus \indicator{R_{N_i}} (A_0(\Pi_{U \setminus \{n\}} \ba_{N_{i-1} \setminus\{n\}})[\interval], \ba_{N_i \setminus\{n\}}) \\
    & \qquad \otimes \formula_n(A_0(\Pi_{U \setminus \{n\}} \ba_{N_{i-1} \setminus\{n\}})[\interval], \Pi_{N^-_{i} \setminus \{n\}} \ba_{N_i \setminus\{n\}}) \\
    & = \zerobf \oplus \onebf \otimes \formula_n(A_0(\Pi_{U \setminus \{n\}} \ba_{N_{i-1} \setminus\{n\}})[\interval], \; \Pi_{N^-_{i} \setminus \{n\}} \ba_{N_i \setminus\{n\}}) \\
    & = \formula_n(A_0(\Pi_{U \setminus \{n\}} \ba_{N_{i-1} \setminus\{n\}})[\interval], \; \Pi_{N^-_{i} \setminus \{n\}} \ba_{N_i \setminus\{n\}})\\ 
    & =  \mu_n(\Pi_{S_{i-1} \setminus \{n\}} \ba_{N_i \setminus\{n\}}) \otimes \formula_n(A_0(\Pi_{U \setminus \{n\}} \ba_{N_{i-1} \setminus\{n\}})[\interval], \;\Pi_{N_{i-1} \setminus \{n\}} \ba_{N_i \setminus\{n\}}) 
    \end{aligned}
\end{equation*}
Hence, the correctness of the first pair follows because
\begin{align*}
    \bigoplus_{\interval \in [\open_{\ba_{N_i \setminus \{n\}}} + 1, \close -1]} \formula_n(A_0(\Pi_{U \setminus \{n\}} \ba_{N_{i-1} \setminus\{n\}})[\interval], \ba_{N_i \setminus\{n\}}) = \mu_n(\Pi_{S_{i-1} \setminus \{n\}} \ba_{N_i \setminus\{n\}}) \otimes M_1.
\end{align*}

As a summary, one scan over the break points of every array decomposition stored in $\Oracle_{i-1}(\bx_{N_{i-1} \setminus \{n\}})$ costs time $O(|\db_n|)$, i.e. the size of the oracle. Furthermore, for a fixed $\ba_{N_{i-1} \setminus \{n\}} \in \bigcup_{i-1 \leq s \leq k} \Pi_{N_{i-1} \setminus \{n\}} R_{N_s}$, we make break points across all the (constructing) array decompositions of $\Oracle_{i}(\ba_{N_{i} \setminus \{n\}})$ where $\ba_{N_{i} \setminus \{n\}} \in \bigcup_{i \leq s \leq k} \Pi_{N_{i}} R_{N_s}$ and $\Pi_{N_{i-1} \setminus \{n\}} \ba_{N_i \setminus\{n\}} = \ba_{N_{i-1} \setminus\{n\}}$ and by construction, the break points are $\interval$ and $\interval - 1$ where
\begin{align*}
    \interval \in \left\{\interval \mid \left( A_0(\Pi_{U \setminus \{n\}} \ba_{N_{i} \setminus \{n\}})[\interval], \ba_{N_{i} \setminus \{n\}}\right) \in \bigcup_{i \leq s\leq k} \Pi_{N_{i}} R_{N_s} \right\}
\end{align*}
Observe that the break points constructed in $\Oracle_{i}(\ba_{N_{i} \setminus \{n\}})$ is a subset of those of $\Oracle_{i-1}(\Pi_{N_{i-1} \setminus \{n\}} \ba_{N_{i} \setminus \{n\}})$, thus we are free from the risk of asking the oracle $\Oracle_{i-1}(\Pi_{N_{i-1} \setminus \{n\}} \ba_{N_{i} \setminus \{n\}})$ with mis-aligned intervals, when constructing $\Oracle_{i}(\ba_{N_{i} \setminus \{n\}})$. The number of break points to be constructed in the array decomposition $\Oracle_{i}(\ba_{N_{i} \setminus \{n\}})$ is at most $\sum_{i \leq s \leq k} \mathsf{deg}_{R_{N_s}}(\ba_{N_i \setminus \{n\}}) $, where $\mathsf{deg}_{R_{N_s}}(\ba_{N_i \setminus \{n\}})$ denotes the degree of $\ba_{N_i \setminus \{n\}}$ in $R_{N_s}$, i.e., the number of tuples in $R_{N_s}$ that coincide with $\ba_{N_i \setminus \{n\}}$ on the variables $\bx_{N_i \setminus \{n\}}$. Therefore, the total number of break points in the oracle $\Oracle_{i}(\bx_{N_{i} \setminus \{n\}})$ is at most 
\begin{align*}
    \sum_{\ba_{N_i \setminus \{n\}} \in \bigcup_{i \leq s \leq k} \Pi_{N_i \setminus \{n\}} R_{N_s}} \sum_{i \leq s \leq k} \mathsf{deg}_{R_{N_s}}(\ba_{N_i \setminus \{n\}}) \leq \sum_{i \leq s \leq k}|R_{N_s}| = O(|\db_n|).
\end{align*}
The new-built oracle $\Oracle_{i}(\bx_{N_{i} \setminus \{n\}})$ is thus of size $O(|\db_n|)$. Furthermore, for each break point, it costs $O(\ackfunc{\db_n})$ to get the two range sums $M_1, M_2$ from the oracle $\Oracle_{i-1}(\bx_{N_{i-1} \setminus \{n\}})$, thus the total time of the construction of $\Oracle_{i}(\bx_{N_{i} \setminus \{n\}})$ is $O(|\db_n| \cdot \ackfunc{\db_n})$. The $\RangeSumOracle$ $\Oracle_{i}(\bx_{N_{i} \setminus \{n\}})$ supports the following range query in time $O(\ackfunc{\db_n})$:

\begin{description}
    \item[oracle] $\Oracle_{i}(\bx_{N_{i} \setminus \{n\}})$
    \item[input] a tuple $\ba_{N_{i} \setminus \{n\}} \in \bigcup_{i \leq s \leq k} \Pi_{N_{i} \setminus \{n\}} R_{N_s}$ (i.e. a query key to identity the array decomposition $\Oracle_{i}(\ba_{N_{i} \setminus \{n\}})$ to be queried) and a query range $[\open+1, \close]$, where $0 \leq \open \leq \close$ are two break points of the array decomposition $\Oracle_{i}(\ba_{N_{i} \setminus \{n\}})$. In particular, the break points of the array decomposition $\Oracle_i(\ba_{N_i \setminus \{n\}})$ are $\interval$ and $\interval-1$, where
    \begin{align*}
        \interval \in \left\{\interval \mid \left( A_0(\Pi_{U \setminus \{n\}} \ba_{N_{i} \setminus \{n\}})[\interval], \ba_{N_{i} \setminus \{n\}}\right) \in \bigcup_{i \leq s\leq k} \Pi_{N_{i}} R_{N_s} \right\}
    \end{align*}
    \item[output] a range sum $\bigoplus_{\interval \in [\open+1, \close]} \formula_n\left(A_0(\Pi_{U \setminus \{n\}} \ba_{N_{i} \setminus \{n\}})[\interval], \; \ba_{N_{i} \setminus \{n\}} \right)$, (or $\zerobf$ if $\open = \close$).
\end{description}

To conclude this section, we present the following theorem, whose proof is immediate from the prior discussion in this section. Here, $k = d(n)$, where we recall that $d(n)$ is number of hyperedges containing the $\pn$-leaf $n$ in $\mH_n$.

\begin{theorem} \label{thm:oracle-main}
    Let $\mH_n$ be a signed hypergraph with a $\pn$-leaf $n$. Let $\formula_n(\bx_{[n]})$ be a $\nestfaqneg$ expression \eqref{eq:derivation} associated with $\mH_n$ (thus recognized by $\CFG_n$ \eqref{eq:cfg_n}) and $\db_n$ be its database instance. Suppose that $U = N_0 \subseteq N_1 \subseteq \cdots \subseteq N_k$ are the only hyperedges containing the $\pn$-leaf $n$ in $\mH_n$. Then, there is an algorithm that takes $(\formula_n(\bx_{[n]}), \db_n)$ as input and constructs in $O(|\formula_n| + k \cdot |\db_n|)$ time, a $\RangeSumOracle$ $\Oracle_i(\bx_{N_i \setminus \{n\}})$, for every $i = 0, 1, \ldots, k$, and the oracle $\Oracle_i(\bx_{N_i \setminus \{n\}})$ is of size $O(|\db_n|)$ and supports the following range query in $O(\ackfunc{\db_n})$ time:
    \begin{description}
        \item[oracle] $\Oracle_{i}(\bx_{N_{i} \setminus \{n\}})$
        \item[input] a query key $\ba_{N_{i} \setminus \{n\}} \in \bigcup_{i \leq s \leq k} \Pi_{N_{i} \setminus \{n\}} R_{N_s}$ to identity the array decomposition $\Oracle_{i}(\ba_{N_{i} \setminus \{n\}})$ to be queried, and a query range $[\open+1, \close]$, where $0 \leq \open \leq \close$ are two break points of the array decomposition $\Oracle_{i}(\ba_{N_{i} \setminus \{n\}})$. In particular, the break points of the array decomposition $\Oracle_i(\ba_{N_i \setminus \{n\}})$ are $\interval$ and $\interval-1$, where
        \begin{align*}
            \interval \in \left\{\interval \mid \left( A_0(\Pi_{U \setminus \{n\}} \ba_{N_{i} \setminus \{n\}})[\interval], \ba_{N_{i} \setminus \{n\}}\right) \in \bigcup_{i \leq s\leq k} \Pi_{N_{i}} R_{N_s} \right\}
        \end{align*}
        \item[output] a range sum $\bigoplus_{\interval \in [\open+1, \close]} \formula_n\left(A_0(\Pi_{U \setminus \{n\}} \ba_{N_{i} \setminus \{n\}})[\interval], \; \ba_{N_{i} \setminus \{n\}} \right)$, (or $\zerobf$ if $\open = \close$).
    \end{description}    
\end{theorem}

\subsection{The $\pn$-elimination step: a summary} \label{sec:putting_things_together}
In this section, we put together the discussions in Appendix~\ref{sec:rewriting}, Appendix~\ref{sec:oracle} and Appendix~\ref{sec:aggregation} to prove the main theorem for the $\pn$-elimination sequence, i.e. \autoref{thm:faqneg-main}. First, we give a proof for \autoref{thm:aggregation-main}, now equipped with the $\RangeSumOracle$ described in \autoref{sec:oracle}.

\begin{proof} [Proof of Theorem \ref{thm:aggregation-main}]
    The aggregation algorithm follows \eqref{eq:derivation-aggregate} line-by-line and ``pushes'' the aggregation operator $\bigoplus_{x_n}$ into the inner-most parenthesis, therefore costing time $O(|\formula_n|)$ to scan over the derivation of $\formula_n(\bx_{[n]})$. Factoring out the term $\formula_n(\bx_{S_i})$, where $i = 0, \ldots, k = d(n)$ and $n \notin S_i$ is a pure syntactic step. The last step aggregates out $x_n$ from the pivot factor and gets a new positive factor $R_{U \setminus \{n\}}(\bx_{U \setminus \{n\}})$ in $O(|R_U|)$ time and space. 

    The only tricky step in \eqref{eq:derivation-aggregate} for pushing the aggregation operator $\bigoplus_{x_n}$ into the inner-most parenthesis is to efficiently compute the list representation of $R_{N_i \setminus \{n\}}(\bx_{N_i \setminus \{n\}})$, where $i = 1, \ldots, k = d(n)$, and by definition,
    \begin{align*}
        R_{N_i \setminus \{n\}}(\ba_{N_i \setminus \{n\}})  = 
        \begin{cases} \bigoplus_{x_n}  \formula_n (x_n, \ba_{N_i \setminus \{n\}})  & \text { if } \ba_{N \setminus \{n\}} \in \Pi_{N \setminus \{n\}} R_N \\ \zerobf & \text { otherwise } \end{cases}
    \end{align*}
    where $ \formula_n (x_n, \ba_{N_i \setminus \{n\}}) =  R_{N_i}({x_n, \ba_{N \setminus \{n\}}}) \oplus \indicator{\neg R_{N_i}}({x_n, \ba_{N_i \setminus \{n\}}}) \otimes \formula_n({x_n, \ba_{N^-_i \setminus \{n\}}})$. Now we apply \autoref{thm:oracle-main} to get the oracles $\Oracle_i(\bx_{N_i \setminus \{n\}})$ for each $i = 1, \ldots, k$. Then, the weight $R_{N_i \setminus \{n\}}(\ba_{N_i \setminus \{n\}})$ can be obtained in $O(\ackfunc{\db_n})$ time, for each $\ba_{N_i \setminus \{n\}} \in \Pi_{N_i \setminus \{n\}} R_{N_i}$, by calling the oracle $\Oracle_i(\bx_{N_i \setminus \{n\}})$ with inputs $\left(\ba_{N_i \setminus \{n\}}, [1, \bot] \right)$. Indeed, the oracle runs in time $O(\ackfunc{\db_n})$ and returns the range sum
    \begin{align*}
        \bigoplus_{\interval \in [1, \bot]} \formula_n \left(A_0(\Pi_{U \setminus \{n\}} \ba_{N_i \setminus \{n\}})[\interval], \; \ba_{N_i \setminus \{n\}}\right) = \bigoplus_{x_n \in \dom(x_n)} \formula_n (x_n, \ba_{N_i \setminus \{n\}}) = R_{N_i \setminus \{n\}}(\ba_{N_i \setminus \{n\}})
    \end{align*}
    Therefore, the list representation of the new factor $R_{N_i \setminus \{n\}}$ can be computed in $O(|R_{N_i}| \cdot \ackfunc{\db_n})$ time and is of size $O(|R_{N_i}|)$. The overall time complexity of the aggregation algorithm is $O(d(n) \cdot |\db_n| \cdot \ackfunc{\db_n})$ since there are $d(n)$ steps in the derivation of $\formula_n(\bx_{[n]})$ that requires the construction of a new factor $R_{N_i \setminus \{n\}}$ and each construction takes at most $O(|\db_n| \cdot \ackfunc{\db_n})$ time (to further push the aggregation operator $\bigoplus_{x_n}$ into the nested subexpression). After we aggregate out the variable $x_n$ from $\formula_n(\bx_{[n]})$, we automatically get a new $\nestfaqneg$ expression $\formula_{n-1}(\bx_{[n-1]})$. Since we keep the derivation of $\formula_n(\bx_{[n]})$ intact, we have $|\formula_{n-1}| = O(|\formula_n|)$ and $\formula_{n-1}(\bx_{[n-1]})$ is indeed a $\nestfaqneg$ expression recognized by $\CFG_{n-1}$. Moreover, as all the new factors $R_{N_i \setminus \{n\}}$ are still of size $O(|R_{N_i}|)$, we have $|\db_{n-1}| = O(|\db_n|)$.
\end{proof}

We finally prove the main theorem \autoref{thm:faqneg-main} for free-connex $\pn$-acyclic $\nestfaqneg$ queries.


\begin{proof} [Proof of \autoref{thm:faqneg-main}]
    Recall that for a $\pn$-leaf, say $n$, we run the following algorithms in order:
    \begin{enumerate}
        \item first, by \autoref{thm:rewrite}, we apply the $\Rewrite$ algorithm in time $O(|\formula|^2 + d(n) \cdot |\db|)$, where $d(n)$ is the number of hyperedges in $\mH$ that contain $n$, to transform $(\formula(\bx_{[n]}), \db)$ into a $\nestfaqneg$ expression $\formula_n(\bx_{[n]})$ recognized by $\CFG_n$ \eqref{eq:cfg_n} and a database instance $\db_n$ such that $\formula_n(\bx_{[n]}) = \formula(\bx_{[n]})$. Moreover, we have that $|\formula_n| \leq |\formula|$ and $|\db_n| \leq |\db|$. 
        \item next, we run the oracle-construction algorithm described in Appendix~\ref{sec:oracle} for $\formula_n(\bx_{[n]})$. By \autoref{thm:oracle-main}, the construction of oracles takes $O(|\formula_n| + |\db_n| \cdot \ackfunc{\db_n})$ time.
        \item finally, we apply the aggregation algorithm on $(\formula_n(\bx_{[n]}), \db_n)$ and by \autoref{thm:aggregation-main}, we obtain in $O(|\formula_n| + d(n) \cdot |\db_n| \cdot \ackfunc{\db_n})$ time the tuple $(\formula_{n-1}(\bx_{[n-1]}), \db_{n-1})$, where $\formula_{n-1}(\bx_{[n-1]})$ is a $\nestfaqneg$ expression associated with $\mH_{n-1} = \removeleaf{\mH_n}{n-1}$ and $\db_{n-1}$ is its database instance such that $\formula_{n-1}(\bx_{[n-1]}) = \bigoplus_{\bx_{[n] \setminus F} \in \dom(\bx_{[n] \setminus F})} \; \formula_n(\bx_{[n]})$. By \autoref{thm:aggregation-main}, we have that $|\formula_{n-1}| \leq |\formula_n|$ and $|\db_{n-1}| \leq |\db_n|$. Thereafter, recall that $\formula_{n-1}(\bx_{n-1})$ is recognized by the following $\CFG_{n-1}$ \eqref{eq:cfg_n-1}
        \begin{equation} 
            \begin{aligned}
               \framebox{$\CFG_{n-1}$} \qquad \qquad \; \\
                \text{ for all } S \subseteq [n-1], \\
                    \nonterm_{n-1}(\bx_{\emptyset}) &::= \; e  
                    && \text{ where } e \in \boldsymbol{D} \setminus \{\zerobf\}  \\
                    \nonterm_{n-1}(\bx_{S}) &::= \; R_K(\bx_K) \otimes \nonterm_{n-1}(\bx_{S^-}),  
                    && \text{ where } K \in \mE^+_{n-1}, S^- \subseteq S \text{ such that } S = K \cup S^- \\
                    & \; \; \mid \; (R_N(\bx_N) \vdash \nonterm_{n-1}(\bx_{N^-})) \otimes \nonterm_{n-1}(\bx_{S^-}), 
                    && \text{ where } N \in \mE^-_{n-1}, N^- \subseteq N, S^- \subseteq S \text{ such that } S = N \cup S^-
            \end{aligned}
        \end{equation}
    \end{enumerate}
    Thus, the $\pn$-elimination step on $n$ takes $O(|\formula|^2 + d(n) \cdot |\db| \cdot \ackfunc{\db})$ time in total. 

    Now we are ready for the next $\pn$-elimination step for $n-2$ with $\pn$-leaf $n-1$. We keep repeating the $\pn$-elimination step for $i = n-2, \dots, f$ and every intermediate $\nestfaqneg$ expression $\formula_{i}(\bx_{[i]})$ and its corresponding database instance $\db_i$ have their sizes bounded by $O(|\formula|)$ and $O(|\db|)$, respectively. In the end, we obtain a full $\nestfaqneg$ expression $\formula_f(\bx_{[f]})$ and a database instance $\db_f$ as desired, i.e.
    \begin{align*}
        \varphi(\bx_F) = \bigoplus_{\bx_{[n] \setminus F} \in \dom(\bx_{[n] \setminus F})} \; \formula(\bx_{[n]}) = \bigoplus_{\bx_{[n-1] \setminus F} \in \dom(\bx_{[n-1] \setminus F})} \; \formula_{n-1}(\bx_{[n-1]}) = \cdots = \formula_f(\bx_{[f]}).
    \end{align*}
    Therefore, the total time for the $\pn$-elimination sequence (for $i = n-1, \dots, f$) is bounded by (by definition, $|\varphi| = |\formula|$)
    \begin{align*}
        \sum_{i = f}^{n-1} O(|\formula|^2 + d(i) \cdot |\db| \cdot \ackfunc{\db}) = O(|\formula|^3 + |\formula| \cdot |\db| \cdot \ackfunc{\db}).
    \end{align*}
    From now on, we follow the enumeration algorithm for full $\nestfaqneg$ queries proposed in Appendix~\ref{sec:enum-full-faqneg} and by Appendix~\ref{cor:enum-full-faqneg}, thus we close the proof.
\end{proof}


\section{Missing proofs in Section~\ref{sec:hardness}} \label{appendix:hardness-proofs}

\subsection{Some Notations and Results}

We apply a characterization for $\alpha$-acyclic hypergraphs by Beeri, Fagin, Maier and Yannakakis~\cite{BeeriFMY83}. The \emph{Gaifman graph} of a hypergraph $\mH$ is the clique graph of $\mH$, i.e., we replace every hyperedge of size $k$ with a clique of size $k$ over the variables in the hyperedge. 

\begin{definition}[Conformal Hypergraph]
A hypergraph $\mH$ is called \emph{conformal} if for every clique in its Gaifman graph, there exists an atom covering the clique. 
\end{definition}

\begin{definition}[Chordal Hypergraph]
A hypergraph $\mH$ is called \emph{chordal} if every cycle of length at least 4 in its Gaifman graph has a chord.
\end{definition}

\begin{theorem}[Beeri et al., 83']\label{thm:non-alpha-acyclic}
A hypergraph $\mathcal{H}$ is $\alpha$-acyclic if and only if it is conformal and chordal.
\end{theorem}

We also reprove a lemma which strengthens Theorem~\ref{thm:non-alpha-acyclic} for signed hypergraphs~\cite{BBThesis}. The result originally appeared in French~\cite{BBThesis}.
Given a signed hypergraph $\mH = ([n], \mE^+,\mE^-)$, for some $S \subseteq [n]$ we denote by $\mH[S]$ the induced signed hypergraph on the vertex set $S$.

\begin{lemma}[Brault{-}Baron, 13']\label{lm:non-alpha-acyclic}
If a signed hypergraph $\mH = ([n], \mE^+,\mE^-)$ is not $\alpha$-acyclic, then we can find a set of vertices $S \subseteq [n]$, with $|S| \geq 3$, such that:
\begin{itemize}
\item[-] the Gaifman graph of the positive hyperedges in $\mH[S]$ is a clique of size $|S|$;
\item[-] there exists a subset $\mE'$ of the negative hyperedges in $\mH[S]$ such that the Gaifman graph of the union of $\mE'$ with the positive atoms in $\mH[S]$ is a $|S|$-cycle.
\end{itemize}
\end{lemma}

\begin{proof}
Take a minimal set of vertices $S$ such that $\mH[S]$ is not $\alpha$-acyclic. Take the minimal set $\mE'  \subseteq \mE^-$ such that the hypergraph with edges $\mE^+[S] \cup \mE' $ is not $\alpha$-acyclic. By Theorem~\ref{thm:non-alpha-acyclic}, there exists a set of vertices $S'$ such that $\mH'[S']$ is a cycle or a clique. In both cases, the minimality of $S$ imposes that $S = S'$. If $\mH[S']$ were a cycle, then we are done. Otherwise, $\mH[S']$ is a clique. If $\mE'  = \emptyset$, then we are also done. Otherwise, let  $e \in \mE' $. By the minimality of $\mE' $, we know that (1) $e \notin \mE^+$ and (2) $e$ participates as an edge in the clique $\mH'[S]$. But then note that $(\mE^+[S] \cup \mE'  \setminus \{e\})[S\setminus x]$, for the unique $x$ where $x \in S$ but $x \notin e$, is a clique of smaller size. This contradicts the minimality of $S$. The proof is now complete.
\end{proof}

\subsection{Missing proofs in Section~\ref{subsec:hardness}} \label{appendix:non-pn-acyclic-lower-bound-proofs}

\begin{proof}[Proof of Theorem~\ref{thm:non-pn-acyclic boolean}]
Let $\mH$ be the signed hypergraph of $Q$. We present the proof of Theorem~\ref{thm:non-pn-acyclic boolean} when the hypergraph $\mH$ is simple, i.e. without duplicate hyperedges. The proof can be easily adapted to multi-hypergraphs. We note that our proof does not work for $Q$ with self-joins, whose linear time boundary remains to be an open problem.

\smallskip

Since $\mH$ is not $\pn$-acyclic, by Lemma~\ref{lm:non-alpha-acyclic}, we can split into two cases.
Suppose it is the first case of Lemma~\ref{lm:non-alpha-acyclic}, then the proof is identical to that in~\cite{BerkholzGS20, BBThesis} which we omit for brevity, since the clique consists of only positive hyperedges (the instance in the reduction will have every negated atom be empty).

Otherwise, we are in the second case of Lemma~\ref{lm:non-alpha-acyclic}. Therefore, there are $\ell \geq 3$ variables $x_1, x_2, \dots, x_\ell$ that form a chordless cycle (or cycle of length 3) in the Gaifman graph of $\mH' = \mH[S]$. Given any input graph $G$ for the \textsc{Triangle} problem, we construct our database instance as follows. For every vertex $v$ in $G$, we have a fresh domain element $a_v$. For any variable $w$ not in the set $\{x_1, x_2, \dots, x_\ell\}$, we define its active domain being $\{\perp_w\}$, a fresh domain element. 
For any $i \in \{1, \dots, \ell\}$, there exists exactly one hyperedge $U_i$ that contains $x_i, x_{i+1}$ (modulo $\ell$) since the cycle $x_1,x_2, \dots,x_\ell$ is chordless. 
%
For $i \in \{1,2,3\}$, for $U_i \in \mE^+$ ($U_i \in \mE^-$ resp.), we define $R_{U_i} :=\left\{\left(f_{u v}\left(w_1\right), \ldots, f_{u v}\left(w_r\right)\right) \mid \{u, v\} \in E(G)\right\}$ ($R_{U_i}:=\left\{\left(f_{u v}\left(w_1\right), \ldots, f_{u v}\left(w_r\right)\right) \mid \{u, v\} \notin E(G)\right\}$ resp.) where $f_{uv}(x_i) = a_u, f_{uv}(x_{i+1})=a_v$ and $f_{uv}(w) = \perp_w$ for all other variables $w$ in the atom. 
For $i \in \{4, \dots, \ell\}$, for $U_i \in \mE^+$ ($U_i \in \mE^-$ resp.), we define $R_{U_i} :=\left\{\left(g_{u v}\left(w_1\right), \ldots, g_{u v}\left(w_r\right)\right) \mid u = v  \in V(G)\right\}$ ($R_{U_i}:=\left\{\left(g_{u v}\left(w_1\right), \ldots, g_{u v}\left(w_r\right)\right) \mid u \neq v \in V(G) \right\}$ resp.) where $g_{uv}(x_i) = a_u, g_{uv}(x_{i+1})=a_v$ and $g_{uv}(w) = \perp_w$ for all other variables $w$ in the atom. 
For the remaining positive relations, we define $R_{U}:=\left\{\left(h_v\left(w_1\right), \ldots, h_v\left(w_r\right)\right) \mid v \in V(G)\right\}$ where $h_v(x_j) = a_v$ for all $j \in \{1, \dots, \ell\}$ and $g_v(w) = \perp$ for all other variables $w$. We set the remaining negative relations to be empty. Note that there could be at most $n^2$ tuples in each atom, where $n$ is the number of vertices in $G$. A moment of reflection should convince the reader that the input graph $G$ contains a triangle if and only if the database constructed returns non-empty result for $Q$. 
\end{proof}

\begin{proof}[Proof of Theorem~\ref{thm:non-free-connex boolean}]
We adapt the proof from~\cite{BaganDG07, BerkholzGS20}. The main difference is how we handle the negated atoms in the query. 

Let $F$ be the set of head variables of $Q$. Let $\mH = ([n], \mE^+, \mE^-)$ be the signed hypergraph of $Q$. Since $\mH$ is not free-connex, there exists a subset $\mE' \subseteq \mE^-$ such that the hypergraph $\mH' = ([n], \mE)$, where $\mE = \mE^+ \cup \mE'$, is not free-connex (but it is $\alpha$-acyclic). By \cite{BaganDG07}, then $\mH'$ admits a sequence of distinct vertices $P = (x, z_1, z_2, \ldots, z_k, y)$ with $k \geq 1$, such that
\begin{itemize}
    \item $x, y \in F$ and $z_1, z_2, \ldots, z_k \notin F$, 
    \item ($P$ is a path) there are $k+1$ hyperedges $e_0, e_1, \cdots, e_{k-1}, e_k \in \mE$ that contain the respective pairs $e_0'= \left\{x, z_1\right\}, e_1^{\prime}=\left\{z_1, z_2\right\}, \ldots, e_{k-1}^{\prime}=\left\{z_{k-1}, z_k\right\}, e_k^{\prime}=\left\{z_k, y\right\}$, and
    \item ($P$ is chordless) for each $e \in \mE$, $|e \cap P| \leq 1$ or $|e\cap P| = e_i'$, for some $0 \leq i\leq k$.
\end{itemize}
W.l.o.g, we assume that in $Q$, $\bx_F = (x, y, \bt)$ and $\bx_{[n]} = (x, y, z_1, z_2, \ldots, z_k, \bt, \bu)$, so each atom of $Q$ is of one of the following types for some $\bv \subseteq \{\bt, \bu\}$: $(1)$ $R(x, z_1, \bv)$; $(2)$ $R(z_k, y, \bv)$; $(3)$ $R(z_i, z_{i+1}, \bv)$, where $1 \leq i < k$; $(4)$ $R(w, \bv)$, where $w \in \{x, z_1, z_2, \ldots, z_k, y\}$ on the path $P$, and $(5)$ $R(\bv)$.
    
    Now we illustrate the reduction from $\bmm$ to the $\cqneg$ $Q$ by constructing an instance $(R_K)_{K \in \mE^+} \cup (R_N)_{N \in \mE^-}$ for $Q$.
    Let $A, B$ be two binary relations that encode the two given $n \times n$ Boolean matrices. Let $I$ be the identity relation on the domain $[n]$, $I= \{(a, a) \mid a \in [n]\}$.
    The active domain for the variables $\{x, z_1, z_2, \ldots, z_k, y\}$ is defined as $[n]$. We introduce $\perp$ as a fresh domain element and define the active domain for variables in $\bt, \bu$ as $\{\perp\}$. Then, for each $K \in \mE^+$, we construct the relational instance following the proof of Lemma 26 in \cite{BaganDG07}. The construction of each negated relation in $\mE^-$ of the corresponding type is simply the complement of its positive counterpart. To be precise, if $K$ has arity $p$, we have:
    \begin{enumerate}
        \item [(1)] $$ R(x, z_1, \bv) \defeq \begin{cases} A \times \{\perp\}^{p - 2}, & \text { if } \{x, z_1, \bv\} \in \mE^+  \\ (\neg A) \times \{\perp\}^{p - 2}, & \text { otherwise } (\{x, z_1, \bv\} \in \mE^-) \end{cases}
        $$
        \item [(2)] $$ R(z_k, y, \bv) \defeq \begin{cases} B \times \{\perp\}^{p - 2}, & \text { if } \{z_k, y, \bv\} \in \mE^+  \\ (\neg B) \times \{\perp\}^{p - 2}, & \text { otherwise } (\{z_k, y, \bv\} \in \mE^-) \end{cases}
        $$
        \item [(3)] $$ R(z_i, z_{i+1}, \bv) \defeq \begin{cases} I \times \{\perp\}^{p - 2}, & \text { if } \{z_i, z_{i+1}, \bv\} \in \mE^+  \\ (\neg I) \times \{\perp\}^{p - 2}, & \text { otherwise } (\{z_i, z_{i+1}, \bv\} \in \mE^-) \end{cases}
        $$
        \item [(4)] $$ R(w, \bv) \defeq \begin{cases} [n] \times \{\perp\}^{p - 1}, & \text { if } \{z_i, z_{i+1}, \bv\} \in \mE^+  \\ \emptyset, & \text { otherwise } (\{z_i, z_{i+1}, \bv\} \in \mE^-) \end{cases}
        $$
        \item [(5)] $$ R(\bv) \defeq \begin{cases} \{\perp\}^{p}, & \text { if } \{\bv\} \in \mE^+  \\ \emptyset, & \text { otherwise } (\{\bv\} \in \mE^-) \end{cases}
        $$
    \end{enumerate}
    It is easy to see that all relations can be constructed in $O(n^2)$ time and also have size $O(n^2)$. We prove that the output of $Q$ on this relational instance (projecting on $x, y$), denoted as $C$, encodes the resulting matrix of the Boolean matrix multiplication. 
    
    Indeed, if $(i, j)$-th entry (in the resulting matrix) is $1$, then there is some $k \in [n]$ such that $(i, k) \in A$ and $(k, j) \in B$. It is easy to verify that the valuation $\nu(\cdot)$, where $\nu(x) = i, \nu(y) = j, \nu(z_1) = \cdots = \nu(z_k) = k$ and $\nu(v)= \; \perp$, for $v \in \{\bt, \bu\}$, is a true valuation ($(i, j) \in C$). If $(i, j)$-th entry (in the resulting matrix) is $0$, for any $k \in [n]$,  either $(i, k) \notin A$ or $(k, j) \notin B$. That is, there is no valuation that satisfies both $R(x, z_1, \bv)$ and $R(z_k, y, \bv)$. Therefore, $(i, j) \notin C$.

    Now that we have proven the reduction to an instance for $Q$ of size $O(n^2)$, a $O(n^2)$ pre-processing time and $O(1)$ delay algorithm for $Q$ implies that the resulting matrix can be obtained in time $O(n^2)$, contradicting $\bmm$.
\end{proof}

\subsection{Missing proofs in Section~\ref{subsec:embedding}}

We begin with the definitions of \emph{$k$-clique embedding} and its \emph{weak edge depth}.


\begin{definition}[$k$-Clique Embedding]
Let $C_k$ be the $k$-clique and $\mathcal{H}$ be a hypergraph. A \emph{$k$-clique embedding} $\psi$ to $\mathcal{H}$, denoted as $C_k \mapsto \mathcal{H}$, maps every vertex $v$ in $C_k$ to a non-empty subset $\psi(v) \subseteq V(\mathcal{H})$ such that the followings hold:
\begin{enumerate}
\item $\psi(v)$ induces a connected subhypergraph;
\item for any $u,v$ in $C_k$, either $\psi(u) \cap \psi(v) \neq \emptyset$ or there exists a hyperedge $e \in E(\mathcal{H})$ such that $e \cap \psi(u) \neq \emptyset $ and $e \cap \psi(v) \neq \emptyset$.
\end{enumerate}
\end{definition}


In this paper, we care about the \emph{weak edge depth} of a $k$-clique embedding $\psi$. Given a $k$-clique embedding $\psi$, we define $\forall e \in E(\mathcal{H})$ the weak edge depth of $e$ as $d_{\psi}(e) := |\{v \in V(C_k) \mid \psi(v) \cap e \neq \emptyset \}|$. Then for a $k$-clique embedding, its weak edge depth is defined to be $\mathsf{wed}(\psi) := \max_e d_{\psi}(e)$. Finally, we define $\mathsf{wed}(\psi)$ to be the minimum weak edge depth of any $k$-clique embedding to $\mathcal{H}$. We are now ready to define the \emph{signed clique embedding power}.

\begin{definition}[Signed Clique Embedding Power]
Given a signed hypergraph $\mathcal{H}$, define  the {\em signed clique embedding power} of $\mathcal{H}$ as  
$$\mathsf{semb}(\mathcal{H}) := \max_{H \in \widehat{\mathcal{H}}}\sup_{k \geq 3} \mathsf{semb}_k(H) = \max_{H \in \widehat{\mathcal{H}}}\sup_{k \geq 3} \frac{k}{\mathsf{wed}(C_k \mapsto H)}$$
where $\widehat{\mathcal{H}}$ is the set of all possible hypergraphs with hyperedges $\mE^+ \cup \mE'$ where $\mE' \subseteq \mE^-$.
\end{definition}


We first show that the singed clique embedding power for non-$\pn$-acyclic queries is strictly greater than 1.

\begin{theorem}\label{thm: non-pn > 1}
If $\mH$ is a non-$\pn$-acyclic signed hypergraph, then its signed clique embedding power is strictly greater than 1.
\end{theorem}

\begin{proof}[Proof of Theorem~\ref{thm: non-pn > 1}]
Since $\mH$ is not $\pn$-acyclic, there exists a subset $\mE' \subseteq \mE^-$ such that the hypergraph $\mH' = ([n], \mE^+ \cup \mE') \in \widehat{\mathcal{H}}$ is not $\alpha$-acyclic. By Theorem~\ref{thm:non-alpha-acyclic}, either $\mH'$ is non-conformal or $\mH'$ contains an induced $\ell$-cycle for some $\ell \geq 4$. 

If $\mH'$ is non-conformal, take a $k$-clique, $k\geq 3$ in its Gaifman graph such that no hyperedge in $q'$ covers the $k$-clique. Now we define the embedding from $C_k$ to $\mH'$ to be any one-to-one mapping between the vertices in $C_k$ and the $k$-clique in $\mH'$. The weak edge depth of this embedding must be strictly smaller than $k$, by the construction. Therefore, $\mH'$ has signed clique embedding power strictly greater than 1, and so does $\mH$.

Otherwise, there exists an induced $\ell$-cycle for some $\ell \geq 4$. We take the same embedding from~\cite{ICALP}. We name the variables of the cycle query as $x_1, \dots, x_{\ell}$. If $\ell$ is odd, let $\lambda = (\ell+1)/2$ and we construct the $\ell$-clique embedding as follows:
\begin{equation}
    \begin{aligned}
 \psi^{-1}(x_1) & = \{ 1,2, \dots, \lambda-1\} \\ 
  \psi^{-1}(x_2) & = \{ 2, 3, \dots, \lambda\} \\ 
  & \dots \\
\psi^{-1}(x_\ell) & = \{ 2\lambda-1, 1, \dots, \lambda-2\} 
    \end{aligned}
\end{equation}

If $\ell$ is even, we define $\lambda = \ell/ 2$ and an embedding from a $(\ell - 1)$-clique as follows:
\begin{align*}
 \psi^{-1}(x_1) & = \{ 1,2, \dots, \lambda-1\} \\ 
  \psi^{-1}(x_2) & = \{ 2, 3, \dots, \lambda\} \\ 
  & \dots \\
   \psi^{-1}(x_{\ell - 1}) & = \{ 2\lambda-2, 2\lambda-1, 1, \dots, \lambda-3\}  \\
\psi^{-1}(x_\ell) & = \psi^{-1}(x_{\ell - 1})
\end{align*}

Observe that the weak edge depth of the embedding is strictly smaller than $\ell$. Therefore in this case $\mH$ also has signed clique embedding power strictly greater than 1.
\end{proof}

We now show that the signed clique embedding power provides a conditional lower bound for the running time of a $\mathsf{SumProd}^{\neg}$ query over the tropical semiring, assuming Conjecture~\ref{conj:tropical_clique}. 

\begin{theorem}\label{thm:tropical_clique_lower bound}
Assuming Conjecture~\ref{conj:tropical_clique}, any $\mathsf{SumProd}^{\neg}$ query with signed hypergraph $\mathcal{H}$ over the tropical semiring cannot be solved in time $O(N^{\mathsf{semb}(\mH)-\epsilon})$ for any constant $\epsilon >0$.
\end{theorem}

\begin{proof}
Let  $\mH' \in \widehat{\mathcal{H}}$ be a hypergraph such that for some $k \geq 3$ there exists a $k$-clique embedding to $\mH'$ with weak edge depth $\lambda = \mathsf{semb}(\mH)$. 
From~\cite{ICALP}, the $\mathsf{SumProd}$ problem on the hypergraph $\mH'$ over the tropical semiring cannot be solved in time  $O(N^{\lambda -\epsilon})$ for any $\epsilon >0 $ assuming Conjecture~\ref{conj:tropical_clique}. Let  $\mH' = ([n], \mE^+ \cup \mE')$. We will show how to reduce $\mathsf{SumProd}$ on $\mH'$ to $\mathsf{SumProd}^\neg$ on $\mH$ (over the tropical semiring).

Given an input $I'$ for $\mathsf{SumProd}$ on $\mH'$, we construct an instance $I$ (of the same size) for  $\mathsf{SumProd}^\neg$ on $\mH$ as follows. If $U \in \mE^+$, $R_U$ is the same positive factor as in $I'$. If $U \in \mE^- \setminus \mE'$, $R_U$ is empty with default value $0$. Finally, if $U \in \mE'$, the negative factor $R_U$ takes the same values as in $I'$ with default value a large enough constant $c>0$ (we pick $c$ such that no sum of $\ell$ values from $I'$ can exceed it, where $\ell$ is the number of atoms in the query). 

It is easy to see that the minimum achieved for the $\mathsf{SumProd}^\neg$ problem on $\mH$ remains the same.
\end{proof}

\begin{proof}[Proof of Theorem~\ref{lb:count}]
Let $Q$ be the $\mathsf{SumProd}^{\neg}$ query with signed hypergraph $([n], \mE^+,  \mE^-)$. Suppose, for the sake of contradiction, that we can
compute $\# Q$ in linear time.

We use the inclusion-exclusion principle. Let $Q_S$ be the Boolean CQ with hypergraph $([n], \mE^+ \cup S)$ for any $S \subseteq \mE^-$.
Then, we can write:
\begin{align*}
\# Q(\db) = \sum_{S \subseteq \mE^-} (-1)^{|S|} \# Q_S(\db) 
\end{align*}
Since $\mH$ is not $\pn$-acyclic, there exists a minimal $U \subset \mE^-$ such that $Q_U$ is not $\alpha$-acyclic.

We can now construct a linear-time reduction from $\# Q_U$ to $\# Q$. Given an instance $\db$ for $\# Q_U$, we construct an instance $\db'$ for $\# Q$ by copying $\db$ for relations in $\mE^+ \cup U$, and making the relation empty if it is in $\mE^- \setminus U$. In this case, we have:
\begin{align*}
\# Q(\db') = \sum_{S \subseteq U} (-1)^{|S|} \# Q_S(\db') =  \sum_{S \subset U} (-1)^{|S|} \# Q_S(\db) + (-1)^{|U|} \# Q_U(\db)
\end{align*}
By minimality of $U$, all the queries $Q_S$ with $S \subset U$ are $\alpha$-acyclic and thus $\# Q_S(\db)$ can be computed in linear time. 
Hence, a linear-time algorithm for $\# Q$ implies a linear-time algorithm for $\# Q_U$, which is not $\alpha$-acyclic.
\end{proof}

%


%
%

\subsection{Missing proofs in Section~\ref{subsec:unconitional}}
\label{appendix:unconditional}



We prove Theorem~\ref{thm:unconditional-lb} in this section. First, as noted in~\cite{ChazelleR91}, it suffices to consider the commutative semigroup $(\mathcal{P}(X), \cup)$ for any set $X$, where $\mathcal{P}$ is the powerset of $X$. Indeed, any solution to a task $T$ with $n$ elements $\{a_i\}_{0\leq i \leq n-1}$ can be interpreted to the same task defined by $(\mathcal{P}([n]),\cup)$ 
. Note that the semigroup $(\mathcal{P}([n]), \cup)$ is faithful for any $n\geq 1$. It leads to the following definition in~\cite{ChazelleR91}.

\begin{definition}
A scheme $S$ is a sequence $s_0, \dots, s_{r-1}$ of subsets of $X$ such that for all $i \in [0,r-1]$, $s_i = s \cup s'$ where $s = s_j$ for some $j<i$ or $s= \{x\}$ for some $x\in X$ or $s = \emptyset$, and likewise for $s'$.
\end{definition}

The following lemma is proved in~\cite{ChazelleR91} for the partial sum case. The exact same proof works for OEPS.
\begin{lemma}
Let $T$ be a task for OEPS over $n$ variables and $S$ a scheme of minimum length solving it. Then, for any faithful semigroup, a solution to $T$ takes time at least $r-n$, where $r$ is the length of the scheme.
\end{lemma}

Following~\cite{ChazelleR91}, we continue with the definition of mappings between schemes. These are often simply maps between sets $f: X \rightarrow Y$ extended to maps between powersets $f: \mathcal{P}(X) \rightarrow \mathcal{P}(Y)$ in the usual way: requiring that $f(A \cup B)=f(A) \cup f(B)$. Other times, we intend that the map be between intervals. We denote by $\mathcal{I}(X)$ the set of all intervals in $\mathcal{P}(X)$, 
$
\mathcal{I}(X):=\left\{\left[x_i, x_j\right] \subseteq X \mid i \leq j\right\}
$
A map $f: X \rightarrow Y$ extends to $f: \mathcal{I}(X) \rightarrow \mathcal{I}(Y)$ by $f\left(\left[x_i, x_j\right]\right)=\left[f\left(x_i\right), f\left(x_j\right)\right]$. It is often convenient to define a map by defining its inverse first. A section of a map $f: X \rightarrow Y$ is a map $g: Y \rightarrow X$ such that $f \circ g=I_T$.

We will construct a family $T_n(t,k)$ of hard tasks parametrized by two integers $t\geq0$ and $k\geq0$, called \emph{time} and \emph{density} respectively; the subscript $n$ is the number of variables and is not a parameter. The construction mimics the one in~\cite{ChazelleR91}, but extends it to be the semantics of OEPS in the obvious way. However, syntactically, readers can think that the tasks for (standard) partial-sum and OEPS are both sets of pairs of elements.

First, define the function $R(t,k)$ for all integers $t\geq1$ and $k\geq0$~\cite{ChazelleR91}:
$$
\begin{aligned}
& R(1, k)=2 k & \text{if }k \geq 0, \\
& R(t, 0)=3 & \text{if }t>1 \\
& R(t, k)=R(t, k-1) \cdot R(t-1, R(t, k-1)) & \text{if }k>0, t>1.
\end{aligned}
$$
This function gives the $n$ needed to construct the hard task $T_n(t, k)$.

The next lemma is crucial.
\begin{lemma}\label{lm:requirements}
For all integers $t \geq 1$ and $k \geq 0$, there is a task $\mathcal{T}_n(t, k)$ for OEPS over the $n$ element set $X=\left\{x_0, \ldots, x_{n-1}\right\}$ satisfying the three requirements:
\begin{enumerate}
\item $\left|\mathcal{T}_n(t, k)\right| \geq k n / 2$, where $n=R(t, k)$.
\item $\left|\left\{(x_i, x_j) \in \mathcal{T}_n(t, k) \mid i=l\right\}\right| \leq k$ for any $l \in[0, n-1]$.
\item If $S=\left\{s_0, s_1, \ldots, s_{r-1}\right\}$ is a scheme solving $\mathcal{T}_n(t, k)$, then $r \geq t\left|T_n(t, k)\right| / 3$.
\end{enumerate}
All intervals in $\mathcal{T}_n(t, k)$ are nontrivial.
\end{lemma}

Our construction of tasks $T_n(t,k)$ for OEPS is by double induction on $t$ and $k$, which is similar to the construction in~\cite{ChazelleR91}.
First, we show the construction for the base cases:
\begin{itemize}
\item if $t \geq 2$ and $k = 0$, define $T_3(t,0) := \emptyset$;
\item if $t=1$ and $k\geq0$, we have $n=R(1,k) = 2k$ and the variable set is $\{x_0,x_1,\dots,x_{n-1}\}$.  Define 
$$T_n(1,k) = \bigcup\limits_{i=0}^{n-k-1} \bigcup\limits_{j=1}^{k} \{(x_i,x_{i+j})\}.$$
\end{itemize}

It can be easily verified that the base case constructions satisfy the three requirements posed in Lemma~\ref{lm:requirements}. For example for the third requirement: when $k=0$, any scheme solving $T_3(t,0)$ trivially has length longer than 0; when $t=1$ and $k\geq0$, any scheme solving $T_n(1,k)$ must contain at least $|T_n(1,k)|$ steps (in order to output $|T_n(1,k)|$ many different sums), which is greater than $t\left|T_n(t, k)\right| / 3 = \left|T_n(1,k)\right|/3$ since $t=1$.

We now construct the tasks when $k>0$ and $t>1$. By induction hypothesis, we have tasks $A=\mathcal{T}_a(t, k-1)$ and $B=\mathcal{T}_b(t-1, a)$ where $a=R(t, k-1)$ and $b=R(t-1, a)$. 
Since $R(t, k)=ab$, we intend to construct task $Q=\mathcal{T}_n(t, k)$ over $n=ab$ variables. 
For clarity, we give different names to all these different variable sets. 
Name the variables in $Q$ by $X=\left\{x_0, \ldots, x_{n-1}\right\}$, those in $A$ by $Y=\left\{y_0, \ldots, y_{a-1}\right\}$, those in $B$ by $Z=\left\{z_0, \ldots, z_{b-1}\right\}$

Divide $X$ into $b$ blocks each containing $a$ consecutive variables. In to each block, we place a copy of the task $A$. To state this formally, we define the map:
$$
\varphi: \begin{array}{clc}
\mathcal{P}(X) & \rightarrow \mathcal{P}(Y) \\
x_i & \mapsto y_{i \bmod a}
\end{array}
$$
and take these sections:
$$
\begin{aligned}
\varphi_j: \mathcal{I}(Y) & \rightarrow \mathcal{I}(X) \\
y_i & \mapsto x_{j a+i}
\end{aligned}
$$
where $j=0, \ldots, b-1$. Each section gives a copy of $A$ placed in $X$, it is the image of $A$ by $\varphi_j$. Though $\varphi$ is a map between subsets of sets, it is defined as if it were a map of sets. As in~\cite{ChazelleR91}, we will not carry out in the notation the details for these distinctions, but the readers are encouraged to reflect on these distinctions.

We then mark some of the $x_i$'s. Let the leftmost variable in each block be marked, that is, $x_{i G}$ for $i=0,1, \ldots, b-1$. Now alter the marking by removing the mark on $x_0$ and placing it on $x_{n-1}$.

We now partition $B$ into a  subsets $B_0, \ldots, B_{a-1}$ so that the partition obeys the following restrictions:
\begin{enumerate}
\item $\left|\left\{\left[z_j, z_k\right] \in B_i \mid j=c\right\}\right| \leq 1$ for all $i \in[0, a-1]$ and $c \in[0, b-1]$.
\item $\left[z_{b-2}, z_{b-1}\right] \notin B_0$.
\end{enumerate}

It is possible to construct this partition thanks to the uniform right-degree condition on $B$ and that fact that there is only one nontrivial interval ending over $z_{b-2}$.
For $i \in[0, a-1]$ define the map,
$$
\begin{aligned}
& \psi_i: \mathcal{I}(Z) \rightarrow \mathcal{I}(X) \\
& {\left[z_j\right] \mapsto \begin{cases}{\left[x_{(j+1) a-i}, x_{(j+1) a}\right]} & \text { for } j \in[0, b-2] \\
{\left[x_{a b-1}\right]} & \text { if } j=b-1.\end{cases} } \\
&
\end{aligned}
$$

Continuing as in~\cite{ChazelleR91}, we now define a map of intervals. To this end we shall ensure that $\psi_i\left(\left[z_j, z_k\right]\right)$ is the smallest interval containing all of $\left\{\psi_i\left(z_j\right), \psi_i\left(z_k\right)\right\}$. Each of these is a section of the map,
$$
\begin{aligned}
& \psi: \mathcal{P}(X) \rightarrow \mathcal{P}(Z) \\
& x_i \mapsto \begin{cases}z_j & \text { if } i=a(j+1), \\
z_{b-1} & \text { if } i=a b-1, \\
\emptyset & \text { otherwise. }\end{cases} \\
&
\end{aligned}
$$

An image of each $B_i$ is placed over $X$ using $\psi_i$. Remark that any such interval over $X$ spans blocks. For this to be true, the restriction that $\left[z_{b-2}, z_{b-1}\right]$ not be in $B_0$ is crucial.

Finally, the task $Q$ is defined as,
$
Q:=\left(\bigcup_{j=0}^{b-1} \varphi_j(A)\right) \cup\left(\bigcup_{i=0}^{a-1} \psi_i\left(B_i\right)\right).
$

It is not hard to show that the task $Q$ satisfies the first two requirements in Lemma~\ref{lm:requirements}. Indeed, by construction, $Q$ is a collection of nontrivial intervals in $X$. In Iact, eacn map $\varphi_j$ and $\psi_i$ is one-to-one. The distinct character of each of these maps assures that each component that went into making $Q$ is disjoint with every other. This implies that
$$
\begin{aligned}
|Q| & =\sum_{j \in[0,b-1]}\left|\varphi_j(A)\right|+\sum_{i \in[0, a-1]}\left|\psi_i\left(B_i\right)\right| \\
& =b|A|+|B| \geq b(k-1) a / 2+a b / 2=k a b / 2
\end{aligned}
$$
and
$$
\begin{aligned}
\left|\left\{\left(x_i, x_j\right) \in Q \mid i=c\right\}\right|= & \left|\left\{\left(y_i, y_j\right) \in A \mid i=c \bmod a\right\}\right| \\
& +\left|\left\{\left(z_i, z_j\right) \in B_{(-c) \bmod a} \mid i=\lfloor(c-1) / a\rfloor\right\}\right| \\
\leq & (k-1)+1=k
\end{aligned}
$$
for all $c \in [0, ab-1]$. Therefore, $Q$ is a task of the correct density and obeys the uniform right-degree condition. 

Our observation is that the lower bound in~\cite{ChazelleR91} still works whitebox-wise. Intuitively, for fixed set of pairs, the OEPS computes strictly more than the standard partial sums, and therefore it should not be easier to compute the OEPS. Mathematically, we need to carefully examine that, indeed, no ``magical shortcut'' is possible. We have the following lemmas, whose proof follows the proofs of Lemma 3 and Lemma 4 in~\cite{ChazelleR91}.

As in~\cite{ChazelleR91}, we stratify $s_j$'s according to whether they consist of ``cross-block'' compositions. Specifically, each of the $s_j$ in scheme $S$ falls into one of $b+1$ categories. Either, for some $i \in[0, b-1]$, $s_j$ lies fully inside block $i, s_j \subseteq\left[x_{i a}, x_{(i+1) a-1}\right]$, or $s_j$ combines elements from different blocks. We partition $S$ into $b+1$ sublists according to this categorization. If $s_j$ lies inside block $i$, place $s_j$ in $S^i$, else place $s_j$ in $S^b$. If $s_j$ is the empty set, place it in the subsequence $S^b$. Maintain the original ordering, but renumber, to obtain sublists $\left\{s_0^i, s_1^i, \ldots, s_{r_i-1}^i\right\}$ where $i=0, \ldots, b$, each $s_j^i$ is an element of $\mathcal{P}(X)$, and the $r_i$ are the lengths of these lists, noting $r=r_0+\cdots+r_b$.

\begin{lemma}\label{lm:solving_A}
For $i=0, \ldots, b-1$, the sequences $\varphi\left(S^i\right)$, defined by $\varphi\left(S^i\right)_j=\varphi\left(s_j^i\right)$, are schemes solving $A$.
\end{lemma}

Before presenting the proof, we note an important difference between the proof of Lemma~\ref{lm:solving_A} in the $\oeps$ setting and the proof of Lemma 3 in~\cite{ChazelleR91} (although the difference is only mentally significant while the proof is almost verbatim). In the proof of Lemma 3 in~\cite{ChazelleR91}, there is an exact correspondence between the sequences $\varphi\left(S^i\right)$ where $i=0, \ldots, b-1$ and the solution of $A$, in the sense that $\varphi\left(S^i\right)$ is a solution for $A$ while its preimage $S^i$ is also a solution for $\varphi_i(A)$ in $Q$. However, in $\oeps$ setting, although $\varphi\left(S^i\right)$ is a solution for $A$, its preimage $S^i$ is not a solution for $\varphi_i(A)$ in $Q$, since by the definition of $\oeps$ a solution for the latter needs to compute sums outside the block $i$. It is in this precise sense that giving the unconditional lower bound for the off-line mode partial sum, no ``magical shortcut'' is possible.

\begin{proof}[Proof of Lemma~\ref{lm:solving_A}]
Let $s_\alpha$ be any member of $\varphi\left(S^i\right)$. Then $s_\alpha$ is the image of some element $s_j$ in $S$. We demonstrate the case $s_j=s_k \cup s_l$ with $k, l<j$. Clearly, $s_k$ and $s_l$ fall into $S^i$ with images $s_\beta$ and $s_\gamma$ in $\varphi\left(S^i\right)$, where $\beta, \gamma<\alpha$. But
$$
s_\alpha=\varphi\left(s_j\right)=\varphi\left(s_k \cup s_l\right)=\varphi\left(s_k\right) \cup \varphi\left(s_l\right)=s_\beta \cup s_\gamma
$$
The other possible precursors of $s_j$ are argued similarly, proving that $\varphi\left(S^i\right)$ is a scheme. We know that $\varphi_i$ is a section of $\varphi$ and that $S$ solves $Q$. This implies $S^i \supseteq \varphi_i\left(\mathcal{T}_a(t, k-1)\right)$, giving
$$
\varphi\left(S^i\right) \supseteq \varphi \circ \varphi_i\left(T_a(t, k-1)\right)=T_a(t, k-1)
$$
So $\varphi\left(S^i\right)$ solves $T_a(t, k-1)$, i.e. $A$.
\end{proof}

\begin{lemma}
The sequence $\psi\left(S^b\right)$, defined by
\begin{enumerate}
\item $\psi\left(S^b\right)_{i+1}=\psi\left(s_i^b\right)$,
\item $\psi\left(S^b\right)_0=\left[z_{b-2}, z_{b-1}\right]$
\end{enumerate}
is a scheme solving $B$. It is not minimal: there is a subsequence of $\psi\left(S^b\right)$, resulting from the removal of $|B|-\left|B_0\right|$ elements from $\psi\left(S^b\right)$, which is also a scheme solving B.
\end{lemma}
\begin{proof}
The fact that $\psi\left(S^b\right)$ is a scheme that solves $B$ is similar as the proof of Lemma~\ref{lm:solving_A}.

Let $q$ be an interval in the task $Q$ of the form $\psi_i(z)$, but $i \neq 0$. That is, $q$ is in the image of $B$, but not of that portion of $B$ placed in the partition $B_0$. Let $i(q)$ be the index of the leftmost $q_i$ in $q$. Let $W(i)$ be the index of the first element in $S^b$ which contains $q_i$ but contains no $q_j$ with $j<i$. Since $S$ solves $Q, W(i)$ is defined. We consider the equation $s_{W(i)}^b=s^{\prime} \cup s^{\prime \prime}$. At least one of $s^{\prime}$ and $s^{\prime \prime}$ contains $q_i$; without loss of generality assume it is $s^{\prime}$. By selection of $W(i), s^{\prime}$ is itself contained in the block containing $q_i$. Because $i$ is not divisible by $a$, the image of this set under $\psi$ must be the empty set. (It cannot be that this set contains $x_{n-1}$.) Hence $\psi\left(s_{W(i)}^b\right)$ either appeared before in $\psi\left(S^b\right)$ or, because $\psi\left(S^b\right)$ is a scheme, it is a singleton. In either case, we can remove this element from the sequence $\psi\left(S^b\right)$ and it still will be a scheme. After removal, it still will solve $B$ since that set contains no singletons.
\end{proof}


Following~\cite{ChazelleR91}, a lower bound on the length of $S$ can be derived. Because $\varphi\left(S^i\right)$ solves $A$, $r_i \geq t|A| / 3, i=0, \ldots, b-1$. Taking the indicated subsequence of $\psi\left(S^b\right)$, we have a scheme of size $r_b+1-|B|+\left|B_0\right|$ solving $B$. Therefore, that sum is bound below by $(t-1)|B| / 3$. Recall that each interval in $B_0$ ends over one of $z_0, \ldots, z_{b-3}$, and conversely, each $z_0, \ldots, z_{b-3}$ has at most one interval in $B_0$ ending over it. Hence $\left|B_0\right| \leq b-2$. Since $|Q|=b|A|+|B|$
$$
\begin{aligned}
r & =r_0+\cdots+r_{b-1}+r_b \\
& \geq b t|A| / 3+(t-1)|B| / 3+|B|-\left|B_0\right|-1 \\
& \geq b t|A| / 3+(t+2)|B| / 3-(b-2)-1 \\
& =t|Q| / 3+(2 / 3)|B|-b+1
\end{aligned}
$$
Because $B=\mathcal{T}_b(t-1, a)$ we have $|B| \geq a b / 2$. Since $a \geq 3$
$$
r \geq t|Q| / 3+a b / 3-b+1>t|Q| / 3
$$
We have completed the verification of the three properties of $T_n(t, k)$ listed in Lemma~\ref{lm:requirements}, and the induction step is complete.

Define the function $A(i,j)$ for all $i \geq 1$ and $j \geq 0$ as in~\cite{Tarjan75}.
$$
\begin{array}{ll}
A(1, j)=2^j, & j \geq 0 \\
A(i, 0)=2, & i>1 \\
A(i, j)=A(i-1, A(i, j-1)), & j>0, i>1 .
\end{array}
$$

The following two lemmas are proved in~\cite{ChazelleR91} relating $R(i,j)$ and $A(i,j)$.

\begin{lemma}
For all $i, j=1,2, \ldots$, we have $R(i+1, j)>A(i, j)$.
\end{lemma}

\begin{lemma}
For all $i, j=1,2, \ldots$, we have $A(i+2, j)>R(i, j)$.
\end{lemma}

Finally, we proceed as in~\cite{ChazelleR91} to finish the preparation for proving Theorem~\ref{thm:unconditional-lb}. Suppose we are given $m$ and $n$ with $m \geq n$. Set $k=\lfloor m / n\rfloor$ and let $t$ be the least integer such that $R(t, k)>n$. We explain why $t \geq 2$. A task cannot repeat a query interval, so $m \leq n(n-1) / 2$, and hence,
$$
R(1, k)=2 k \leq 2\lfloor m / n\rfloor \leq n-1
$$

By the lower bound lemma, there exists a task $T=T_{n^{\prime}}(t-1, k)$ with
\begin{itemize}
\item $n^{\prime}=R(t-1, k)$, and it follows by the definition of $t$ that $n^{\prime}<n$.
\item $T$ has size $|T|$ between $k n^{\prime} / 2$ and $k n^{\prime}$.
\item Any solution to $T$ has length at least $(t-1)|T| / 3$.
\end{itemize}
Place $\left[n / n^{\prime}\right]$ copies of $T$ side by side. Add extra variables and queries to correct to form of this resultant task, that is, it will be over $n$ variables and have $m$ queries. The size of any solution is bounded from below by:
$$
\begin{aligned}
(t-1)\left\lfloor n / n^{\prime}\right\rfloor|T| / 3 & \geq(t-1)\left\lfloor n / n^{\prime}\right\rfloor\lfloor m / n\rfloor n^{\prime} / 6 \\
& \geq m(t-1) / 24 \\
& \geq m t / 48 .
\end{aligned}
$$
The following chain of inequalities shows that $t+2 \geq \alpha(m, n)$,
$$
A(t+2, k)>R(t, k)>n>\log n
$$
Because $t \geq 2$, it follows that $t \geq \alpha(m, n) / 2$. Therefore the cost of solving $T$ is at least $m \alpha(m, n) / 96$.

Consider now $m$ and $n$ given, where $m<n$. The previous paragraphs shows the existence of a task $T$ over $m$ variables with $m$ queries which takes time at least $(m \alpha(m, m)) / 96$ to solve. If all the variables $x_0, \ldots, x_{m-1}$ do not appear in $T$, eliminate the unused variables and renumber as $x_0, \ldots, x_{m-1}$. Add variables $x_{m^{\prime}}, \ldots, x_{n-1}$. Find an interval of the form $\left[i, m^{\prime}-1\right]$ in $T$, as $x_{m^{\prime}-1}$ is used in $T$ such an interval exists, and replace it with the interval $[i, n-1]$. Any solution to the resulting task can be made to solve the original task $T$ by a transformation which includes removing the at least $n-m$ computation steps referencing variables with indices inside the interval $[m, n-1]$. So, the time to solve this new task must be,
$$
\begin{aligned}
n-m+m \alpha(m, m) / 96 & \geq(n-m) / 192+m \alpha(m, m) / 96 \\
& \geq n / 192+m(2 \alpha(m, m)-1) / 192 \\
& \geq(n+m \alpha(m, m)) / 192.
\end{aligned}
$$

\rebut{We are now ready to prove Theorem~\ref{thm:unconditional-lb}.}

\begin{proof}[Proof of Theorem~\ref{thm:unconditional-lb}]

We show that any constant delay enumeration algorithm for $\varphi$ with preprocessing time $T$ leads to an algorithm for \oeps\ that runs in time $T + O(m+n)$.

Let $a_1, a_2, \dots, a_n$ and $(l_1, r_1)$, $(l_2, r_2)$, $\dots$, $(l_m, r_m)$ be inputs to the $\mathsf{OpenEndedPartialSum}$ problem where $m \leq n$. 

We first construct factors $A$, $B$, and $R$ as follows: Fix a domain $D = \{1,2,\dots,n\}$, and
\begin{itemize}
\item for every $i \in D$, $A(i) = \mathbf{1}$, the multiplicative identity under the semiring;
\item for every $j \in D$, $B(j) = a_j$;
\item for every $k \in \{1,2,\dots, m\}$, $R(k, l_k) = \mathbf{1}$ and $R(k, r_k) = \mathbf{1}$.
\end{itemize}
This construction runs in time $O(n + 2m)$.
Then, we simply compute $\varphi$ in time $T + O(n)$. 
Note that for any $i \in \{1,2,\dots,m\}$, the open ended sum for $(l_i, r_i)$ is stored in $\phi(i)$: 
$$\varphi(i) = \bigoplus_{1\leq j\leq n} A(i) \otimes B(j) \otimes \indicator{\lnot R}(i, j) 
= \bigoplus_{j \in [n] \setminus \{l_i,  r_i\}} a_j.$$

Hence the overall running time of this algorithm is $T + O(n + 2m)$.
By Theorem~\ref{thm:unconditional-lb}, we have that $T + O(n + 2m) = \Omega(n + m \cdot \alpha(m, m))$.
Therefore, $T = \Omega(m \cdot \alpha(m, m))$, as desired.
\end{proof}

\section{Missing proofs in Section~\ref{sec:beyond}}
\label{appendix:beyond}

\begin{proof}[Proof of Theorem~\ref{thm:cq:diff}]
    Following~\cite{CQDiff}, we can write $Q$ as follows:
    $$ Q = \bigcup_{R_e \in Q_2} (\bigwedge_{S_e \in Q_1} S_e \wedge \neg R_e)$$
    Hence, $Q$ can be viewed as a union of $\cqneg$, each $\cqneg$ of the form $Q_1 \wedge \neg R_e$. Since $Q_1$ is $\alpha$-acyclic and $Q_1 \wedge R_e$ is $\alpha$-acyclic,  $Q_1 \wedge \neg R_e$ is $\pn$-acyclic and free-connex (trivially, since it is full) and thus by Theorem~\ref{thm:boolean} can be enumerated with linear-time preprocessing and constant delay. We can now apply the Cheater's Lemma~\cite{cheater21} to obtain linear-time preprocessing and constant-delay enumeration for $Q$.
\end{proof}

   

\begin{proof}[Proof of Corollary~\ref{cor:cq:diff}]
    From~\cite{CQDiff}, we can reduce $Q$ to a difference $Q_1'- Q_2'$ in linear time such that $Q_1', Q_2'$ are full CQs and satisfy the properties of Theorem~\ref{thm:cq:diff}. We then simply apply Theorem~\ref{thm:cq:diff}.
\end{proof}

\section{$\cqneg$ with Bounded Width} \label{appendix:width}
What can we say about the data complexity of evaluating a $\cqneg$ if it is not $\pn$-acyclic? Lanzinger~\cite{Lanzinger21} defined a width measure of a $\cqneg$ called {\em nest-set width ($\mathsf{nsw}$)} and showed that a Boolean $\cqneg$ with bounded $\mathsf{nsw}$ can be computed in polynomial time (combined complexity). One consequence of \eqref{eq:inc-excl} is that it is possible to obtain in a straightforward way a much tighter upper bound in terms of data complexity.

\begin{definition}
Let $\varphi$ be a $\cqneg$ with signed hypergraph $\mH = ([n], \mE^+, \mE^-)$. Then, $\beta$-$\#\subw(\varphi) = \max_{S \subseteq \mE^-} \#\subw(\varphi_S)$, where $\varphi_S$ is the CQ with hypergraph $([n], \mE^+ \cup S)$ and $\#\subw(\varphi)$ is the \textit{counting} version of the submodular width defined in \cite{FAQ-AI}.
\end{definition}

Thanks to ~\cite{PANDA, FAQ-AI}, for any Boolean CQ $\varphi$, we can compute $\#\varphi$ in time $\tilde{O}(|\db|^{\#\subw(\varphi)})$, where $\tilde{O}$ hides a polylogarithmic factor in $|\db|$. Combining this with \eqref{eq:inc-excl}, we have:  

\begin{theorem}
Let $\varphi$ be a Boolean $\cqneg$. Then, we can compute $\# \varphi$ (and thus $\varphi$) in time $\tilde{O}(|\db|^{\beta\text{-}\#\subw(\varphi)})$
\end{theorem}

From~\cite{Lanzinger21}, we know that $\beta$-$\#\subw$ is always at most the nest-set width; moreover, there are queries were nest-set width is unbounded but $\beta$-$\#\subw$ is bounded.

\end{document}